\documentclass[11pt,a4paper]{article}
\usepackage{jheppub,pifont}

\newcommand{\tick}{{\color{blue}\ding{52}}}
\newcommand{\cross}{{\color{red}\ding{55}}}

\title{Parton distribution function dependence of benchmark Standard Model total cross sections at the 7~TeV LHC}

\author{G.~Watt}
\affiliation{Theory Group, Physics Department, CERN, CH-1211 Geneva 23, Switzerland}
\emailAdd{Graeme.Watt@cern.ch}

\abstract{We compare predictions for the $W$, $Z$, $gg\to H$ and $t\bar{t}$ total cross sections at the Large Hadron Collider (LHC), for a centre-of-mass energy of 7~TeV, using the most recent publicly available next-to-leading order and next-to-next-to-leading order parton distribution functions (PDFs) from all PDF fitting groups.  In particular, we focus on the dependence on the different values of the strong coupling, $\alpha_S(M_Z^2)$, used by each group.  We also perform a comparison of the relevant quark--antiquark and gluon--gluon luminosity functions.  We make some comments on the recent PDF4LHC recommendations.  Finally, we discuss the comparison of data and theory for $W$ and $Z$ cross sections at the LHC.}


\arxivnumber{1106.5788}

\begin{document}

\begin{flushright}
  CERN-PH-TH/2011-149 \\
  28th June 2011
\end{flushright}

\maketitle

\section{Introduction} \label{sec:introduction}

The first proton run at a centre-of-mass energy ($\sqrt{s}$) of 7~TeV at the Large Hadron Collider (LHC) took place from 30th March 2010 until 4th November 2010, with almost 50~pb$^{-1}$ of integrated luminosity delivered to the ATLAS and CMS detectors, half of it in one week at the end of October.  The first measurement of $W$ and $Z$ cross sections was reported by both ATLAS and CMS at the ICHEP conference in July 2010, and subsequently published with integrated luminosities of approximately 320~nb$^{-1}$ for ATLAS~\cite{Aad:2010yt} and 2.9~pb$^{-1}$ for CMS~\cite{Khachatryan:2010xn}.  Preliminary results with the full 2010 data set were presented in March 2011 by both ATLAS~\cite{ATLAS:WandZ} and CMS~\cite{CMS:WandZ}.  (The LHCb experiment has also presented preliminary results for the $W$ and $Z$ cross sections in the forward region~\cite{LHCb:WandZ}.)  The observation of the first few top candidate events was also reported at the ICHEP conference, and first $t\bar{t}$ cross-section measurements were published with integrated luminosities of 2.9~pb$^{-1}$ for ATLAS~\cite{Aad:2010ey} and 3.1~pb$^{-1}$ for CMS~\cite{Khachatryan:2010ez}.  Again, preliminary results with the full 2010 data set were presented in March 2011~\cite{ATLAS:ttbar,CMS:ttbar} and CMS have since published their measurements in the dilepton channel~\cite{Chatrchyan:2011nb} and the lepton+jets channel~\cite{Chatrchyan:2011ew}.  The precision of the cross sections for $W$, $Z$ and $t\bar{t}$ production will improve substantially in the near future as the 2011 LHC run progresses at an impressive pace, with the remarkable milestone of 1~fb$^{-1}$ of integrated luminosity already delivered to the ATLAS and CMS detectors by 17th June 2011.  Searching for, or excluding, the Standard Model Higgs boson ($H$), produced mainly from gluon--gluon fusion through a top-quark loop, requires precise knowledge of the theoretical cross section.  All these benchmark processes ($W$, $Z$, $t\bar{t}$ and $gg\to H$) are sensitive to the parton distribution functions (PDFs) of the proton.  The proton PDFs are determined by several groups from (global) analysis of a wide range of deep-inelastic scattering (DIS) and related hard-scattering data, taken at the HERA and Tevatron colliders and from fixed-target experiments at lower centre-of-mass energies; see refs.~\cite{Forte:2010dt,Alekhin:2011sk,DeRoeck:2011na} for recent reviews.

The aim of this paper is to compare next-to-leading order (NLO) and next-to-next-to-leading order (NNLO) predictions, calculated within the standard framework of collinear factorisation, for the benchmark processes using the most recent publicly available PDFs from each group, all available using the current \textsc{lhapdf} V5.8.5~\cite{Whalley:2005nh} interface.  We will pay particular attention to the values of the strong coupling, $\alpha_S$, used by each fitting group and to the corresponding $\alpha_S$ dependence of the benchmark cross sections.  We do \emph{not} aim to come up with a single ``best'' prediction for each of the benchmark cross sections together with a complete evaluation of all sources of theoretical uncertainty.  We consider only uncertainties due to PDFs and $\alpha_S$ on fixed-order (NLO and NNLO) predictions.  We do \emph{not} consider, for example, optimal (factorisation and renormalisation) scale choices and variations, electroweak corrections, or the effect of threshold resummation.  For the $gg\to H$ process, we do \emph{not} consider $(C_A\,\pi\,\alpha_S)^n$-enhanced terms, use of a finite top-quark mass in the calculation of higher-order corrections, bottom-quark loop contributions, etc.  Discussion of these issues can be found in the recent \emph{Handbook of LHC Higgs Cross Sections}~\cite{LHCHiggsCrossSectionWorkingGroup:2011ti}.  The findings regarding PDFs and $\alpha_S$ reported here will be relevant also for more complete calculations, and the PDF and $\alpha_S$ uncertainties often make up a sizeable or dominant part of the total theoretical uncertainty.  We will present results only for total cross sections rather than differential distributions, and we do not attempt to account for the precise experimental cuts, although for fairly inclusive cuts the results should be similar.  The findings for $\sqrt{s}=7$~TeV will be similar also for slightly higher energies, should this happen before the planned LHC shutdown in 2013.  Significant updates to the PDFs are expected by the time the LHC runs at up to $\sqrt{s}=14$~TeV, expected from 2014, therefore we do not present results at that centre-of-mass energy in this paper.

The background leading up to this paper is as follows.  The \emph{LHC Higgs Cross Section Working Group}~\cite{LHCHiggs:2010} was formed in January 2010 and requested a recommendation from the \emph{PDF4LHC Working Group}~\cite{PDF4LHC} on PDFs and $\alpha_S$ values (and their uncertainties) to be used for cross-section calculations.  A proposal was made, initially by J.~Stirling, to use different PDFs to calculate LHC benchmark processes at 7~TeV, specifically production of $W^\pm$, $Z^0$, $t\bar{t}$ and $gg\to H$ for $M_H=120,180,240$~GeV.  The aims were to establish the degree of compatibility and identify outliers amongst PDF sets, and to compare cross sections at the \emph{same} $\alpha_S$ values, thereby showing to what extent differences in predictions are due to the different $\alpha_S$ values adopted by each group, rather than differences in the PDFs themselves.  The results at NLO, initially presented by G.W.~in a PDF4LHC meeting at CERN on 26th March 2010~\cite{PDF4LHC:March2010}, formed the basis for the subsequent PDF4LHC \emph{Interim Report}~\cite{Alekhin:2011sk} and PDF4LHC \emph{Interim Recommendations}~\cite{Botje:2011sn} used in the \emph{Handbook of LHC Higgs Cross Sections}~\cite{LHCHiggsCrossSectionWorkingGroup:2011ti}.

In this paper, we extend\footnote{Much of the new material was initially presented in some recent talks~\cite{LHQCD11:February2011,PDF4LHC:March2011,LHCEWWG:April2011}.} the comparison of benchmark cross sections in a number of ways compared to the original PDF4LHC exercise~\cite{PDF4LHC:March2010,Alekhin:2011sk}.  We will account for new NLO PDF sets released since March 2010, particularly the new CT10~\cite{Lai:2010vv} and NNPDF2.1~\cite{Ball:2011mu} analyses, and we extend all comparisons to NNLO,\footnote{See also ref.~\cite{Alekhin:2010dd} for NNLO comparisons of $W$, $Z$ and Higgs production with different PDFs.} making use of the recently released \textsc{hathor} public code~\cite{Aliev:2010zk} for an approximate NNLO $t\bar{t}$ calculation.  A companion webpage~\cite{mstwpdf} will be used as a repository for supplementary plots and may be updated after the publication of this paper, for example, as new PDF sets continue to be released.  It is hoped that this paper and the companion webpage will prove to be a useful resource to accompany the publication of precise measurements of LHC cross sections from the 7~TeV proton runs.  In a separate but related paper~\cite{higgs}, we comment on the ability of the Tevatron jet data to discriminate between PDF sets with different high-$x$ gluon distributions and we respond to some recent criticism~\cite{Baglio:2010um,Baglio:2010ae,Baglio:2011wn,Alekhin:2010dd,Alekhin:2011ey,Alekhin:2011cf} regarding theoretical uncertainties due to PDFs (and $\alpha_S$) in the $gg\to H$ cross section at the Tevatron (and LHC).

The structure of this paper is as follows.  In section~\ref{sec:pdfsets} we briefly describe the PDF sets considered in the study and specify formulae used for uncertainty calculations, in section~\ref{sec:luminosities} we compare partonic luminosities, in section~\ref{sec:xsasmz} we plot total cross sections versus $\alpha_S(M_Z^2)$, in section~\ref{sec:pdf4lhc} we make some comments on the PDF4LHC recommendations, in section~\ref{sec:wandz} we discuss $W$ and $Z$ production\footnote{The PDF dependence of $W$ and $Z$ production at the LHC was also previously discussed in refs.~\cite{Martin:1999ww,Thorne:2008am}.} in more detail, then finally we conclude in section~\ref{sec:conclusions}.

\section{PDF sets considered} \label{sec:pdfsets}
\begin{table}
\centering
{\tiny
  \begin{tabular}{|l|c|c|c|c|c|c|}
    \hline
    & {\bf MSTW08} & {\bf CTEQ6.6/CT10} & {\bf NNPDF2.1} & HERAPDF1.0/1.5 & ABKM09 & GJR08/JR09 \\
    \hline
        { HERA DIS} & \tick & \tick & \tick & \tick & \tick & \tick \\
        { Fixed-target DIS} & \tick & \tick & \tick & \cross & \tick & \tick \\
        { Fixed-target DY} & \tick & \tick & \tick & \cross & \tick & \tick \\
        { Tevatron $W$,$Z$} & \tick & \tick & \tick & \cross & \cross & \cross \\
        { Tevatron jets} & \tick & \tick & \tick & \cross & \cross & \tick \\ \hline
        { GM-VFNS} & \tick & \tick & \tick & \tick & \cross & \cross \\
        { NNLO} & \tick & \cross & \cross & \tick & \tick & \tick \\
        \hline
  \end{tabular}
}
  \caption{Comparison of major PDF sets considered, and their gross features distinguished by the main classes of data included (upper part of table) and important aspects of the theoretical treatment (lower part of table), specifically regarding the treatment of heavy quarks in DIS and the provision of NNLO PDFs.  More refined differences between PDF sets are described in the text.\label{tab:compare}}
\end{table}
We will consider only \emph{public} sets, defined to be those available for use with the latest \textsc{lhapdf} V5.8.5~\cite{Whalley:2005nh}.  We consider NLO PDFs from MSTW08~\cite{Martin:2009iq}, CTEQ6.6~\cite{Nadolsky:2008zw}, CT10~\cite{Lai:2010vv}, NNPDF2.1~\cite{Ball:2011mu}, HERAPDF1.0~\cite{HERA:2009wt}, HERAPDF1.5 (preliminary)~\cite{HERA:2010}, ABKM09~\cite{Alekhin:2009ni} and GJR08~\cite{Gluck:2007ck,Gluck:2008gs}.  We consider NNLO PDFs from MSTW08~\cite{Martin:2009iq}, HERAPDF1.0~\cite{HERA:2009wt}, ABKM09~\cite{Alekhin:2009ni} and JR09~\cite{JimenezDelgado:2008hf,JimenezDelgado:2009tv}.  The broad distinctions between data sets fitted and aspects of the theoretical treatment are summarised in table~\ref{tab:compare}; see also the descriptions in refs.~\cite{Forte:2010dt,Alekhin:2011sk,DeRoeck:2011na}.  Only three groups (MSTW08, CTEQ6.6/CT10 and NNPDF2.1) make fully \emph{global} fits to HERA and fixed-target DIS data, fixed-target Drell--Yan production, and Tevatron data on $W$, $Z$ and jet production, although GJR08 includes all these processes other than Tevatron $W$ and $Z$ production.  The HERAPDF1.0 fit includes \emph{only} the combined HERA I inclusive data, while the HERAPDF1.5 fit additionally includes the preliminary combined HERA II inclusive data.  The CT10 and NNPDF2.1 global fits include the combined HERA I inclusive data~\cite{HERA:2009wt}, while the other global fits include the older separate data from H1 and ZEUS.  The MSTW08, CT10, NNPDF2.1 and GJR08 fits include Tevatron Run II data, while CTEQ6.6 uses only Tevatron Run I data.  Most groups now treat the heavy-quark contribution to DIS structure functions using a general-mass variable flavour number scheme (GM-VFNS), other than ABKM09 and GJR08/JR09 who use a fixed flavour number scheme (FFNS).  The change from the inadequate zero-mass variable flavour number scheme (ZM-VFNS) to the GM-VFNS was the major improvement between NNPDF2.0~\cite{Ball:2010de} and NNPDF2.1~\cite{Ball:2011mu}, now allowing a meaningful comparison to other NLO global fits.  The NNPDF fits parameterise the starting distributions at $Q_0^2=2$~GeV$^2$ as neural networks and use Monte Carlo methods for experimental error propagation.  The other groups all use the more traditional approach of parameterising the input PDFs as some functional form in $x$, each with a handful of free parameters, and use the Hessian method for experimental error propagation with differing values of the tolerance $\Delta\chi^2$, that is, the change in the global goodness-of-fit measure relative to the best-fit value.  Contrary to the ``standard'' input parameterisation at $Q_0^2\ge1$~GeV$^2$, the GJR08/JR09 sets use a ``dynamical'' input parameterisation of valence-like input distributions at an optimally chosen $Q_0^2<1$~GeV$^2$, which gives a slightly worse fit quality and lower $\alpha_S$ values than the corresponding ``standard'' parameterisation, but is nevertheless favoured by the GJR08/JR09 authors.  Public NNLO fits are only available from MSTW08, HERAPDF1.0, ABKM09 and JR09.  (The first NNLO fits from the CTEQ/CT and NNPDF groups should be available soon, together with HERAPDF1.5 NNLO, but these analyses are not yet published.)  The Tevatron jet cross sections are excluded from the JR09 fit, where complete NNLO corrections are unavailable, whereas they are included in the MSTW08 NNLO fit by making the approximation of using the NLO partonic cross section supplemented by 2-loop threshold corrections~\cite{Kidonakis:2000gi}.

\subsection{Definition of PDF uncertainties}

All groups other than NNPDF propagate experimental errors on the fitted data points by diagonalising the $n\times n$ covariance (or Hessian) matrix, with $n$ free parameters, and provide $2n$ eigenvector PDF sets, $S_k^\pm$ ($k=1,\ldots,n$); see the discussion in ref.~\cite{Alekhin:2011sk} for more details.  We use asymmetric uncertainties (apart from for ABKM09, where $S_k^+ = S_k^-$) to calculate the PDF uncertainty, denoted $(\Delta F)_\pm$, on an observable $F$ such as a hadronic cross section:
\begin{align}
  (\Delta F)_+ &= \sqrt{\sum_{k=1}^n \left\{{\rm max}\left[\;F(S_k^+)-F(S_0),\;F(S_k^-)-F(S_0),\;0\right]\right\}^2}, \label{eq:plusHess} \\
  (\Delta F)_- &= \sqrt{\sum_{k=1}^n \left\{{\rm max}\left[\;F(S_0)-F(S_k^+),\;F(S_0)-F(S_k^-),\;0\right]\right\}^2}, \label{eq:minusHess}
\end{align}
where $n=20,22,26,10,25,13$ for MSTW08, CTEQ6.6, CT10, HERAPDF1.0/1.5, ABKM09, GJR08/JR09, respectively.  The HERAPDF1.0/1.5 fit provides additional ``model'' and ``parameterisation'' asymmetric errors, which are added in quadrature to the ``experimental'' errors.  The HERAPDF1.0 NNLO fit does not include any uncertainties, only the central fit for two values of $\alpha_S(M_Z^2) = 0.1145$ and $0.1176$.

For NNPDF2.1, instead the average $\langle F\rangle$ and standard deviation $\Delta F$ are taken over the observable $F$ calculated with each PDF replica set, $S_k$ ($k=1,\ldots,N_{\rm rep}=100$), that is,
\begin{align}
  \langle F\rangle = \frac{1}{N_{\rm rep}}\sum_{k=1}^{N_{\rm rep}}F(S_k), \label{eq:MCav} \\
  \Delta F = \sqrt{\frac{N_{\rm rep}}{N_{\rm rep}-1}\left(\langle F^2\rangle-\langle F\rangle^2\right)}. \label{eq:MCsd}
\end{align}

For MSTW08, eigenvector PDF sets are consistently provided for both the estimated 68\% confidence-level (C.L.)~and 90\% C.L.~uncertainties.  For other groups, where usually only 68\% is provided (but only 90\% for CTEQ6.6/CT10) we assume that the 68\% C.L.~and 90\% C.L.~uncertainties are related by a factor of 1.64485.  For NNPDF2.1, while it is possible to calculate exact confidence levels (within the limited statistics), we will assume that a standard deviation, $\Delta F$ given by eq.~\eqref{eq:MCsd}, represents a 68\% C.L.~uncertainty and then rescale by 1.64485 to get a 90\% C.L.~uncertainty where necessary.  We will sometimes only show 68\% C.L.~plots in this paper, but the corresponding plots at 90\% C.L.~can be seen at ref.~\cite{mstwpdf}.  The PDF uncertainties from HERAPDF1.0/1.5 and ABKM09 include contributions from variation of the heavy-quark masses, $m_c$ and $m_b$, while these are not included for CTEQ6.6, CT10 and GJR08/JR09.  For MSTW08 and NNPDF2.1, the uncertainty from the choice of $m_c$ and $m_b$ will not be included here, although it \emph{can} be included in principle according to the prescriptions presented in refs.~\cite{Martin:2010db,Ball:2011mu}.  Variations of $m_c=1.40\pm0.15$~GeV led to a change of just over 1\% in the $W$ and $Z$ total cross sections for MSTW08, while varying $m_b=4.75\pm0.25$~GeV gave a negligible change (0.1\%), such that the $m_{c,b}$ uncertainties are not important when added in quadrature with the 2--3\% ``PDF+$\alpha_S$'' uncertainty~\cite{Martin:2010db}; see also tables~\ref{tab:WZuncertainties} and \ref{tab:PMuncertainties} in section~\ref{sec:wandz}.  Similar variations of $m_c$ and $m_b$ for Standard Model Higgs boson production ($M_H=120$~GeV) at the 7~TeV LHC had essentially no impact~\cite{Martin:2010db}.  Additional uncertainties from the particular GM-VFNS variant are estimated~\cite{Thorne:2010pa} to be $\sim2\%$ at NLO, but less than 1\% at NNLO, and will not be considered here.  Combined HERA I data~\cite{HERA:2009wt}, not included in the MSTW08 fit, gave an upwards shift in the $W$ and $Z$ total cross sections by 1--2\%, with a smaller change in the Higgs cross section~\cite{Thorne:2010kj}.  The combined HERA I and II inclusive data, together with the combined charm structure function $F_2^{\rm charm}$, will be included in a future update to MSTW08, after it is finally published.

\subsection{Definition of PDF correlations} \label{sec:correlations}

Defining a correlation cosine between two quantities $F$ and $G$ in the Hessian approach,
\begin{equation} \label{eq:cosphiHess}
  \cos\phi_{FG} = \frac{1}{4\,\Delta F\,\Delta G}\,\sum_{k=1}^n\,\left[F(S_k^+)-F(S_k^-)\right]\,\left[G(S_k^+)-G(S_k^-)\right],
\end{equation}
where the uncertainties $\Delta F$ and $\Delta G$ are calculated using the symmetric formula:
\begin{equation} \label{eq:symmunc}
  \Delta F = \frac{1}{2}\sqrt{\sum_{k=1}^n \left[F(S_k^+)-F(S_k^-)\right]^2},
\end{equation}
then values of $\cos\phi_{FG}\approx 1$ mean that $F$ and $G$ are correlated, values of $\approx -1$ mean that they are anticorrelated, while values of $\approx 0$ mean that they are uncorrelated.  In the Monte Carlo approach used by the NNPDF group, the equivalent of eq.~\eqref{eq:cosphiHess} is given by
\begin{equation}
  \cos\phi_{FG} = \frac{N_{\rm rep}}{N_{\rm rep}-1}\frac{\langle FG\rangle-\langle F\rangle\langle G\rangle}{\Delta F\,\Delta G},
\end{equation}
where the average $\langle F\rangle$ and standard deviation $\Delta F$ are defined in eqs.~\eqref{eq:MCav} and \eqref{eq:MCsd}, respectively.  In both the Hessian and Monte Carlo approaches, an ellipse in the $F$--$G$ plane can then be defined by the two parametric equations:
\begin{align}
  F &= F(S_0) + \Delta F\,\cos\theta, \label{eq:ellipseF} \\
  G &= G(S_0) + \Delta G\,\cos\left(\theta+\phi_{FG}\right), \label{eq:ellipseG}
\end{align}
where $\theta\in [0,2\pi]$ and $F(S_0)$ should be replaced by $\langle F\rangle$ in the Monte Carlo approach.  We give explicit examples in section~\ref{sec:wandz} for the $W$ and $Z$ total cross sections at the LHC.  Finally, we note that the uncertainty in a general function of the two quantities, $H(F,G)$, is given by (see also section 4 of ref.~\cite{Alekhin:2011sk}):
\begin{equation} \label{eq:uncertaintyH}
  \left(\Delta H\right)^2 = \left(\Delta F\frac{\partial H}{\partial F}\right)^2+\left(\Delta G\frac{\partial H}{\partial G}\right)^2+2\,\Delta F\,\Delta G\,\cos\phi_{FG}\,\frac{\partial H}{\partial F}\frac{\partial H}{\partial G}.
\end{equation}
As a specific example, the fractional uncertainty in the ratio of the two quantities $R\equiv F/G$ is therefore given by
\begin{equation} \label{eq:uncertaintyR}
  \left(\frac{\Delta R}{R}\right)^2 = \left(\frac{\Delta F}{F}\right)^2+\left(\frac{\Delta G}{G}\right)^2-2\,\frac{\Delta F\,\Delta G}{F\,G}\,\cos\phi_{FG},
\end{equation}
so the uncertainty in the ratio can be obtained from the correlation cosine (or vice versa).  Of course, eqs.~\eqref{eq:ellipseF}, \eqref{eq:ellipseG}, \eqref{eq:uncertaintyH} and \eqref{eq:uncertaintyR} hold much more generally than just for PDF uncertainties and can be applied to define an ellipse also for experimental data; see ref.~\cite{LHCEWWG:April2011}.

\subsection{Values\texorpdfstring{ and uncertainties}{} of strong coupling\texorpdfstring{ $\alpha_S$}{}} \label{sec:alphaS}

\begin{figure}
  \centering
  \begin{minipage}{0.5\textwidth}
    (a)\\
    \includegraphics[width=\textwidth]{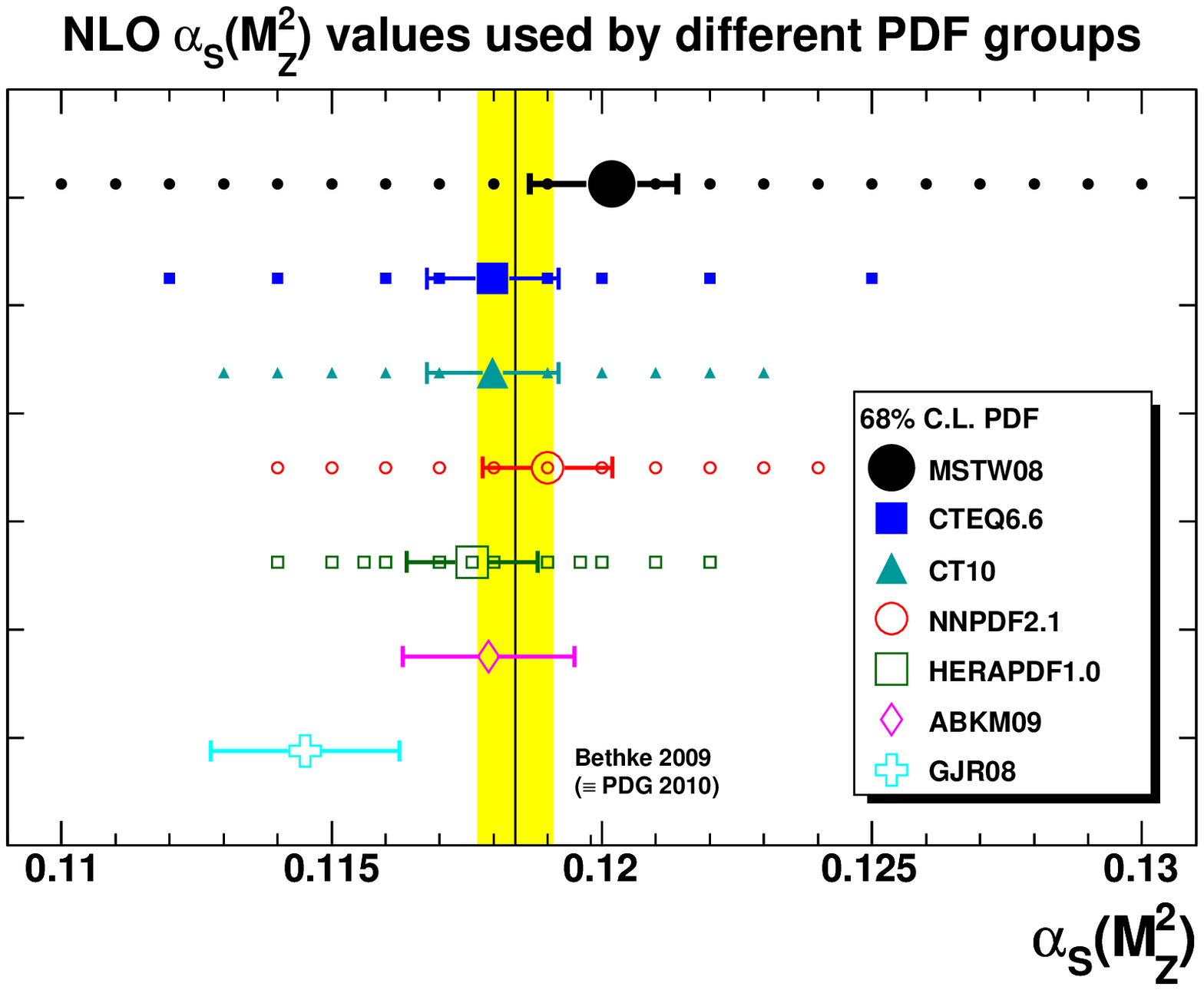}
  \end{minipage}%
  \begin{minipage}{0.5\textwidth}
    (b)\\
    \includegraphics[width=\textwidth]{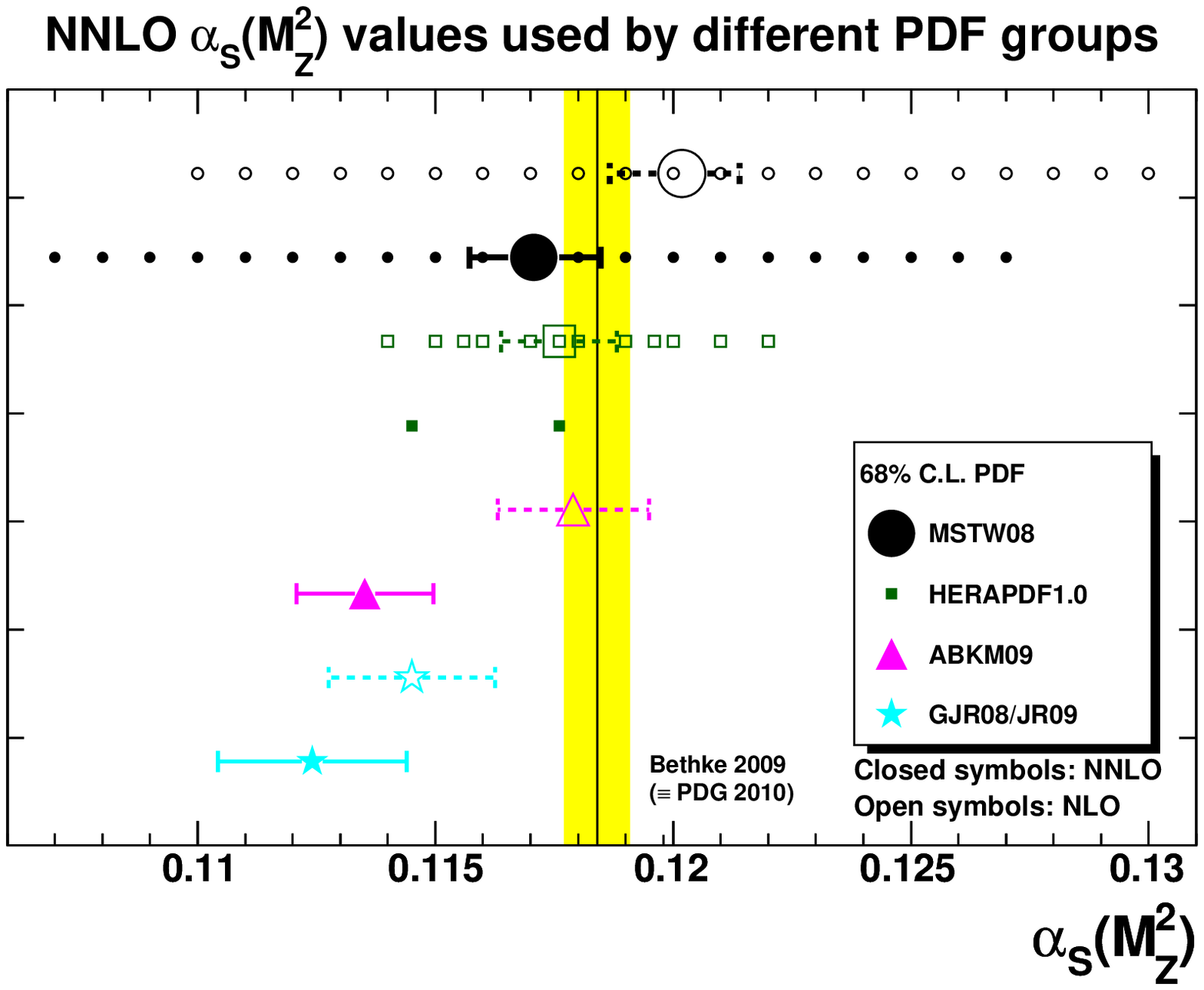}
  \end{minipage}
  \caption{Values of $\alpha_S(M_Z^2)$, and their 1-$\sigma$ uncertainties, used by different PDF fitting groups at (a)~NLO and (b)~NNLO.  The smaller symbols indicate the PDF sets with alternative values of $\alpha_S(M_Z^2)$ provided by each fitting group.  The shaded band indicates the world average $\alpha_S(M_Z^2)$~\cite{Bethke:2009jm}.}
  \label{fig:asmzvalues}
\end{figure}
The values of $\alpha_S(M_Z^2)$, and the corresponding uncertainties, for MSTW08, ABKM09 and GJR08/JR09 are obtained from a simultaneous fit with the PDF parameters.  Other groups choose a fixed value, generally close to the world average~\cite{Bethke:2009jm}, and for those groups we assume an uncertainty of $\pm0.0012$ ($\pm0.0020$) for a 68\% (90\%) C.L.~\cite{Alekhin:2011sk}.  The central values and 1-$\sigma$ uncertainties are depicted in figure~\ref{fig:asmzvalues} as the larger symbols and error bars, while the smaller symbols indicate the PDF sets with alternative values of $\alpha_S(M_Z^2)$ provided by each group.  The fitted NLO $\alpha_S(M_Z^2)$ value is always larger than the corresponding NNLO $\alpha_S(M_Z^2)$ value in an attempt by the fit to mimic the missing higher-order corrections, which are generally positive.  Note that most of the fitted NLO $\alpha_S(M_Z^2)$ values (other than GJR08) agree with the world average~\cite{Bethke:2009jm}, while MSTW08 is the only set in agreement at NNLO.

\begin{table}
  \centering
  \begin{tabular}{|l|r|r|r|r|r|r|r|r|}
    \hline
    \multicolumn{3}{|c|}{MSTW 2008} & \multicolumn{3}{c|}{68\% C.L.~uncertainties} & \multicolumn{3}{c|}{90\% C.L.~uncertainties} \\ \hline
    & Pert. & Central & PDF & \multicolumn{2}{c|}{PDF+$\alpha_S$} & PDF & \multicolumn{2}{c|}{PDF+$\alpha_S$} \\ \cline{5-6} \cline{8-9}
    \raisebox{1.5ex}[0pt]{Process} & order & value & & Exact & Quad. & & Exact & Quad. \\ \hline
    & NLO & 0.931 nb & $^{+2.1\%}_{-1.5\%}$ & $^{+2.4\%}_{-1.8\%}$ & $^{+2.3\%}_{-1.8\%}$ & $^{+3.5\%}_{-3.2\%}$ & $^{+4.6\%}_{-3.8\%}$ & $^{+4.0\%}_{-4.0\%}$ \\
    \raisebox{1.5ex}[0pt]{$Z^0\to\ell^+\ell^-$} & NNLO & 0.959 nb & $^{+1.7\%}_{-1.6\%}$ & $^{+2.5\%}_{-1.8\%}$ & $^{+2.0\%}_{-1.9\%}$ & $^{+3.4\%}_{-3.1\%}$ & $^{+4.4\%}_{-3.9\%}$ & $^{+4.1\%}_{-4.0\%}$ \\ \hline
    & NLO & 10.15 nb & $^{+2.2\%}_{-1.6\%}$ & $^{+2.4\%}_{-1.8\%}$ & $^{+2.3\%}_{-1.9\%}$ & $^{+3.5\%}_{-3.3\%}$ & $^{+4.6\%}_{-3.9\%}$ & $^{+4.1\%}_{-4.1\%}$ \\
    \raisebox{1.5ex}[0pt]{$W^\pm\to\ell^\pm\nu$} & NNLO & 10.47 nb & $^{+1.7\%}_{-1.6\%}$ & $^{+2.6\%}_{-1.9\%}$ & $^{+2.0\%}_{-1.9\%}$ & $^{+3.4\%}_{-3.1\%}$ & $^{+4.6\%}_{-3.9\%}$ & $^{+4.2\%}_{-4.1\%}$ \\ \hline
    & NLO & 5.96 nb & $^{+2.2\%}_{-1.6\%}$ & $^{+2.4\%}_{-2.0\%}$ & $^{+2.3\%}_{-1.9\%}$ & $^{+3.6\%}_{-3.3\%}$ & $^{+4.7\%}_{-4.0\%}$ & $^{+4.2\%}_{-4.1\%}$ \\
    \raisebox{1.5ex}[0pt]{$W^+\to\ell^+\nu$} & NNLO & 6.16 nb & $^{+1.8\%}_{-1.6\%}$ & $^{+2.6\%}_{-1.9\%}$ & $^{+2.1\%}_{-1.9\%}$ & $^{+3.6\%}_{-3.2\%}$ & $^{+4.9\%}_{-4.0\%}$ & $^{+4.4\%}_{-4.2\%}$ \\ \hline
    & NLO & 4.19 nb & $^{+2.2\%}_{-1.7\%}$ & $^{+2.5\%}_{-1.9\%}$ & $^{+2.4\%}_{-1.9\%}$ & $^{+3.6\%}_{-3.5\%}$ & $^{+4.7\%}_{-4.1\%}$ & $^{+4.2\%}_{-4.3\%}$ \\
    \raisebox{1.5ex}[0pt]{$W^-\to\ell^-\nu$} & NNLO & 4.31 nb & $^{+1.7\%}_{-1.7\%}$ & $^{+2.5\%}_{-1.9\%}$ & $^{+2.0\%}_{-2.0\%}$ & $^{+3.5\%}_{-3.3\%}$ & $^{+4.4\%}_{-4.2\%}$ & $^{+4.1\%}_{-4.3\%}$ \\ \hline
    & NLO & 10.91 & $^{+0.2\%}_{-0.2\%}$ & $^{+0.2\%}_{-0.2\%}$ & $^{+0.2\%}_{-0.2\%}$ & $^{+0.3\%}_{-0.3\%}$ & $^{+0.4\%}_{-0.3\%}$ & $^{+0.4\%}_{-0.3\%}$ \\
    \raisebox{1.5ex}[0pt]{$W^\pm/Z^0$ ratio} & NNLO & 10.92 & $^{+0.2\%}_{-0.1\%}$ & $^{+0.3\%}_{-0.2\%}$ & $^{+0.2\%}_{-0.1\%}$ & $^{+0.4\%}_{-0.3\%}$ & $^{+0.5\%}_{-0.4\%}$ & $^{+0.4\%}_{-0.3\%}$ \\ \hline
    & NLO & 1.422 & $^{+1.0\%}_{-0.8\%}$ & $^{+1.0\%}_{-0.8\%}$ & $^{+1.0\%}_{-0.8\%}$ & $^{+2.2\%}_{-1.6\%}$ & $^{+2.3\%}_{-1.6\%}$ & $^{+2.2\%}_{-1.6\%}$ \\
    \raisebox{1.5ex}[0pt]{$W^+/W^-$ ratio} & NNLO & 1.429 & $^{+1.0\%}_{-0.7\%}$ & $^{+1.1\%}_{-0.7\%}$ & $^{+1.0\%}_{-0.7\%}$ & $^{+2.2\%}_{-1.5\%}$ & $^{+2.3\%}_{-1.5\%}$ & $^{+2.2\%}_{-1.5\%}$ \\ \hline
    & NLO & 12.41 pb & $^{+1.4\%}_{-1.7\%}$ & $^{+3.3\%}_{-2.7\%}$ & $^{+2.3\%}_{-2.8\%}$ & $^{+2.8\%}_{-3.1\%}$ & $^{+7.6\%}_{-6.6\%}$ & $^{+5.5\%}_{-6.6\%}$ \\
    \raisebox{1.5ex}[0pt]{$H$ (120 GeV)} & NNLO & 15.71 pb & $^{+1.1\%}_{-1.6\%}$ & $^{+3.7\%}_{-2.8\%}$ & $^{+2.9\%}_{-3.0\%}$ & $^{+2.6\%}_{-3.2\%}$ & $^{+7.3\%}_{-7.0\%}$ & $^{+6.7\%}_{-7.1\%}$ \\ \hline
    & NLO & 5.20 pb & $^{+1.7\%}_{-2.1\%}$ & $^{+3.4\%}_{-2.7\%}$ & $^{+2.5\%}_{-3.0\%}$ & $^{+3.3\%}_{-3.5\%}$ & $^{+7.6\%}_{-6.4\%}$ & $^{+5.8\%}_{-6.8\%}$ \\
    \raisebox{1.5ex}[0pt]{$H$ (180 GeV)} & NNLO & 6.53 pb & $^{+1.5\%}_{-1.8\%}$ & $^{+3.8\%}_{-3.1\%}$ & $^{+3.1\%}_{-3.1\%}$ & $^{+3.2\%}_{-3.6\%}$ & $^{+7.5\%}_{-7.3\%}$ & $^{+7.2\%}_{-7.4\%}$ \\ \hline
    & NLO & 2.81 pb & $^{+2.1\%}_{-2.5\%}$ & $^{+3.6\%}_{-3.0\%}$ & $^{+2.8\%}_{-3.3\%}$ & $^{+4.0\%}_{-4.1\%}$ & $^{+7.7\%}_{-6.6\%}$ & $^{+6.2\%}_{-7.2\%}$ \\
    \raisebox{1.5ex}[0pt]{$H$ (240 GeV)} & NNLO & 3.52 pb & $^{+1.9\%}_{-2.1\%}$ & $^{+3.9\%}_{-3.4\%}$ & $^{+3.4\%}_{-3.4\%}$ & $^{+4.0\%}_{-4.3\%}$ & $^{+7.9\%}_{-7.7\%}$ & $^{+7.8\%}_{-7.8\%}$ \\ \hline
    & NLO & 167.9 pb & $^{+2.7\%}_{-3.3\%}$ & $^{+4.3\%}_{-3.9\%}$ & $^{+3.5\%}_{-4.3\%}$ & $^{+5.3\%}_{-5.4\%}$ & $^{+8.8\%}_{-8.3\%}$ & $^{+7.9\%}_{-9.0\%}$ \\
    \raisebox{1.5ex}[0pt]{$t\bar{t}$} & NNLO & 173.0 pb & $^{+2.7\%}_{-2.7\%}$ & $^{+4.4\%}_{-4.2\%}$ & $^{+4.2\%}_{-4.1\%}$ & $^{+5.2\%}_{-5.4\%}$ & $^{+9.5\%}_{-9.0\%}$ & $^{+9.8\%}_{-9.1\%}$ \\
    \hline
  \end{tabular}
  \caption{NLO and NNLO predictions for total cross sections using MSTW 2008 PDFs~\cite{Martin:2009iq,Martin:2009bu} comparing PDF-only uncertainties and PDF+$\alpha_S$ uncertainties computed using either the exact prescription~\cite{Martin:2009bu} or simply adding $\alpha_S$ uncertainties in quadrature with the PDF-only uncertainties.\label{tab:pdfasquad}}
\end{table}
The prescription for ``PDF+$\alpha_S$'' uncertainties in the MSTW08 case was given in ref.~\cite{Martin:2009bu}.  We compare this prescription with the alternative prescription of adding in quadrature in table~\ref{tab:pdfasquad} for all the benchmark processes.  (The settings for the cross-section calculations are given later in section~\ref{sec:xsasmz}.)  Adding in quadrature for MSTW08 was proposed as a simplification in refs.~\cite{Botje:2011sn,LHCHiggsCrossSectionWorkingGroup:2011ti} and expected to give slightly smaller uncertainties.  For CTEQ6.6/CT10 and HERAPDF1.0/1.5 we simply add $\alpha_S$ uncertainties in quadrature with the PDF uncertainties, as recommended by the fit authors.  For NNPDF2.1, we use the prescription given in refs.~\cite{Binoth:2010ra,Demartin:2010er}, that is, we take the number of replicas for each $\alpha_S(M_Z^2)$ value from a Gaussian distribution with mean of 0.1190 and standard deviation 0.0012.  Rounding to the nearest integer values, this gives $N_{\rm rep}^{(j)} = 4,25,71,100,71,25,4$, for $\alpha_S^{(j)}(M_Z^2) = 0.116,0.117,0.118,0.119,0.120,0.121,0.122$, then we take the averages and standard deviations over $N_{\rm rep} = \sum_j N_{\rm rep}^{(j)} = 300$ replicas.  For each $\alpha_S^{(j)}(M_Z^2)$ value, the $N_{\rm rep}^{(j)}$ replicas are chosen randomly from the full 100 replicas available.  To eliminate dependence on the particular subset of $N_{\rm rep}^{(j)}$ replicas chosen, the averages and standard deviations over the $N_{\rm rep}=300$ replicas are themselves calculated many ($10^5$) times, then the results averaged.  This final averaging is supplementary to the official NNPDF prescription and can be considered as an improvement at little computational cost.  For ABKM09 and GJR08/JR09, $\alpha_S(M_Z^2)$ is one of the fit parameters used to produce the eigenvector sets, which implies mixing of $\alpha_S(M_Z^2)$ with PDF parameters, so a ``PDF+$\alpha_S$'' uncertainty is obtained automatically and it is not straightforward to calculate a ``PDF only'' uncertainty.  Care must then be taken that a different $\alpha_S$ value is used in the partonic cross section for each eigenvector PDF set.

\section{\texorpdfstring{Comparison of partonic}{Partonic} luminosities} \label{sec:luminosities}
To understand properties of hadronic cross sections, such as PDF uncertainties or the dependence on collider energy, it is useful to consider the relevant parton--parton luminosities (see, for example, refs.~\cite{Eichten:1984eu,Ellis:1991qj,Campbell:2006wx,StirlingLumiPlots,Quigg:2009gg,Quigg:2011zu}).  We define the dimensionless variable, $\tau\equiv \hat{s}/s$, where $\tau \in (0,1]$ and $\sqrt{\hat{s}}$ is the partonic centre-of-mass energy.  A PDF of flavour $a=q,\bar{q},g$ is denoted as $f_a(x,\mu^2)$, with $x$ the momentum fraction and $\mu$ the factorisation scale.  We will consider the following combinations of partonic luminosity:
\begin{align}
  \frac{\partial {\cal L}_{\Sigma_q (q\bar{q})}}{\partial \hat{s}} &= \frac{1}{s} \int_\tau^1\frac{{\rm d}x}{x}\;\sum_{q=d,u,s,c,b} \left[f_q(x,\hat{s})f_{\bar{q}}(\tau/x,\hat{s}) + f_{\bar{q}}(x,\hat{s})f_{q}(\tau/x,\hat{s})\right], \label{eq:qqbarlumi} \\
  \frac{\partial {\cal L}_{gg}}{\partial \hat{s}} &= \frac{1}{s} \int_\tau^1\frac{{\rm d}x}{x}\;f_{g}(x,\hat{s})f_{g}(\tau/x,\hat{s}), \label{eq:gglumi} \\
  \frac{\partial {\cal L}_{GG}}{\partial \hat{s}} &= \frac{1}{s} \int_\tau^1\frac{{\rm d}x}{x}\;f_{G}(x,\hat{s})f_{G}(\tau/x,\hat{s}), \quad f_{G} \equiv f_{g} + \frac{4}{9}\sum_{q=d,u,s,c,b} \left(f_q + f_{\bar{q}}\right), \label{eq:GGlumi}
\end{align}
where $q\bar{q}$ is relevant for $W$ and $Z$ production, $gg$ is relevant for Higgs and $t\bar{t}$ production, and $GG$ is relevant for dijet production~\cite{Ellis:1991qj,Combridge:1983jn}.  Note that the dependence of $\partial {\cal L}/\partial \tau$ only on the scaling variable $\tau \equiv \hat{s}/s$ rather than separately on $\hat{s}$ and $s$ is (mildly) broken by the choice of factorisation scale $\mu^2 = \hat{s}$.  This means, for example, that ratios of partonic luminosities at $\sqrt{s} = 7$~TeV plotted as a function of $\sqrt{\tau}$ will not be identical to the same ratios at $\sqrt{s} = 14$~TeV.  The logarithmic integration over $x$ in eqs.~\eqref{eq:qqbarlumi}, \eqref{eq:gglumi} and \eqref{eq:GGlumi} can be interpreted as an integration over the rapidity $y=\ln(x/\sqrt{\tau})$ of the final state.  We will only consider rapidity-integrated partonic luminosities relevant to the total cross section in this paper, but it would be straightforward to also study the rapidity-dependent quantities.  Note that the $q\bar{q}$ luminosity does not specifically include the correct flavour combinations for $W^\pm$ or $Z^0$ production.  More detailed studies would, for example, include the correct couplings of the vector bosons to quarks and antiquarks, or consider the specific $u\bar{d}$ (for $W^+$) or $d\bar{u}$ (for $W^-$) partonic luminosities.  Nevertheless, the generic $q\bar{q}$ luminosity defined in eq.~\eqref{eq:qqbarlumi} should exhibit the general features of the more precise partonic luminosities relevant for the $W$ and $Z$ total cross sections.

\begin{figure}
  \centering
  \begin{minipage}{0.5\textwidth}
    (a)\\
    \includegraphics[width=\textwidth]{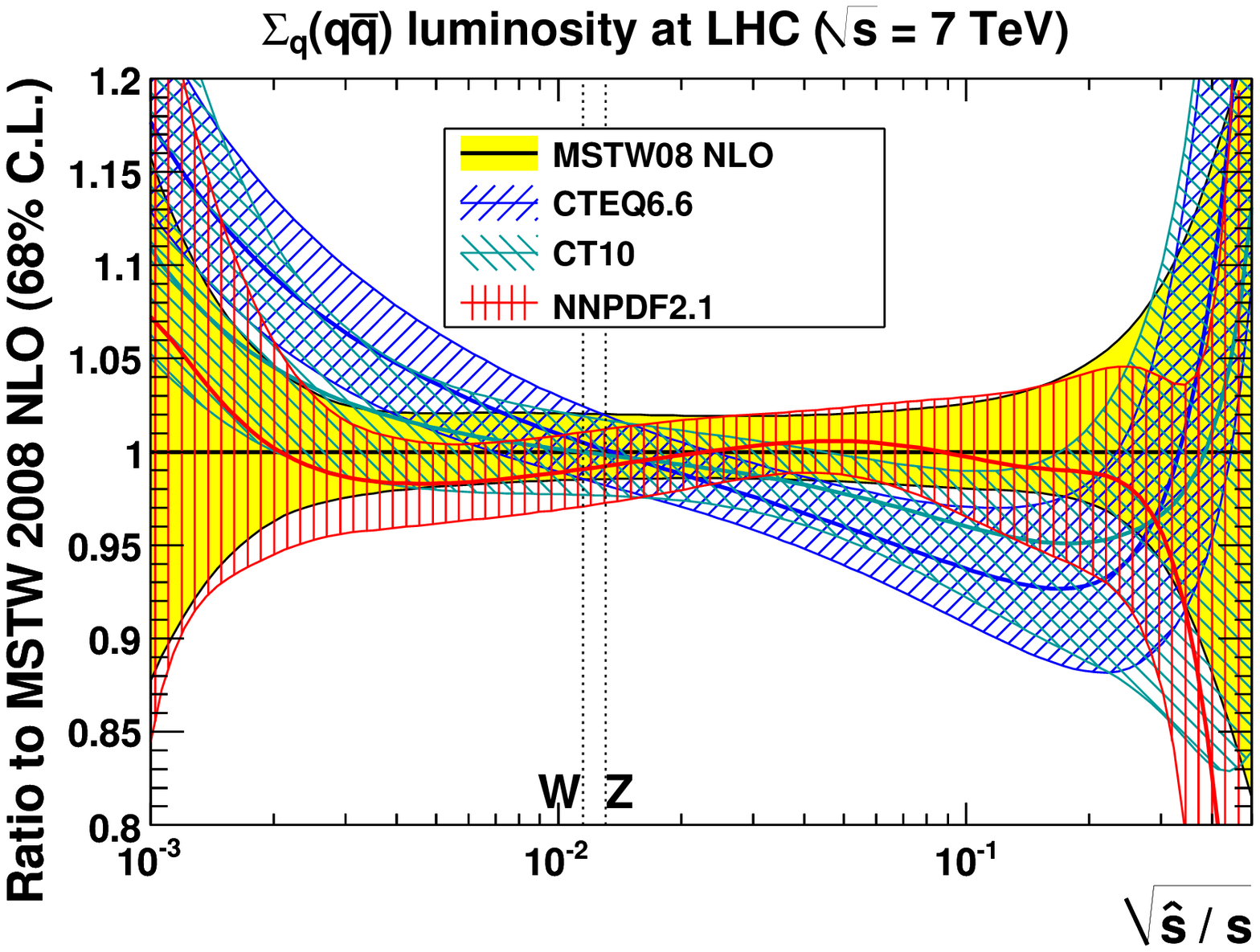}
  \end{minipage}%
  \begin{minipage}{0.5\textwidth}
    (b)\\
    \includegraphics[width=\textwidth]{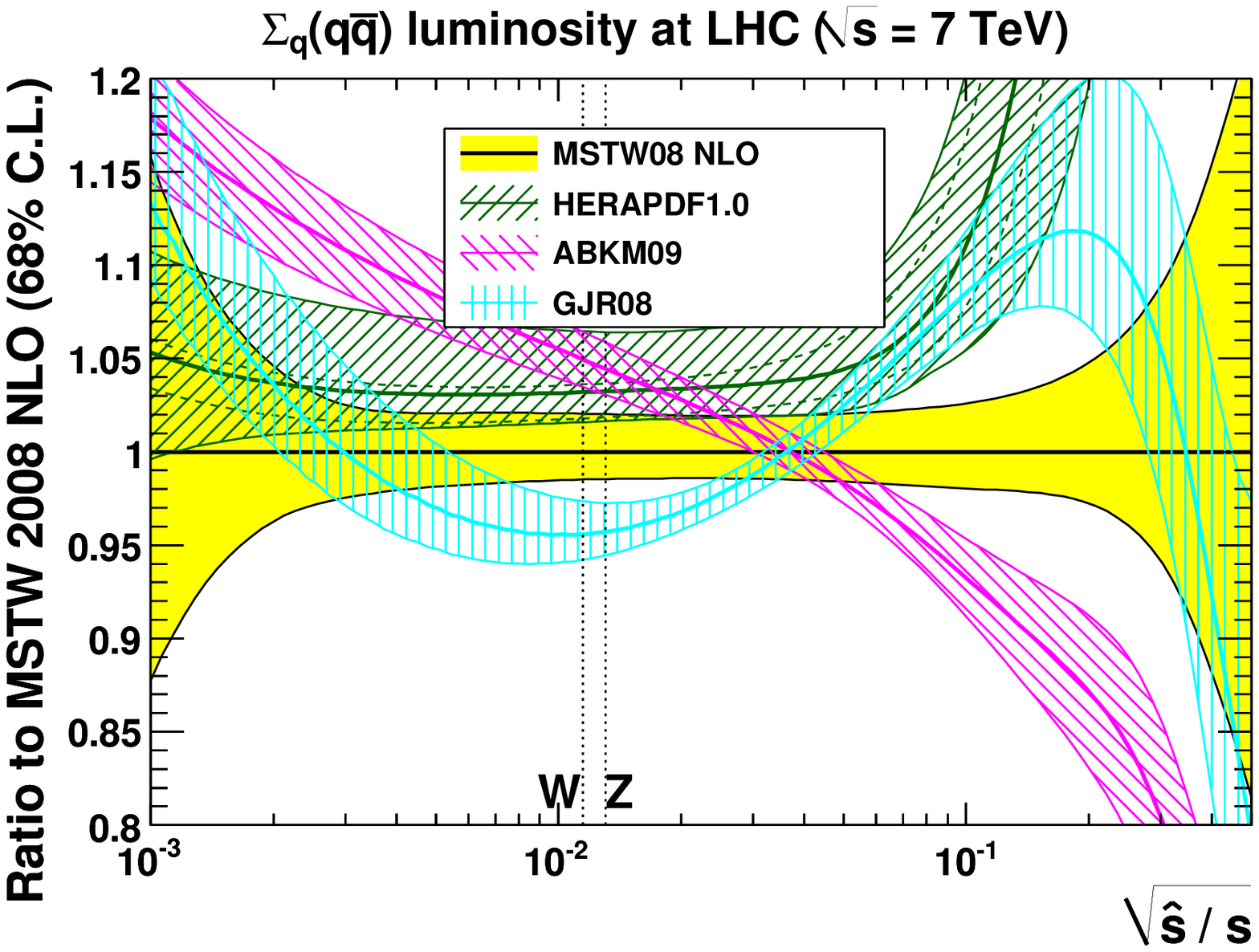}
  \end{minipage}
  \begin{minipage}{0.5\textwidth}
    (c)\\
    \includegraphics[width=\textwidth]{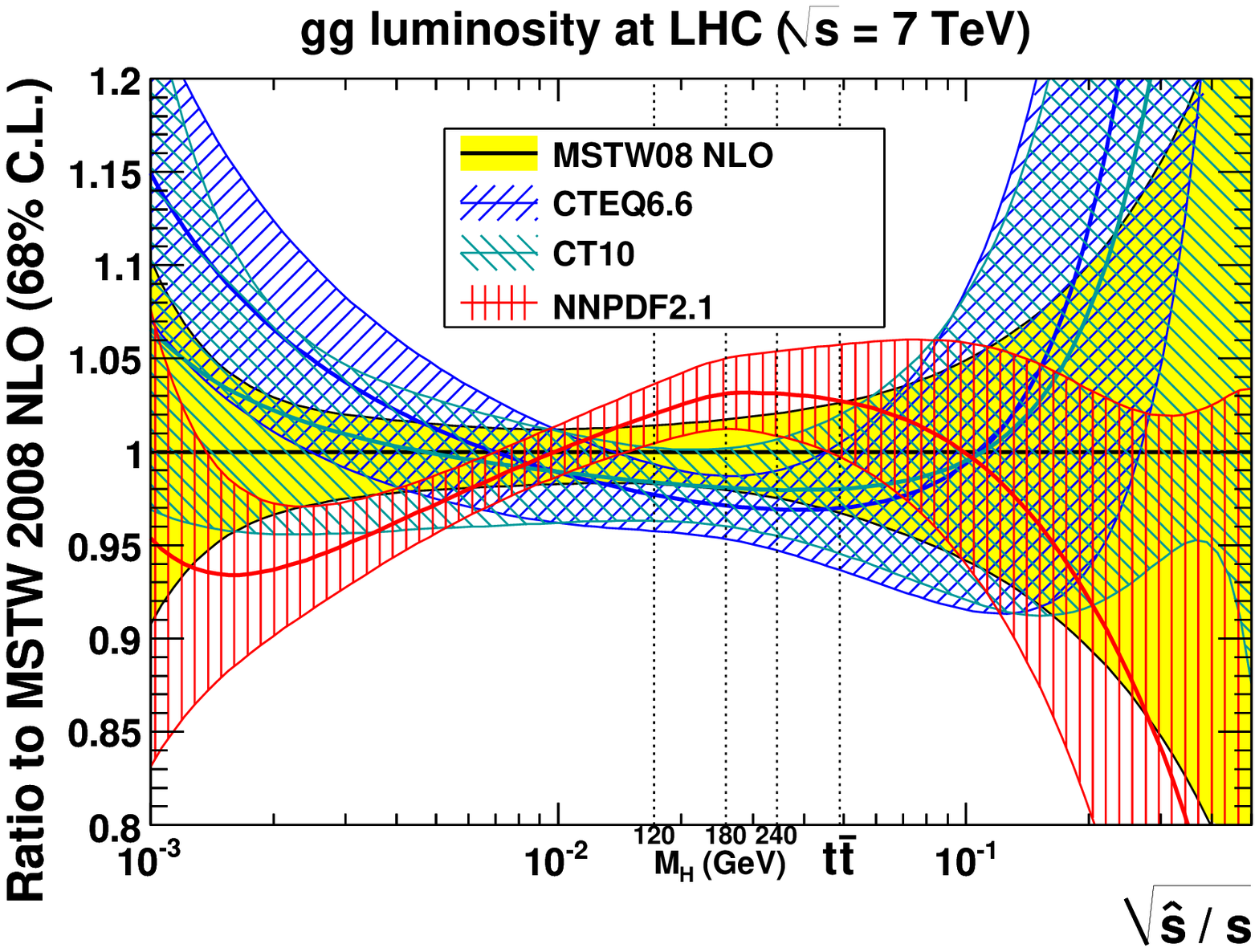}
  \end{minipage}%
  \begin{minipage}{0.5\textwidth}
    (d)\\
    \includegraphics[width=\textwidth]{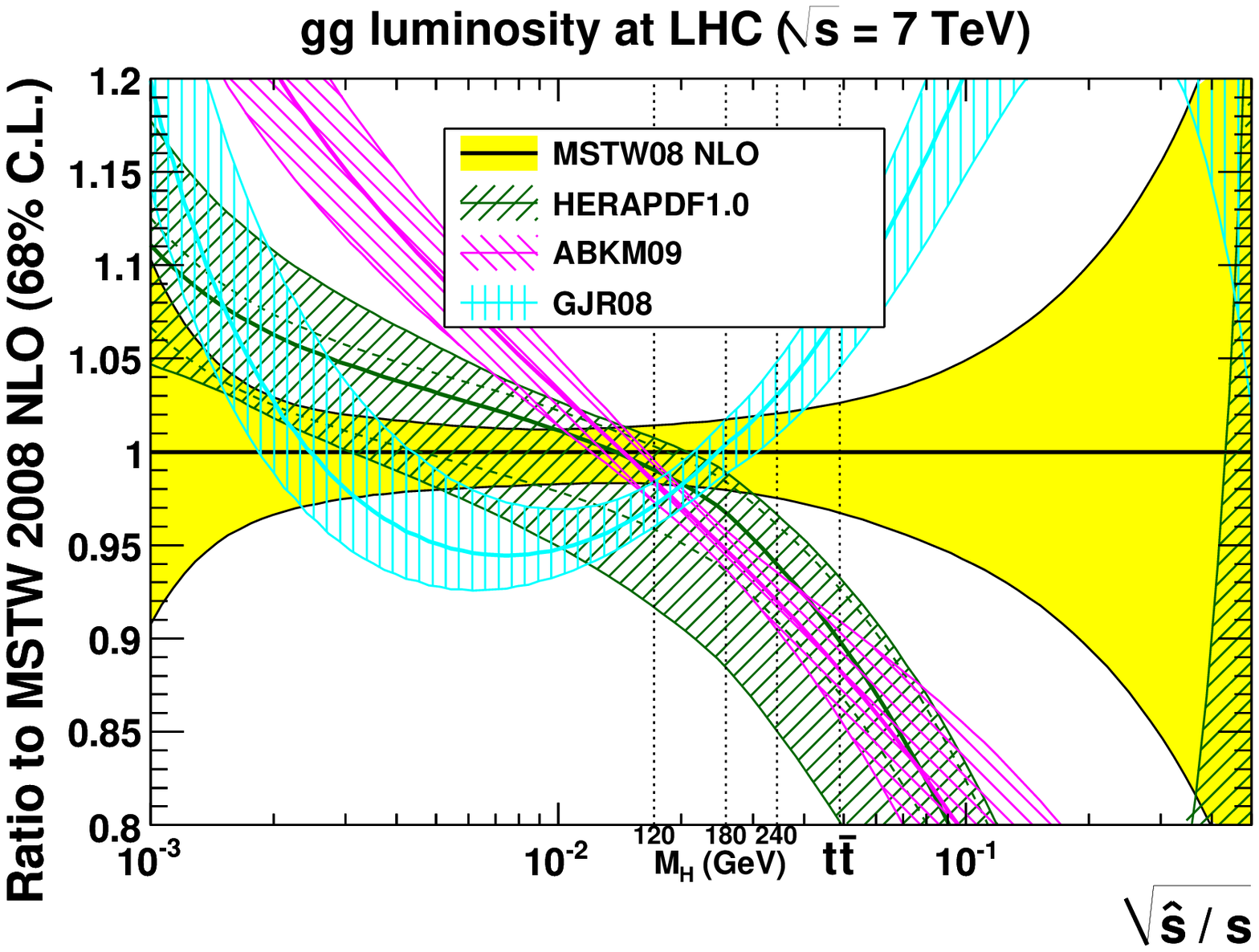}
  \end{minipage}
  \begin{minipage}{0.5\textwidth}
    (e)\\
    \includegraphics[width=\textwidth]{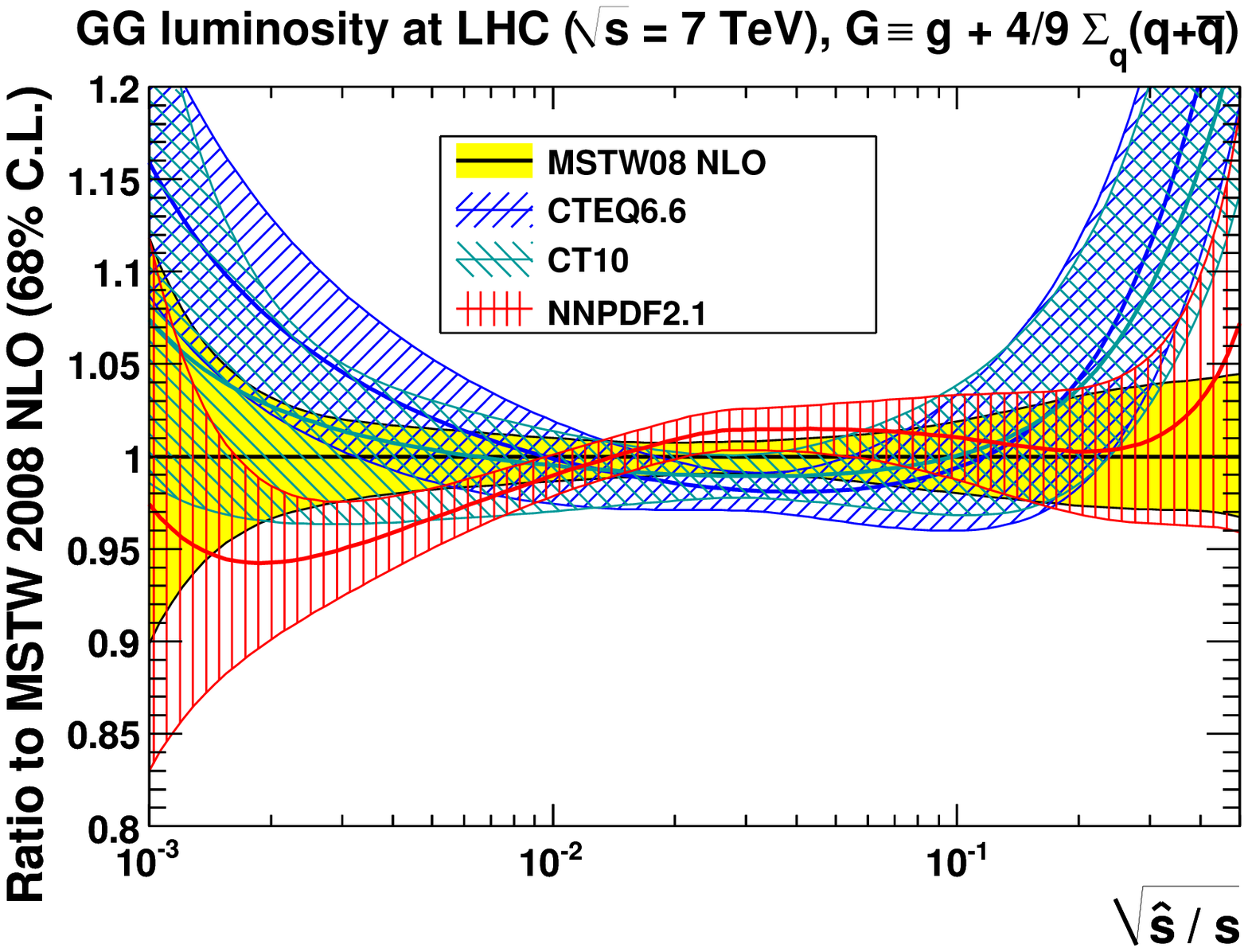}
  \end{minipage}%
  \begin{minipage}{0.5\textwidth}
    (f)\\
    \includegraphics[width=\textwidth]{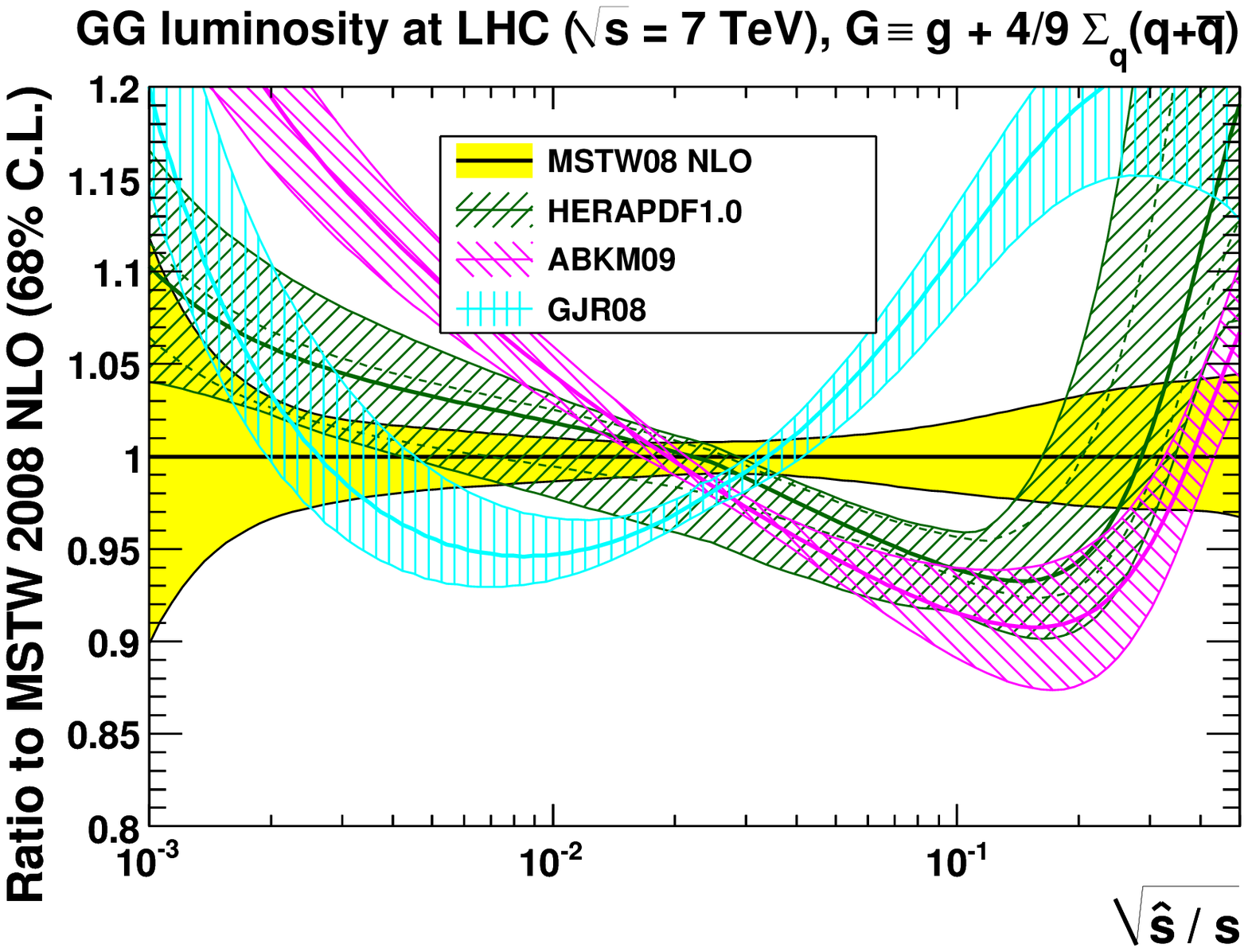}
  \end{minipage}
  \caption{NLO parton--parton luminosities as the ratio with respect to MSTW 2008.}
  \label{fig:rationlolumi}
\end{figure}
In figure~\ref{fig:rationlolumi} we show the NLO parton--parton luminosities as the ratio with respect to the MSTW 2008 NLO luminosities, for the LHC at $\sqrt{s} = 7$~TeV.  The inner uncertainty bands (dashed lines) for HERAPDF1.0 correspond to the (asymmetric) experimental errors, computed using eqs.~\eqref{eq:plusHess} and \eqref{eq:minusHess}, while the outer uncertainty bands (shaded regions) also include the model and parameterisation errors, including uncertainties on heavy-quark masses but not on $\alpha_S$.  We use the default $\alpha_S$ values for each set, shown in figure~\ref{fig:asmzvalues}.  Recall that it is not possible to separate the ``PDF only'' uncertainty for ABKM09 and GJR08/JR09, therefore the uncertainty bands for those sets also include the $\alpha_S$ uncertainty, and the uncertainty bands for ABKM09 also include uncertainties on heavy-quark masses.  This is undesirable but unavoidable given that these groups do not provide PDF sets for fixed $\alpha_S$ (and fixed $m_{c,b}$ for ABKM09).  These additional sources of uncertainty may be quite significant; see, for example, refs.~\cite{Martin:2009bu,Martin:2010db} for the $\alpha_S$ and $m_{c,b}$ dependence of the MSTW08 PDFs.  The relevant values of $\sqrt{\hat{s}} = M_{W,Z}$ are indicated on the plots of the $q\bar{q}$ luminosities in figure~\ref{fig:rationlolumi}(a,b).  There is fairly good agreement for the global fits, but more variation for the other sets, particularly when moving away from the central $\sqrt{\hat{s}}\sim M_{W,Z}$.  For the $gg$ luminosities in figure~\ref{fig:rationlolumi}(c,d), the relevant values of $\sqrt{\hat{s}} = M_{H}, 2m_t$ are indicated, where $\sqrt{\hat{s}} = 2m_t$ (with $m_t=171.3$~GeV) is the threshold for $t\bar{t}$ production.  Again, there is reasonable agreement for the global fits, but more variation for the other sets, particularly at large $\hat{s}$, where HERAPDF1.0 and ABKM09 have much softer high-$x$ gluon distributions.  The somewhat larger $gg$ luminosity of CTEQ6.6/CT10 at large $\hat{s}$ compared to MSTW08 may be due to the inclusion of Tevatron Run I jet data, which are known to prefer a larger high-$x$ gluon distribution than the more reliable Run II data~\cite{Martin:2009iq}, while the larger $gg$ luminosity at small $\hat{s}$ may be due to the positive-definite input gluon constraint of CTEQ6.6/CT10, which is not imposed in the MSTW08 or NNPDF2.1 fits.  By the momentum sum rule, the CTEQ6.6/CT10 $gg$ luminosity at the intermediate $\hat{s}$ values relevant for $gg\to H$ and $t\bar{t}$ production is then slightly smaller than that of MSTW08 and NNPDF2.1.  The $GG$ luminosities in figure~\ref{fig:rationlolumi}(e,f), relevant for dijet production, interpolate between being $gg$-dominated at small $\hat{s}$ and $qq$-dominated at large $\hat{s}$.
\begin{figure}
  \centering
  \begin{minipage}{0.5\textwidth}
    (a)\\
    \includegraphics[width=\textwidth]{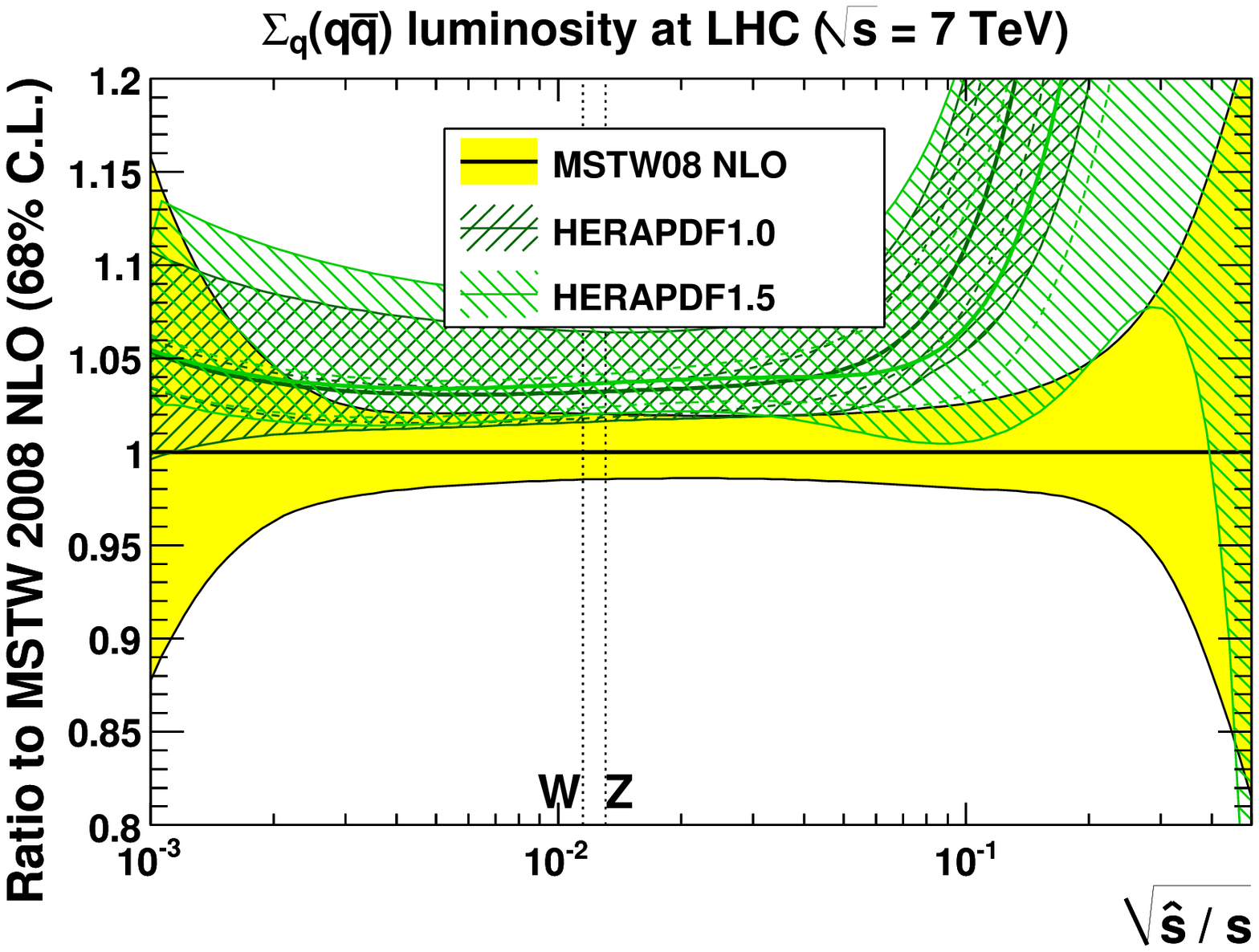}
  \end{minipage}%
  \begin{minipage}{0.5\textwidth}
    (b)\\
    \includegraphics[width=\textwidth]{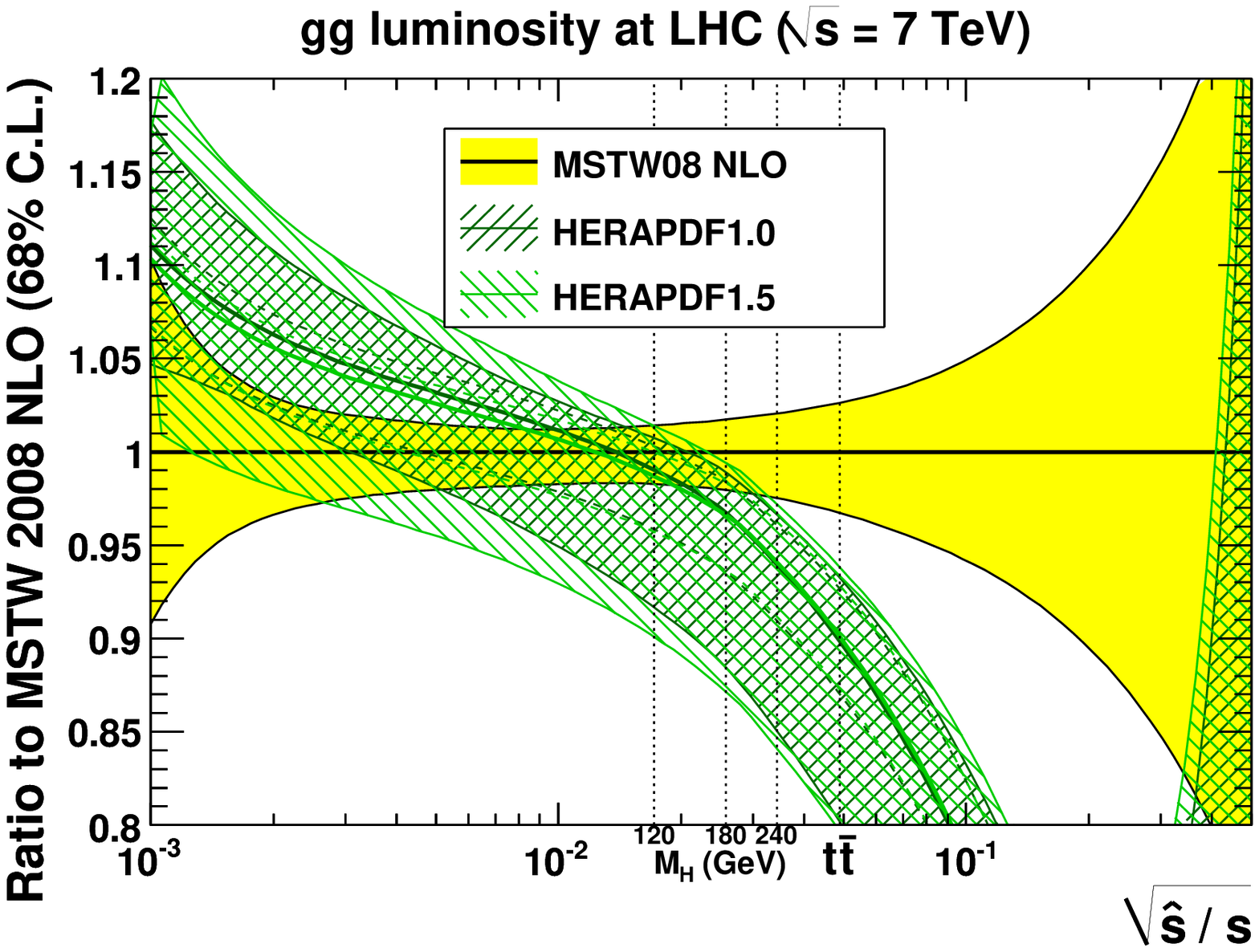}
  \end{minipage}
  \caption{NLO parton--parton luminosities from HERAPDF1.0/1.5 as the ratio to MSTW 2008.}
  \label{fig:rationlolumihera}
\end{figure}
Plots comparing the $q\bar{q}$ and $gg$ luminosities from MSTW 2008 with HERAPDF1.0 and HERAPDF1.5 are shown in figure~\ref{fig:rationlolumihera}.  The $q\bar{q}$ luminosity from HERAPDF1.5 is somewhat closer to MSTW08 at large $\hat{s}$ than HERAPDF1.0, while there is little difference in the $gg$ luminosity in going from HERAPDF1.0 to HERAPDF1.5.

\begin{figure}
  \centering
  \begin{minipage}{0.5\textwidth}
    (a)\\
    \includegraphics[width=\textwidth]{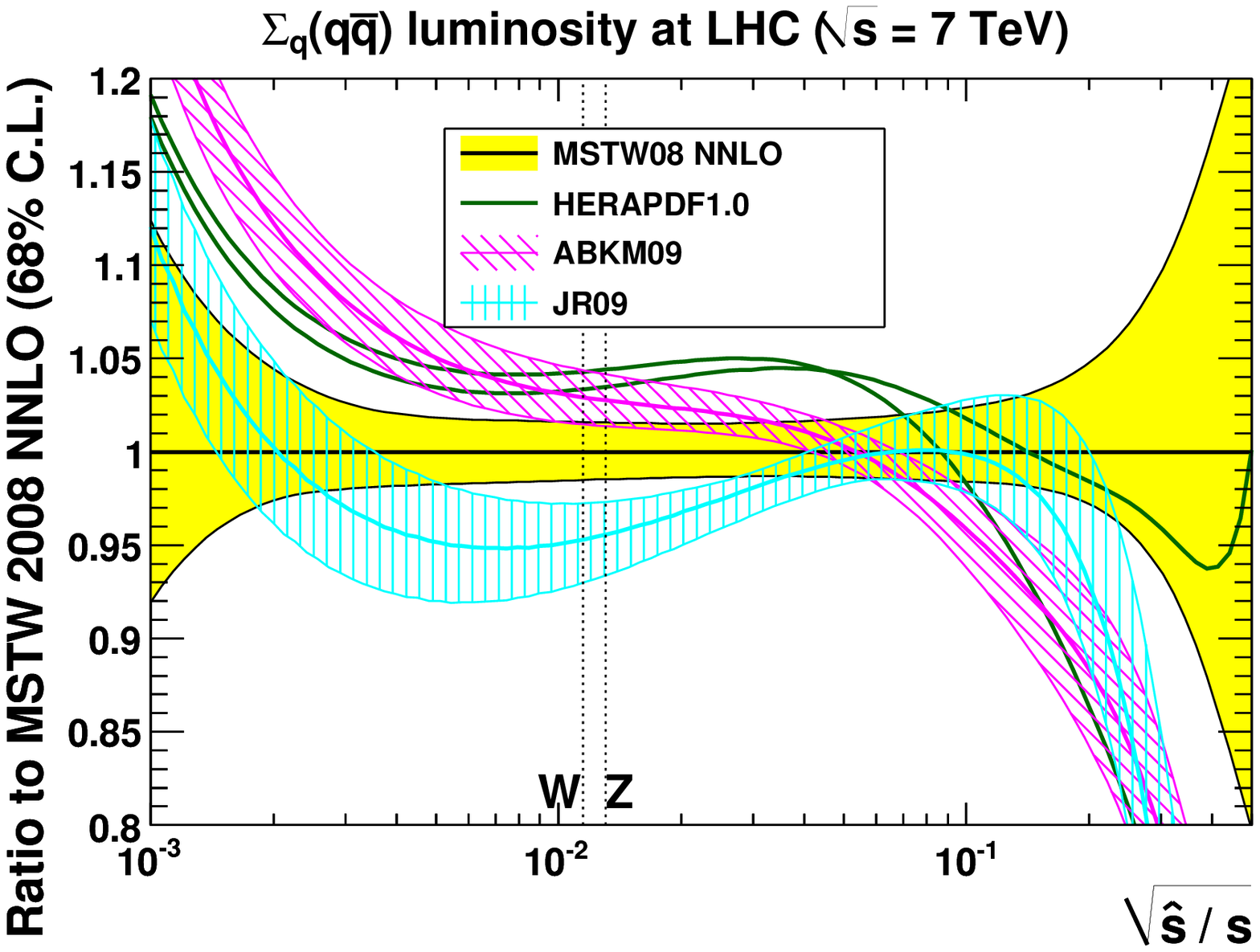}
  \end{minipage}%
  \begin{minipage}{0.5\textwidth}
    (b)\\
    \includegraphics[width=\textwidth]{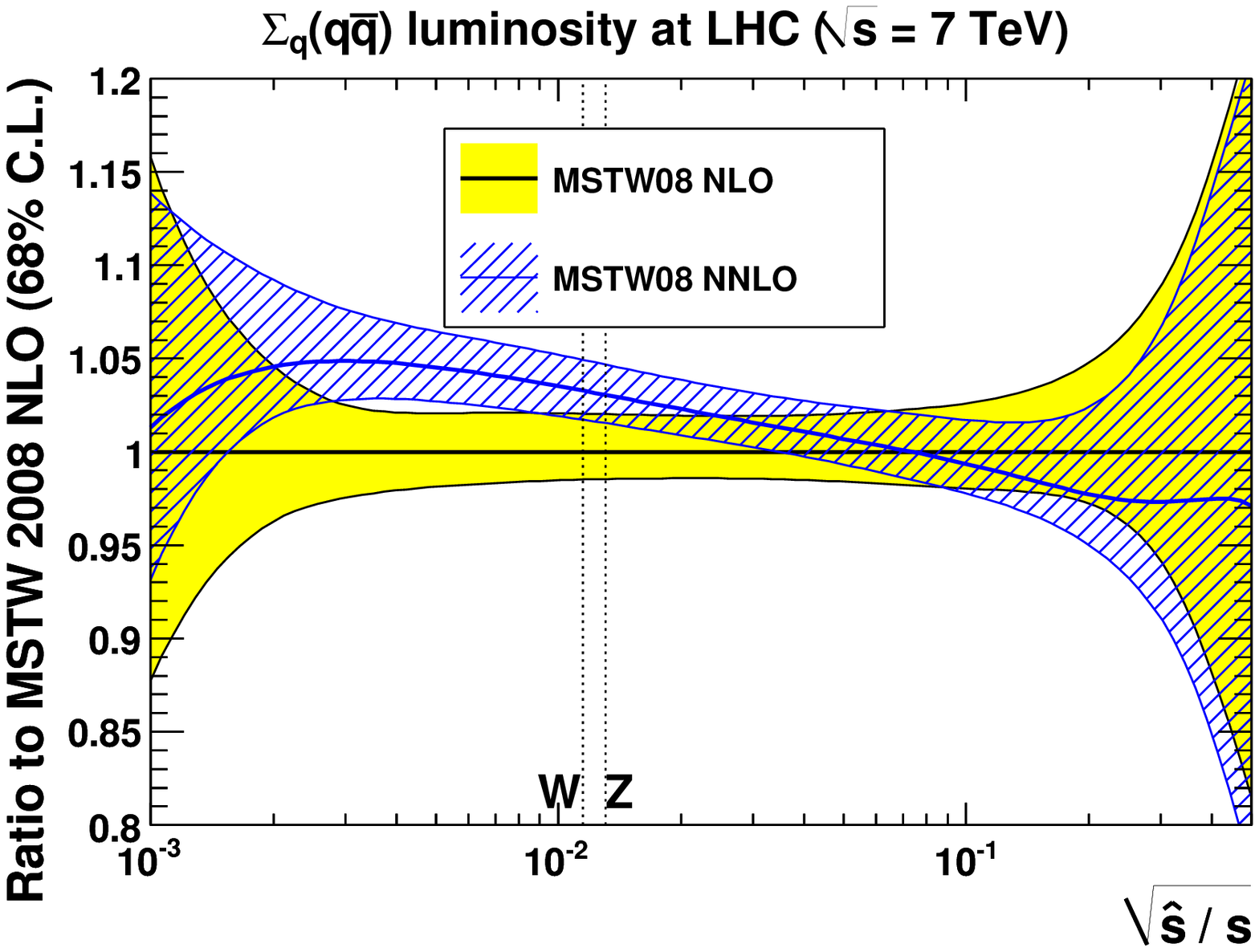}
  \end{minipage}
  \begin{minipage}{0.5\textwidth}
    (c)\\
    \includegraphics[width=\textwidth]{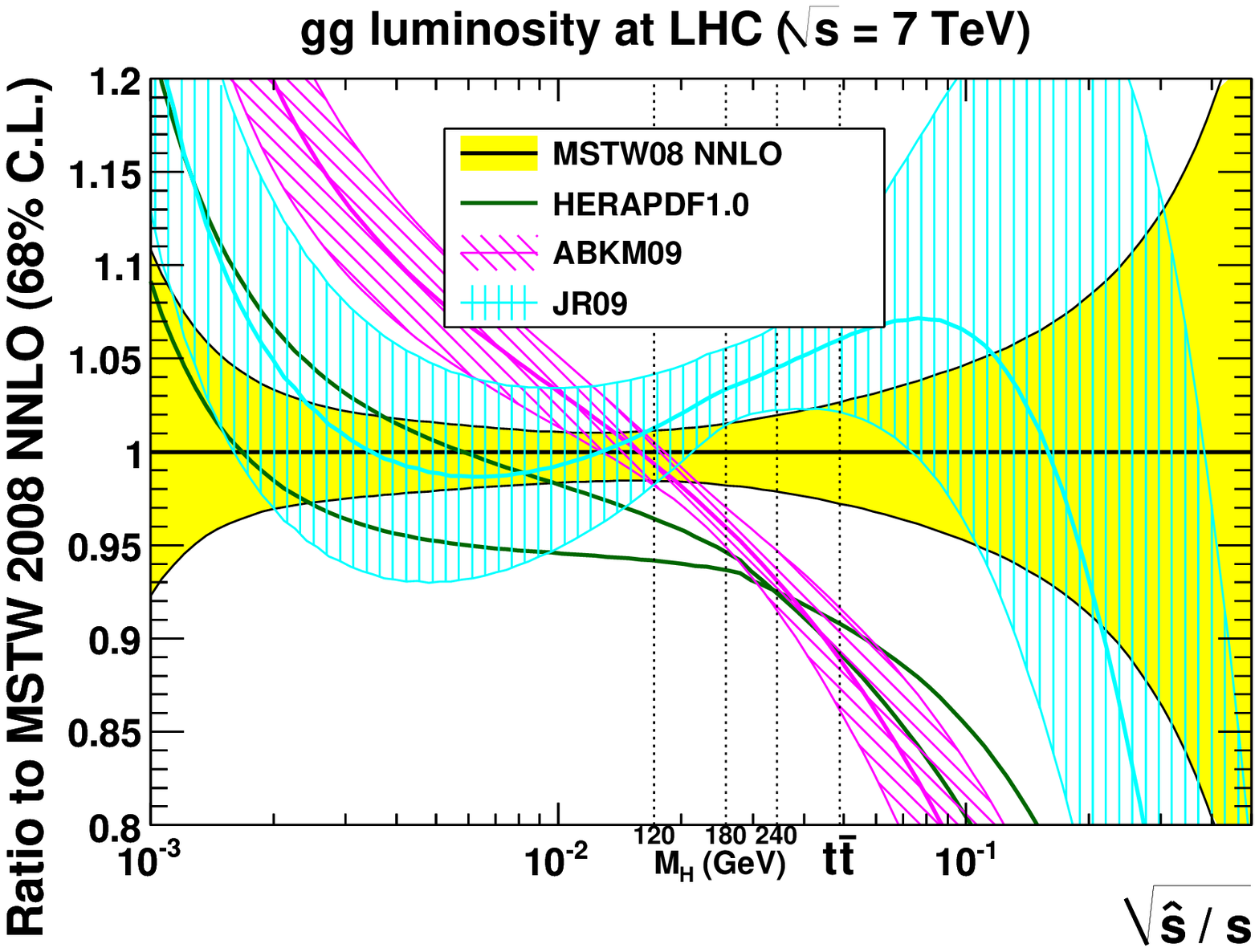}
  \end{minipage}%
  \begin{minipage}{0.5\textwidth}
    (d)\\
    \includegraphics[width=\textwidth]{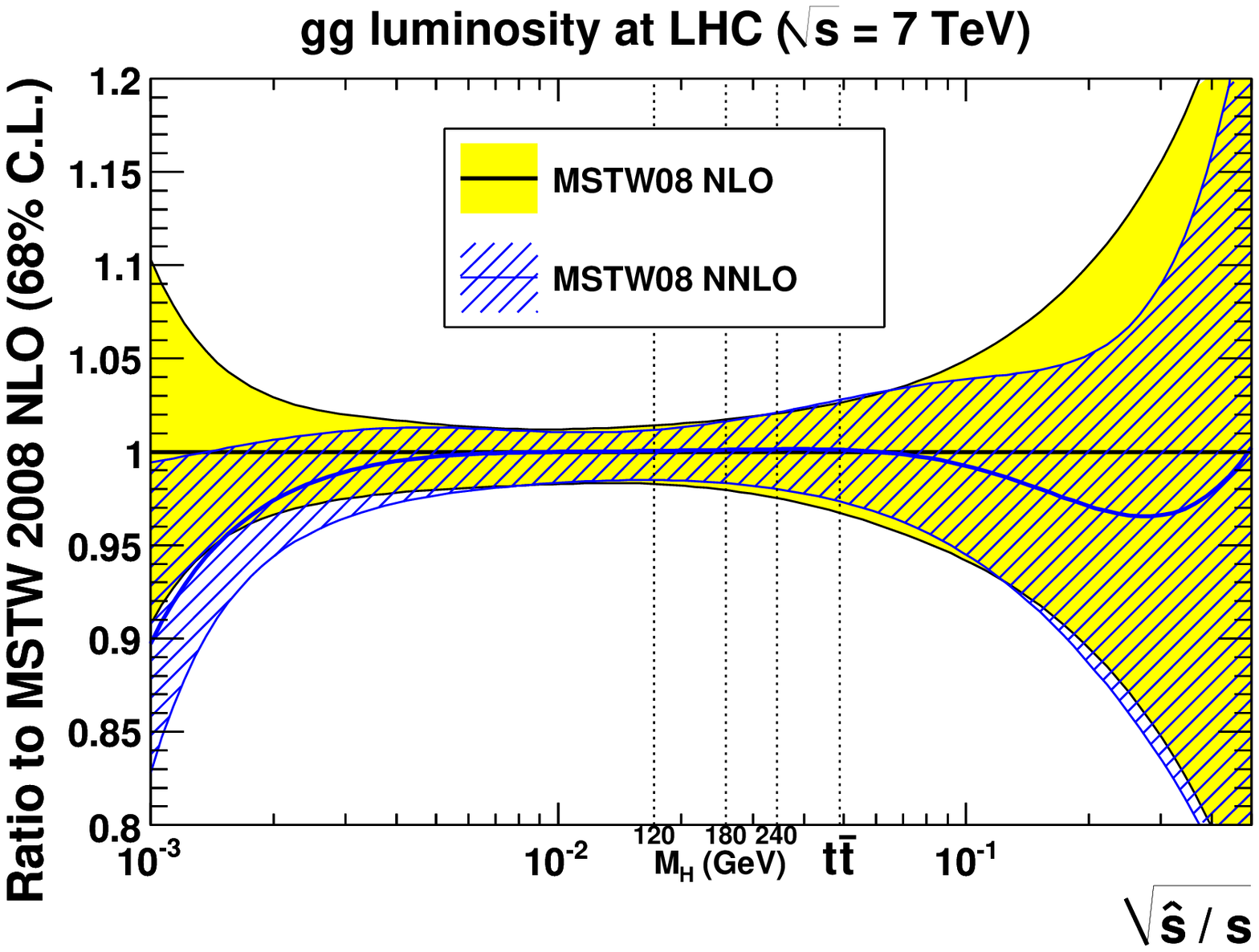}
  \end{minipage}
  \begin{minipage}{0.5\textwidth}
    (e)\\
    \includegraphics[width=\textwidth]{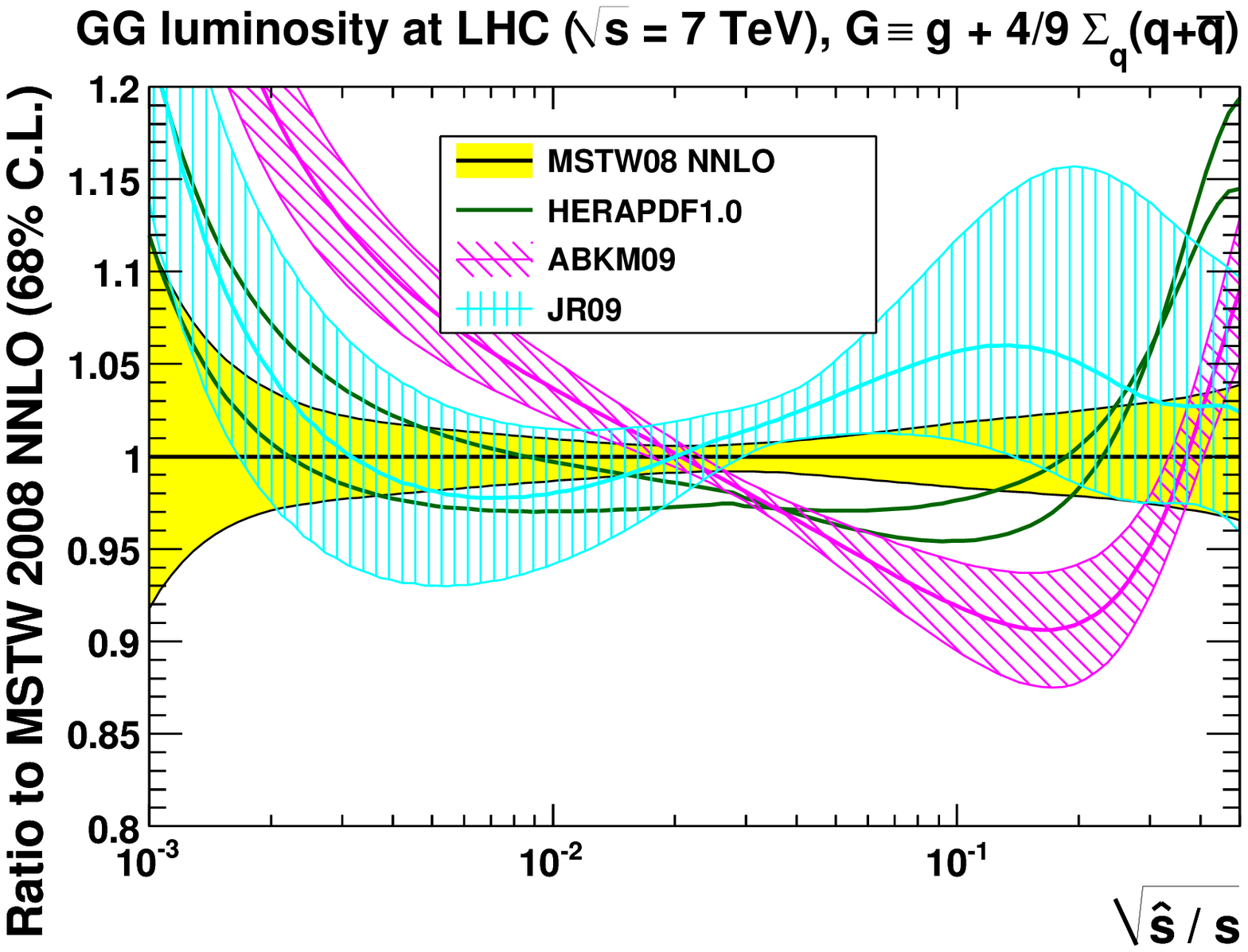}
  \end{minipage}%
  \begin{minipage}{0.5\textwidth}
    (f)\\
    \includegraphics[width=\textwidth]{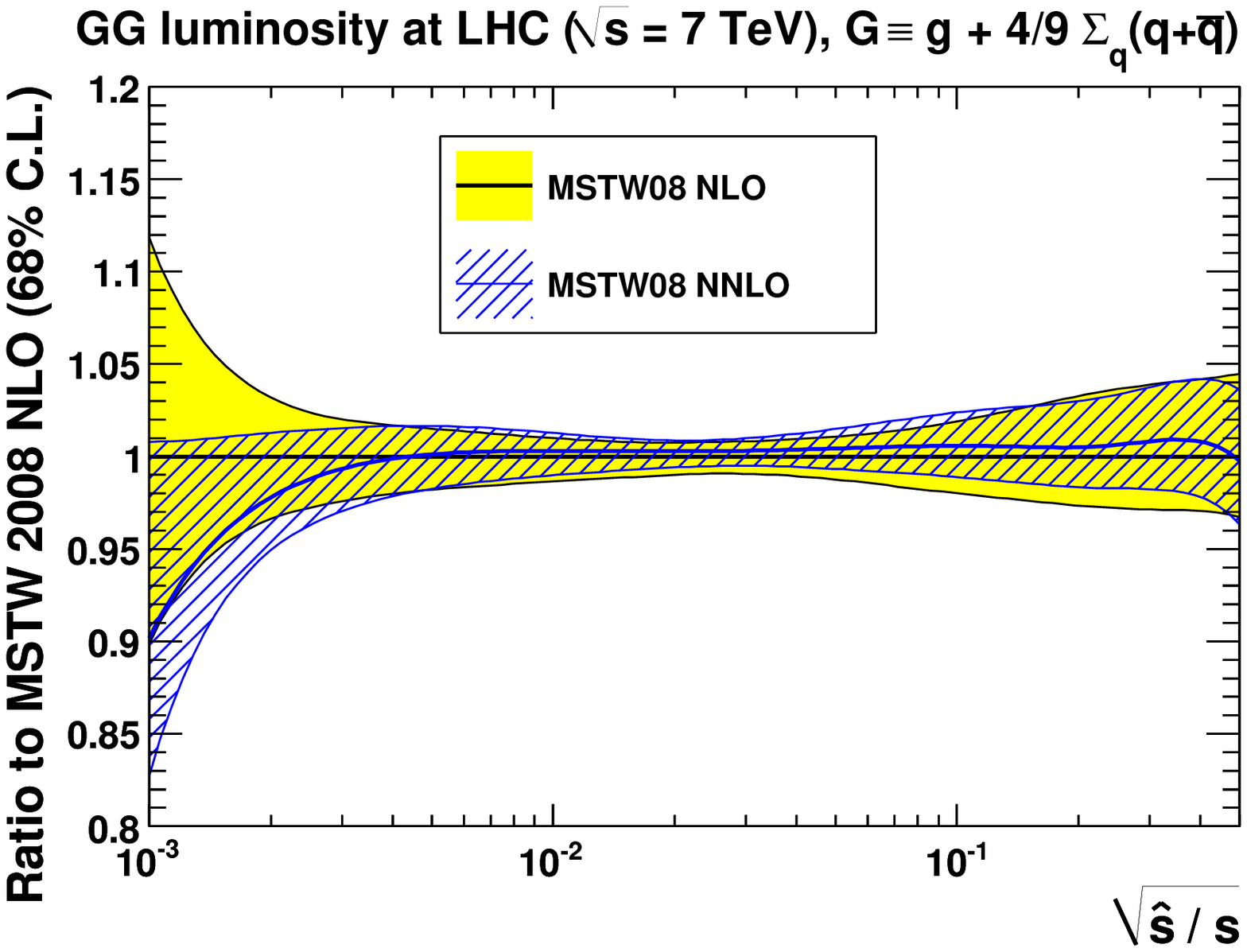}
  \end{minipage}
  \caption{NNLO parton--parton luminosities as the ratio with respect to MSTW 2008.}
  \label{fig:rationnlolumi}
\end{figure}
In figure~\ref{fig:rationnlolumi}(a,c,e) we show the NNLO parton--parton luminosities as the ratio with respect to the corresponding MSTW 2008 NNLO luminosities.  The two HERAPDF1.0 NNLO curves shown are for both $\alpha_S(M_Z^2)=0.1145$ and $0.1176$, where the latter value gives the smaller $gg$ and $GG$ luminosity, and the larger $q\bar{q}$ luminosity, at low $\hat{s}$ values.  The NNLO trend between groups is similar to at NLO.  In figure~\ref{fig:rationnlolumi}(b,d,f) we also compare the MSTW08 NNLO parton--parton luminosities to the corresponding MSTW08 NLO luminosities.  The $q\bar{q}$ luminosity differs between NLO and NNLO by an amount comparable to the experimental uncertainty, while there is less difference in the $gg$ and $GG$ luminosities, partially due to the larger $\alpha_S(M_Z^2)$ value used at NLO compensating for the missing NNLO corrections to the evolution; see figure~\ref{fig:asmzvalues}(b).

\begin{figure}
  \centering
  \begin{minipage}{0.5\textwidth}
    (a)\\
    \includegraphics[width=\textwidth]{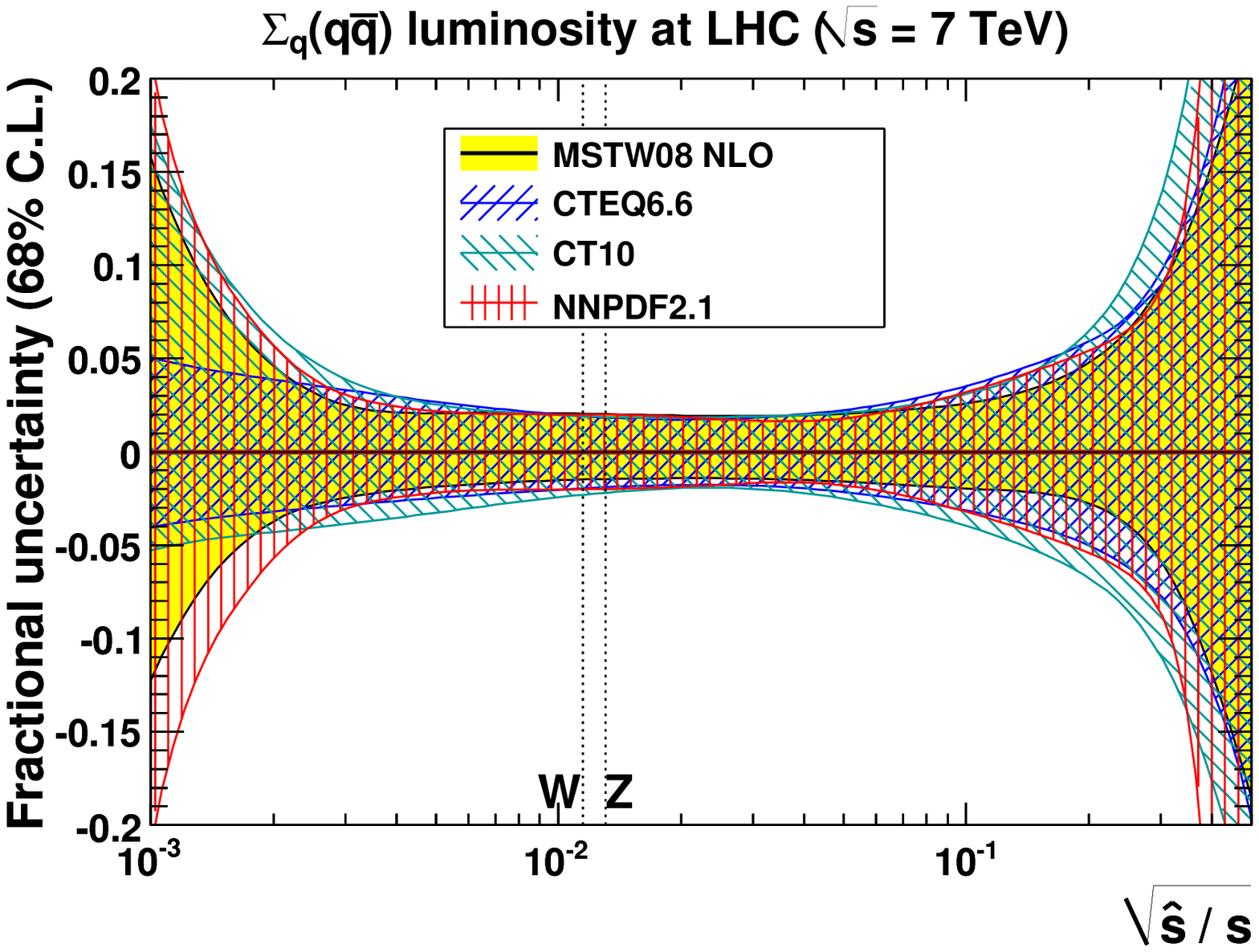}
  \end{minipage}%
  \begin{minipage}{0.5\textwidth}
    (b)\\
    \includegraphics[width=\textwidth]{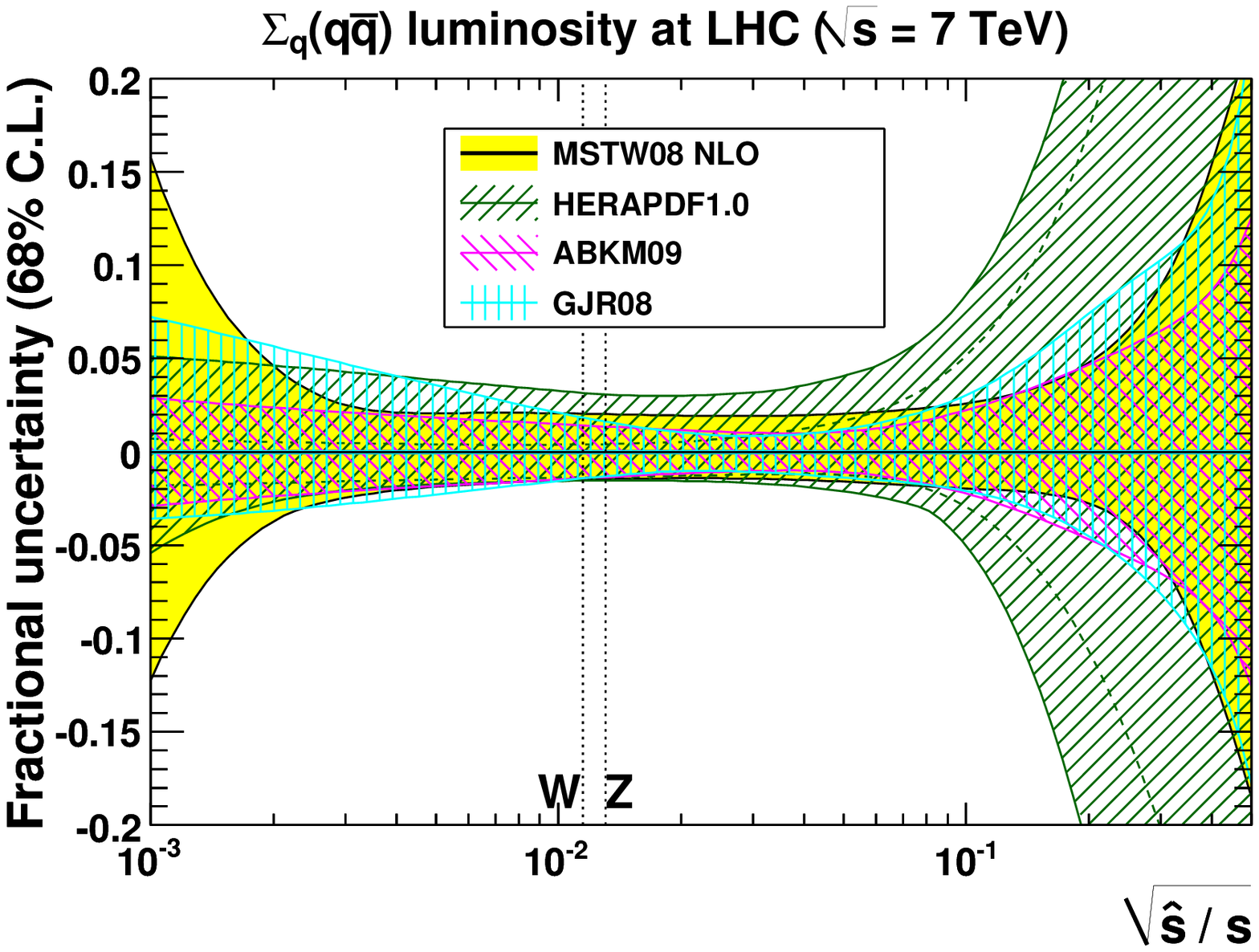}
  \end{minipage}
  \begin{minipage}{0.5\textwidth}
    (c)\\
    \includegraphics[width=\textwidth]{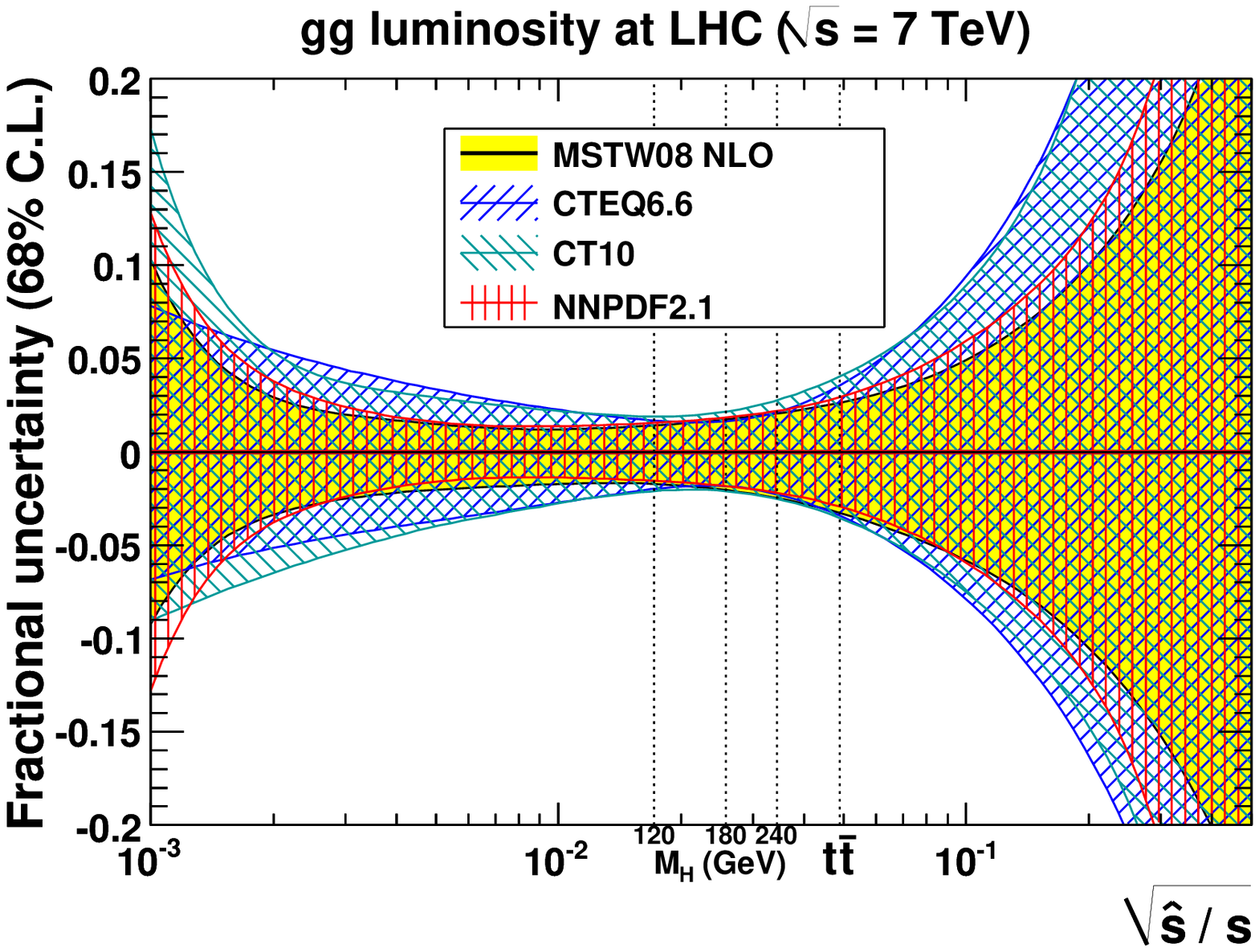}
  \end{minipage}%
  \begin{minipage}{0.5\textwidth}
    (d)\\
    \includegraphics[width=\textwidth]{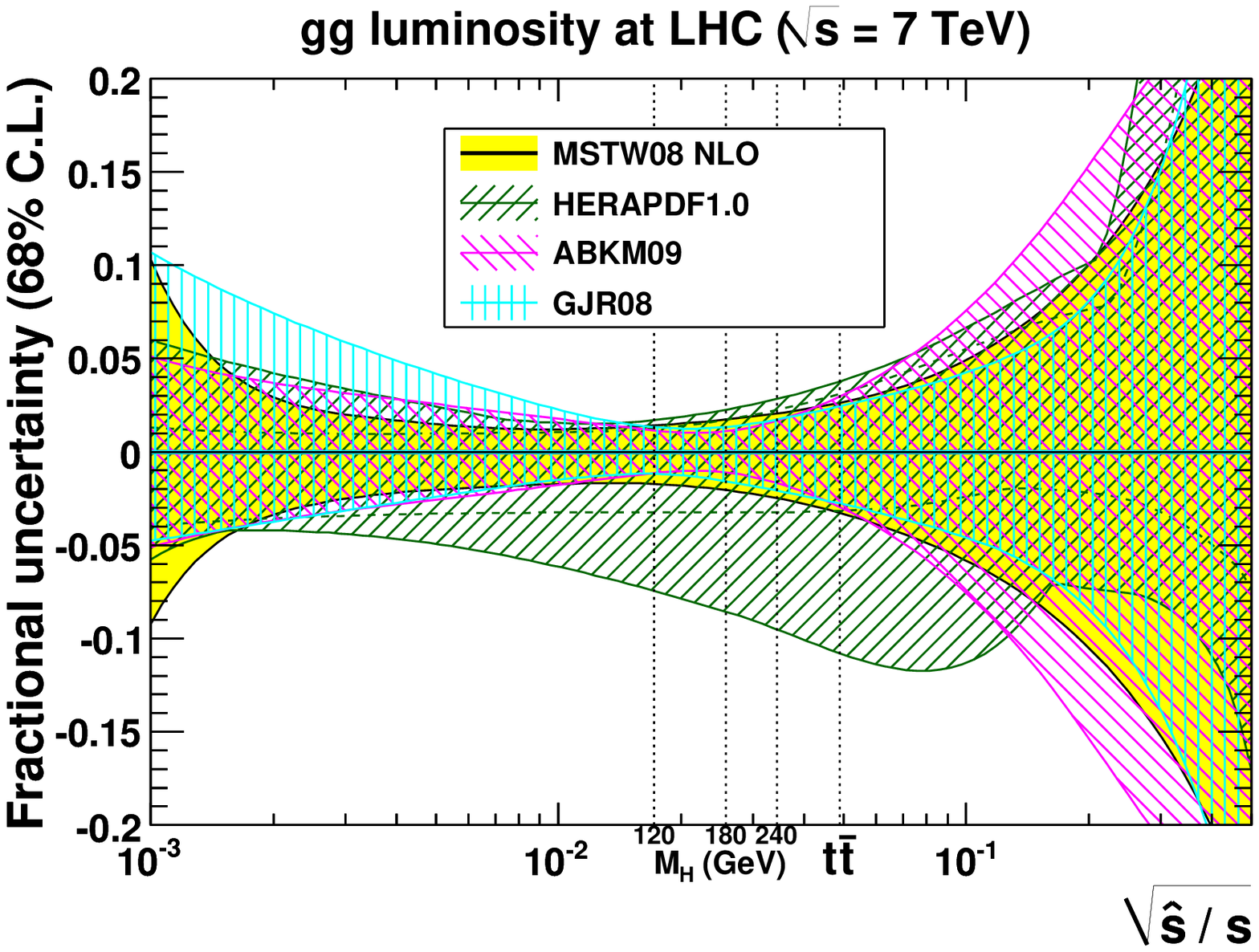}
  \end{minipage}
  \begin{minipage}{0.5\textwidth}
    (e)\\
    \includegraphics[width=\textwidth]{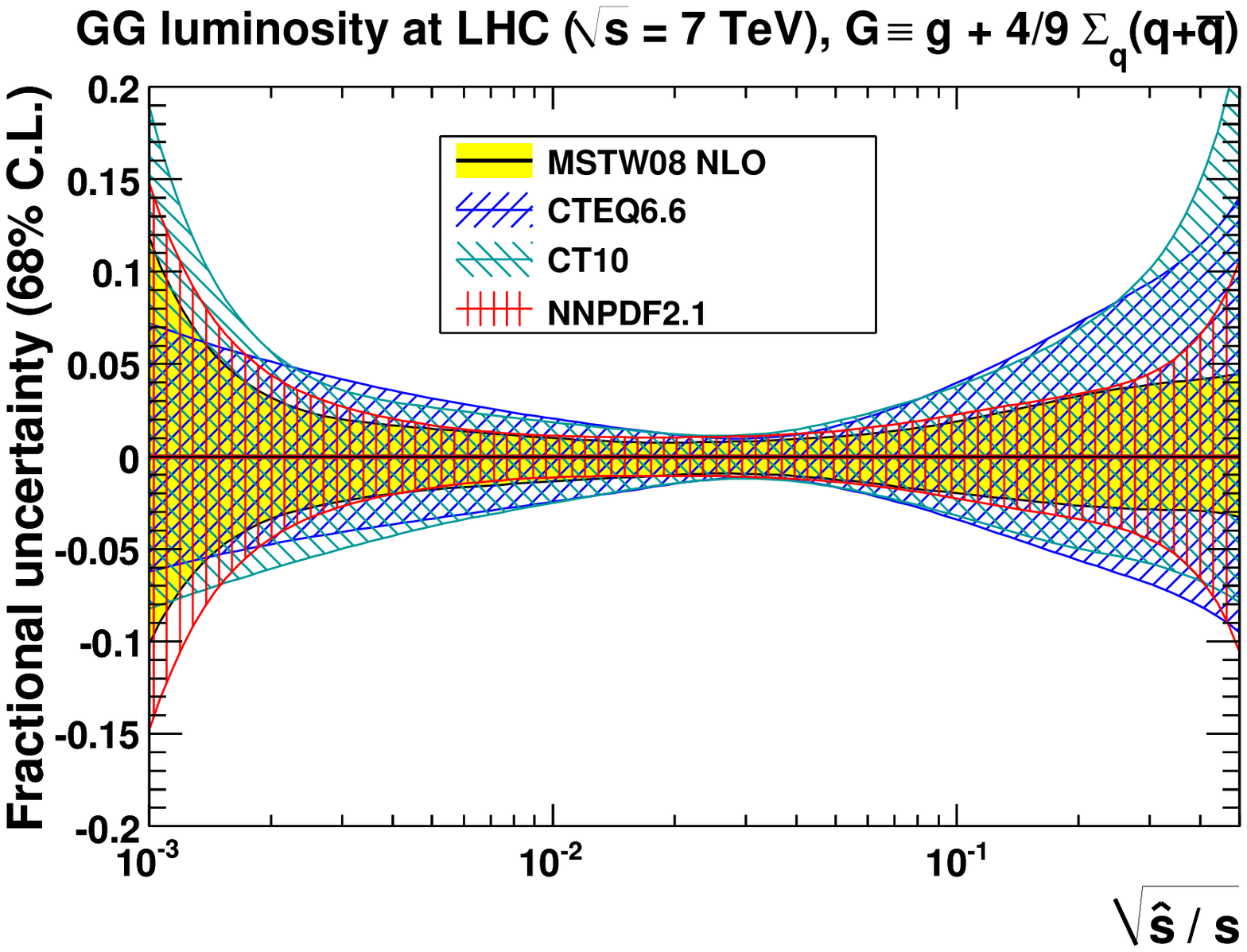}
  \end{minipage}%
  \begin{minipage}{0.5\textwidth}
    (f)\\
    \includegraphics[width=\textwidth]{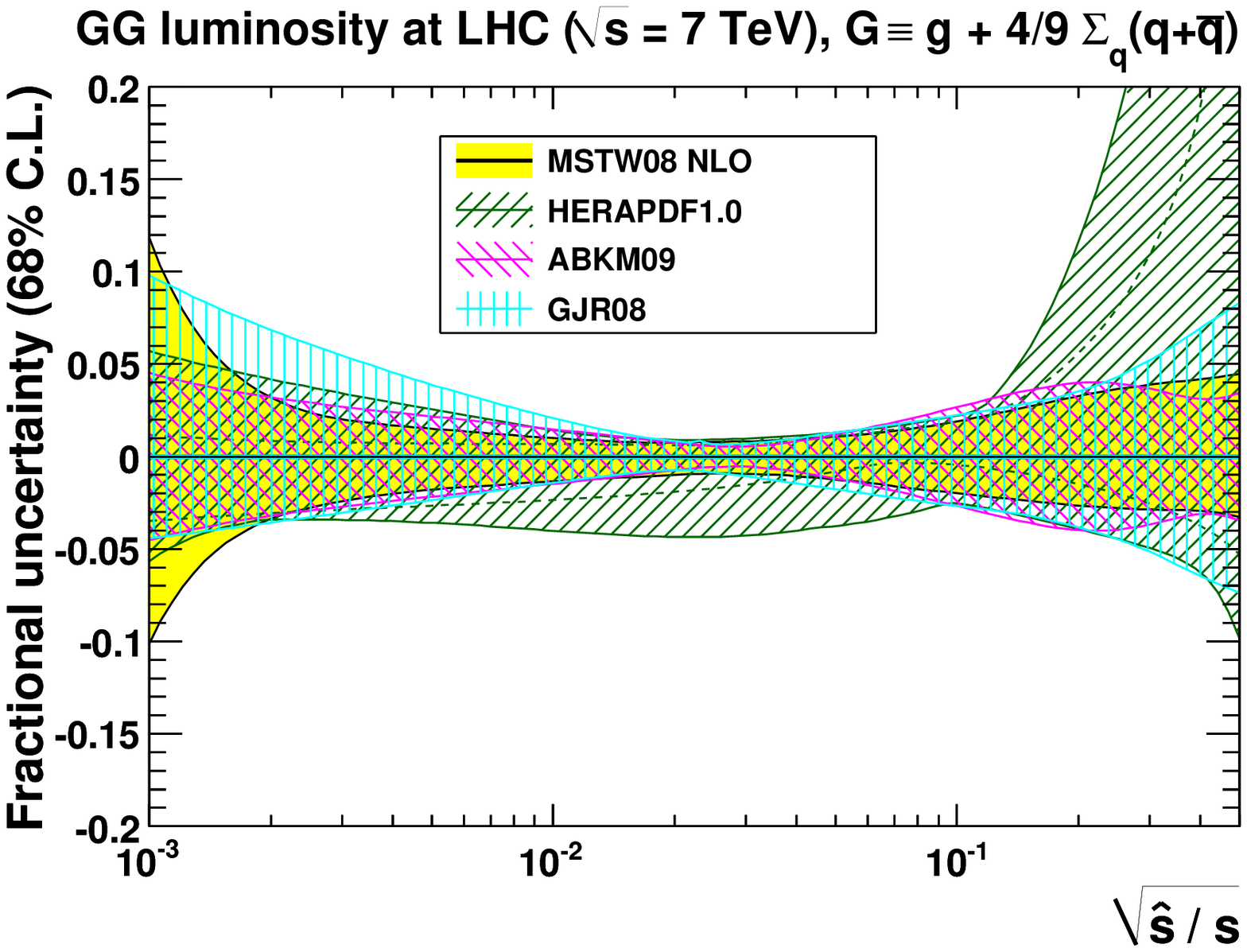}
  \end{minipage}
  \caption{Fractional uncertainties of the NLO parton--parton luminosities.}
  \label{fig:fracnlolumi}
\end{figure}
In figure~\ref{fig:fracnlolumi} we show the fractional uncertainties of the NLO parton--parton luminosities.  For the global fits (MSTW08, CTEQ6.6/CT10, NNPDF2.1) there is good agreement between the fractional uncertainties from different groups, particularly at intermediate values of $\hat{s}$.  This is quite remarkable considering the very different methods used for parameterisation and experimental error propagation, especially comparing the novel NNPDF approach to the more traditional methods used by the other groups.
\begin{figure}
  \centering
  \begin{minipage}{0.5\textwidth}
    (a)\\
    \includegraphics[width=\textwidth]{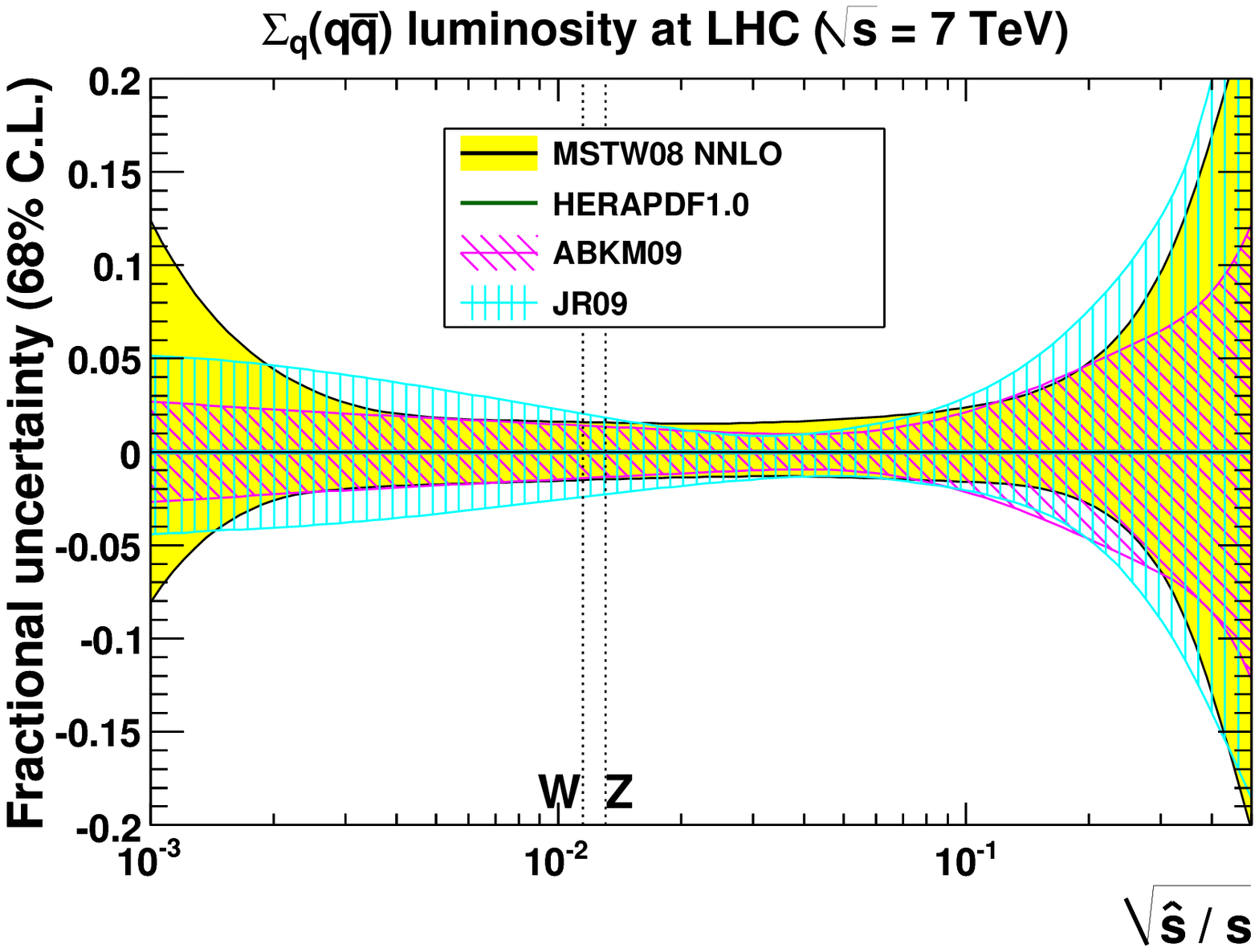}
  \end{minipage}%
  \begin{minipage}{0.5\textwidth}
    (b)\\
    \includegraphics[width=\textwidth]{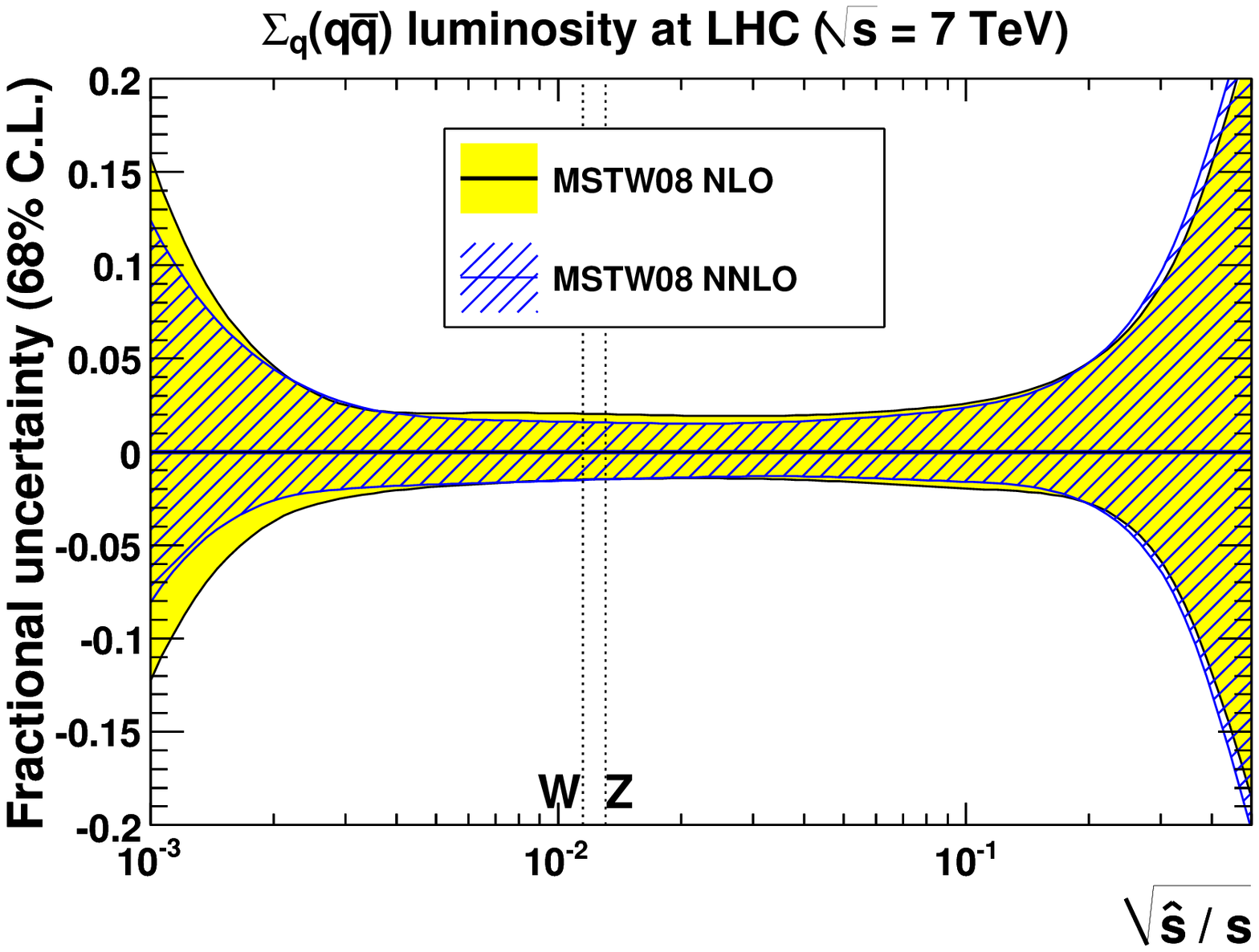}
  \end{minipage}
  \begin{minipage}{0.5\textwidth}
    (c)\\
    \includegraphics[width=\textwidth]{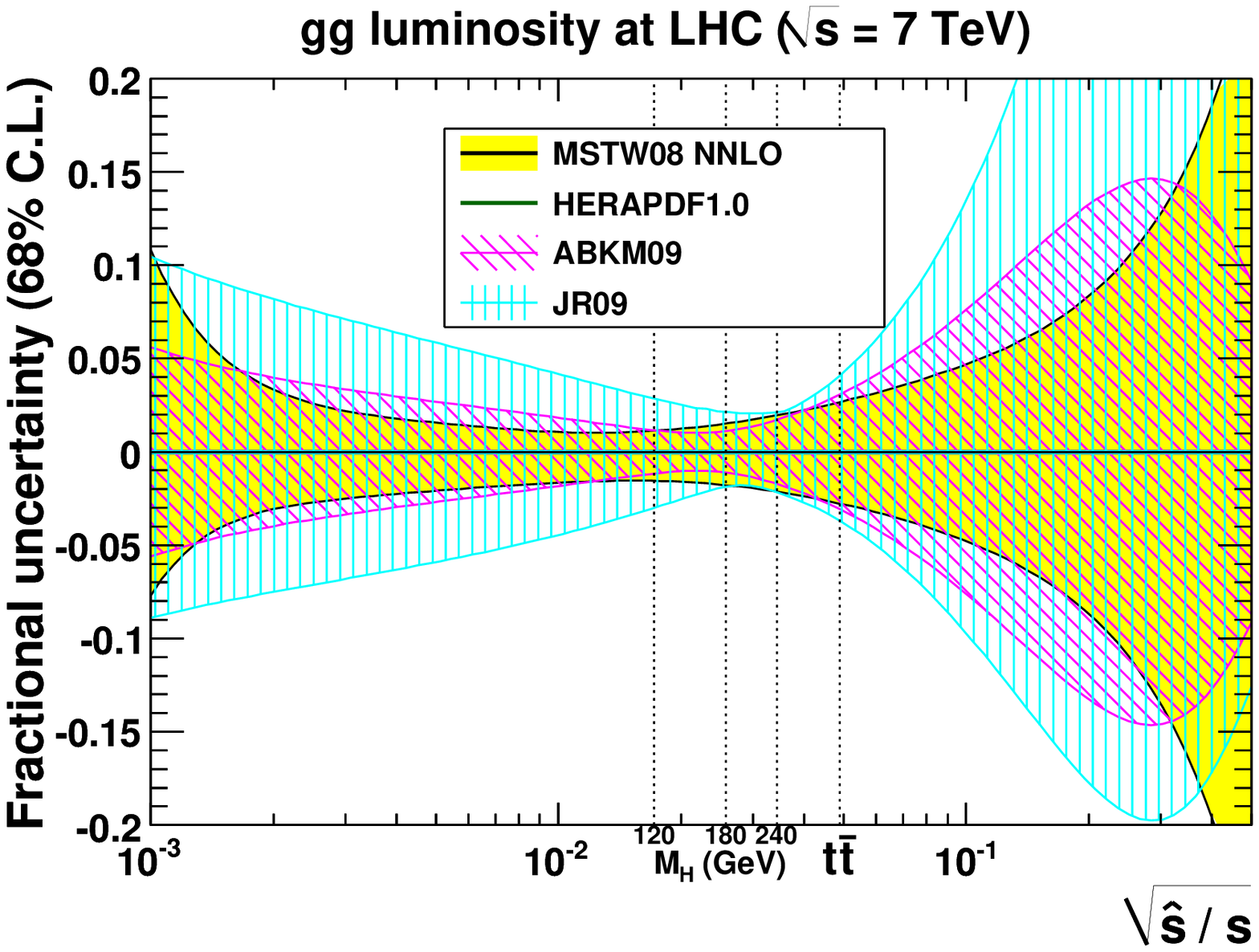}
  \end{minipage}%
  \begin{minipage}{0.5\textwidth}
    (d)\\
    \includegraphics[width=\textwidth]{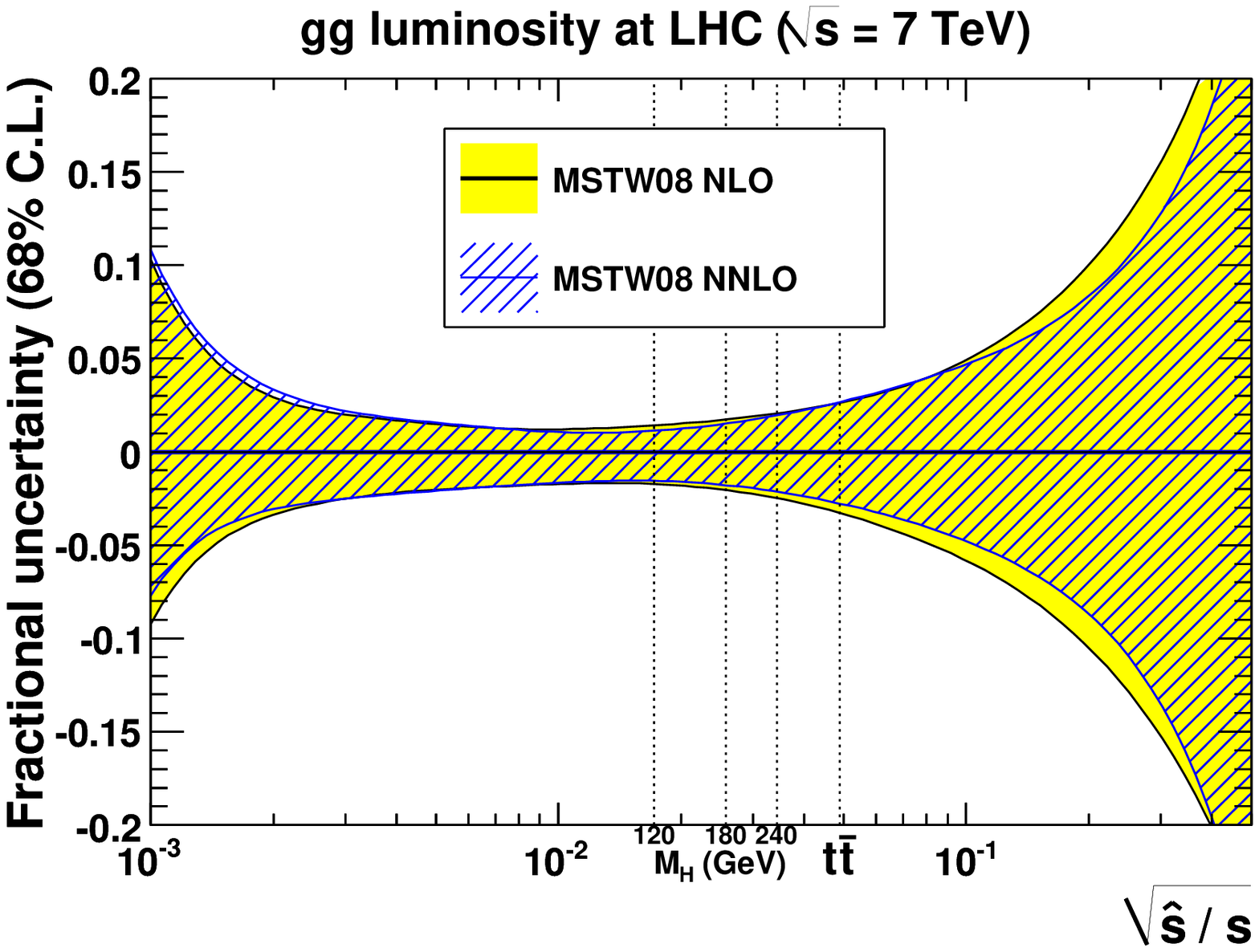}
  \end{minipage}
  \begin{minipage}{0.5\textwidth}
    (e)\\
    \includegraphics[width=\textwidth]{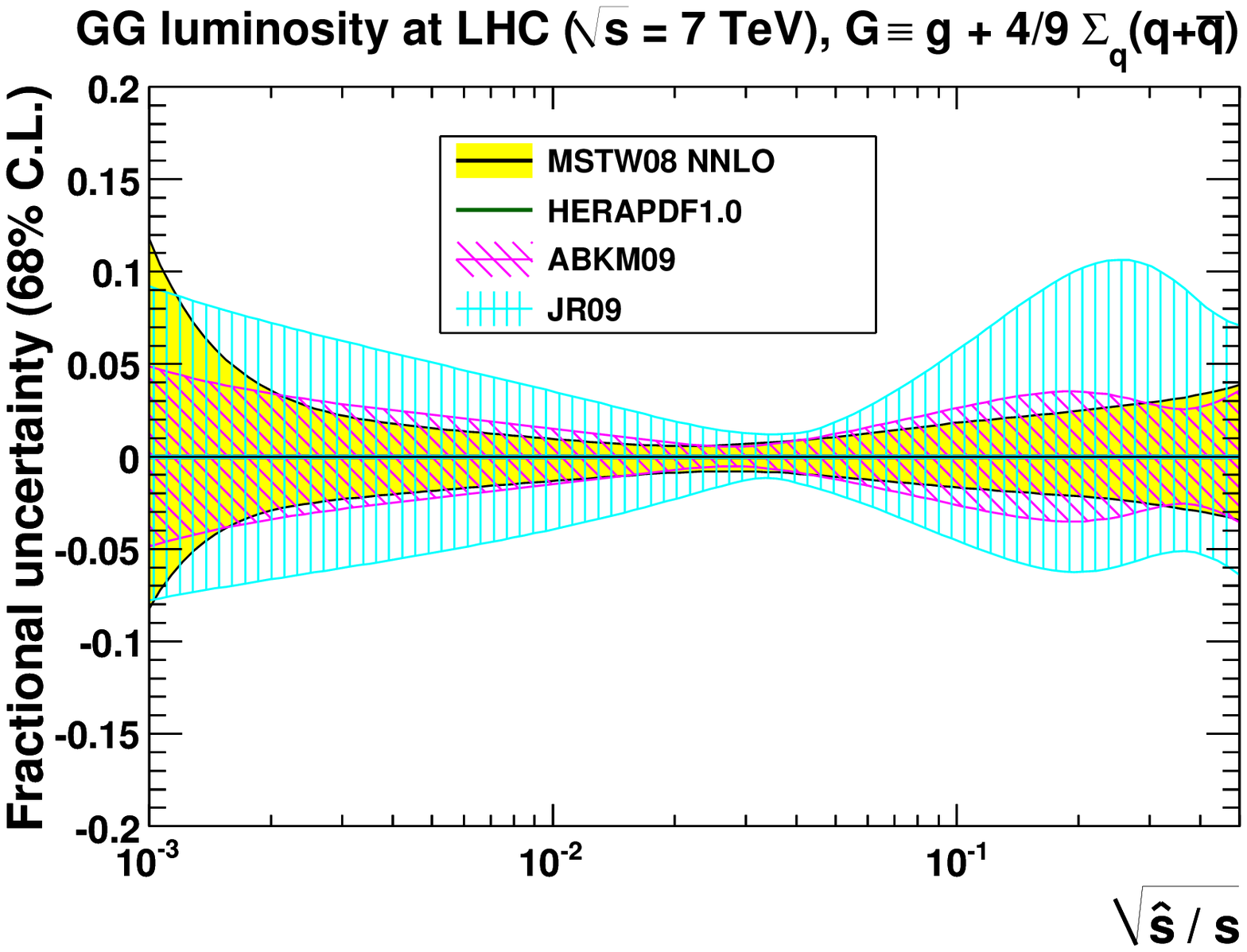}
  \end{minipage}%
  \begin{minipage}{0.5\textwidth}
    (f)\\
    \includegraphics[width=\textwidth]{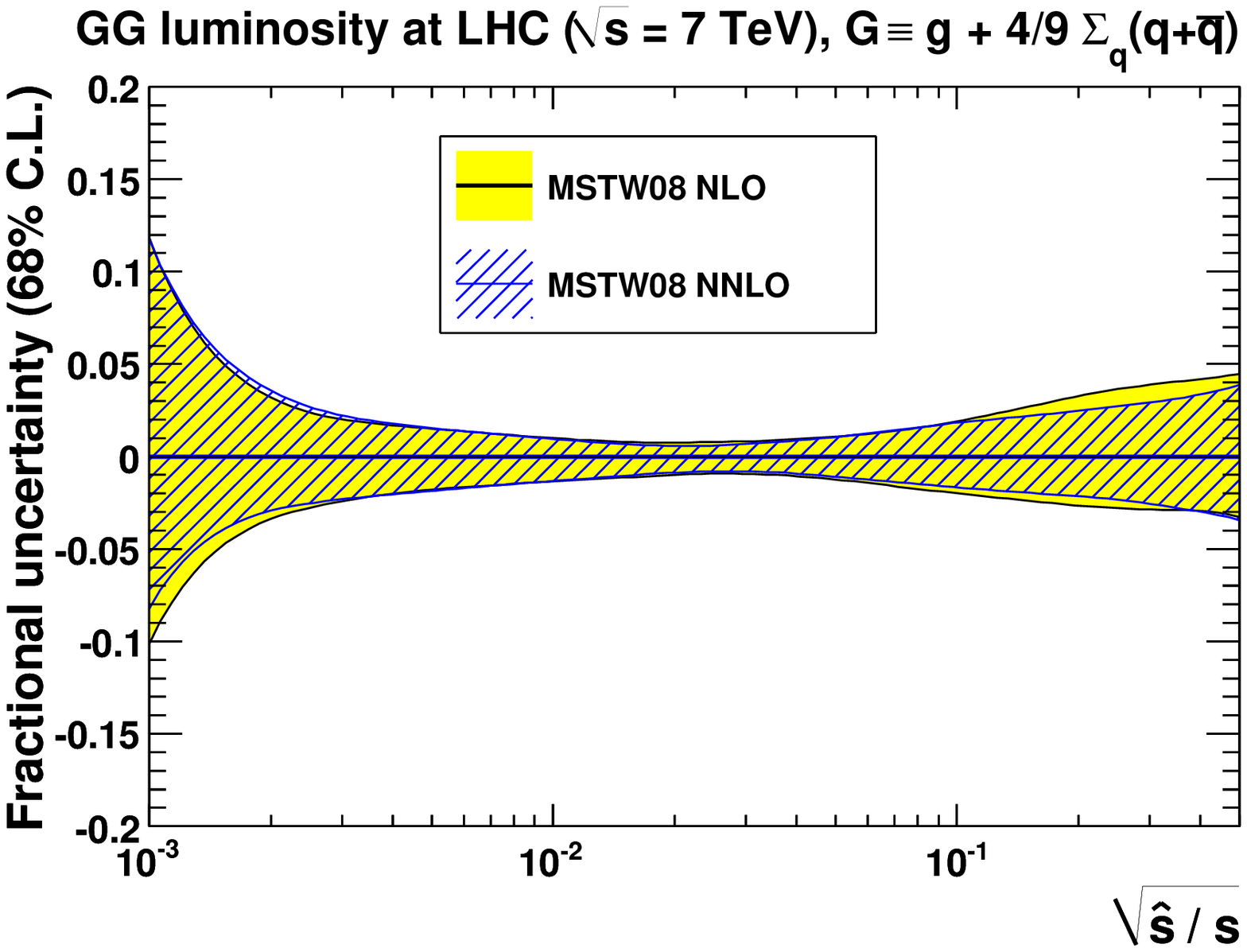}
  \end{minipage}
  \caption{Fractional uncertainties of the NNLO parton--parton luminosities.}
  \label{fig:fracnnlolumi}
\end{figure}
In figure~\ref{fig:fracnnlolumi} we show the fractional uncertainties of the NNLO parton--parton luminosities, again comparing MSTW08 NNLO to (a,c,e)~other NNLO sets and to (b,d,f)~MSTW08 NLO.  The NNLO fractional uncertainties are similar to those at NLO.  Uncertainties from ABKM09 and GJR08/JR09 are generally larger than for the global fits, but this may partly be due to the additional sources of uncertainty from $\alpha_S$ (GJR08/JR09) and both $\alpha_S$ and $m_{c,b}$ (ABKM09).

\section{\texorpdfstring{Total cross sections as a function of $\alpha_S(M_Z^2)$}{sigma versus alphaS(MZ)}} \label{sec:xsasmz}

The cross-section calculations are all performed in a 5-flavour ZM-VFNS using corresponding PDFs evolved with a maximum of 5 flavours,\footnote{The default PDF grids provided by NNPDF2.1 are evolved with 6 flavours for $Q^2>m_t^2$.  Moreover, the nearest grid points to $m_t=175$~GeV are at $Q=156$~GeV ($n_f=5$) and $Q=187$~GeV ($n_f=6$), meaning that interpolation between these two grid points does not have a well-defined $n_f$ value.  Although a variant of NNPDF2.1 with the default $\alpha_S(M_Z^2)=0.119$ is now provided with 5-flavour evolution, grids for other $\alpha_S$ values are only provided with 6-flavour evolution.  Therefore we use only the 6-flavour NNPDF2.1 grids.} as used also for the parton--parton luminosity plots in section~\ref{sec:luminosities}.  We want to isolate the PDF (and $\alpha_S$) dependence, therefore we use precisely the same code for all PDF sets with common settings.  The PDF+$\alpha_S$ uncertainties are computed using the recommended prescription of each fitting group, summarised in section~\ref{sec:alphaS}.  No attempt is made to evaluate other theoretical uncertainties and a single scale choice ($\mu_R=\mu_F=M_{W},M_{Z},M_{H},m_{t}$) is adopted for each process.  We only present \emph{total} cross sections, not differential distributions.  For the PDF4LHC benchmark exercise at NLO~\cite{Alekhin:2011sk}, a modified \textsc{mcfm} V5.7 package~\cite{Campbell:2010ff,terascale} with agreed parameters and input files was produced by J.~Campbell.  Instead we use private code (obtained from J.~Stirling) for total cross sections at NLO, also used for the majority of plots in ref.~\cite{Alekhin:2011sk}, which is much faster and was checked against the official \textsc{mcfm} package.  Although fairly precise total cross sections for $W^\pm$, $Z^0$ and $t\bar{t}$ production have now been measured at the LHC, we resist the temptation to directly compare with data, since in many cases a more extensive evaluation of other theoretical uncertainties is necessary, and most of the cross-section measurements with the full 2010 data set are still preliminary~\cite{ATLAS:WandZ,CMS:WandZ,ATLAS:ttbar,CMS:ttbar}.

\subsection{\texorpdfstring{$W^\pm$ and $Z^0$}{W and Z} production}

\begin{figure}
  \centering
  \begin{minipage}{0.5\textwidth}
    (a)\\
    \includegraphics[width=\textwidth]{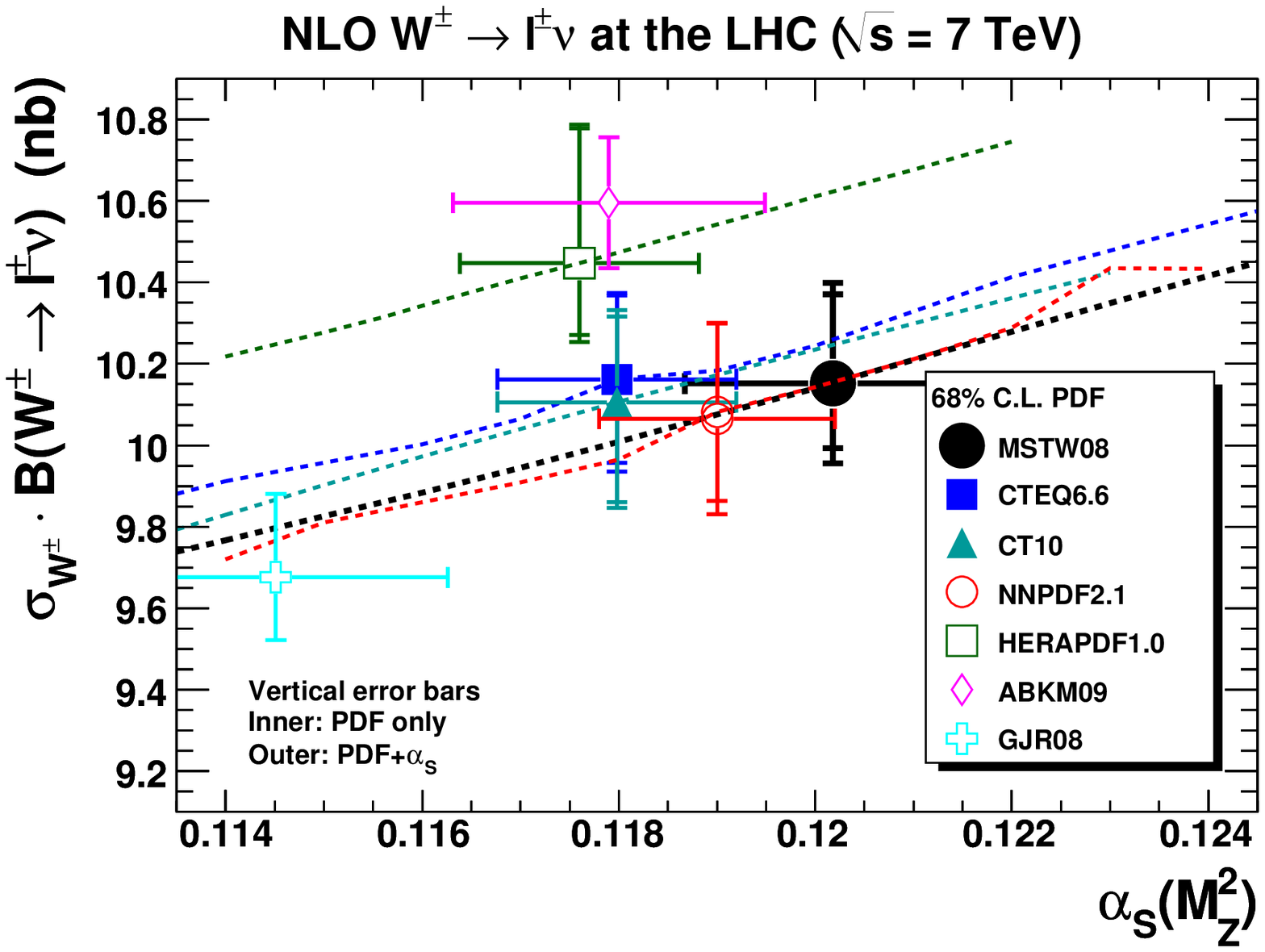}
  \end{minipage}%
  \begin{minipage}{0.5\textwidth}
    (b)\\
    \includegraphics[width=\textwidth]{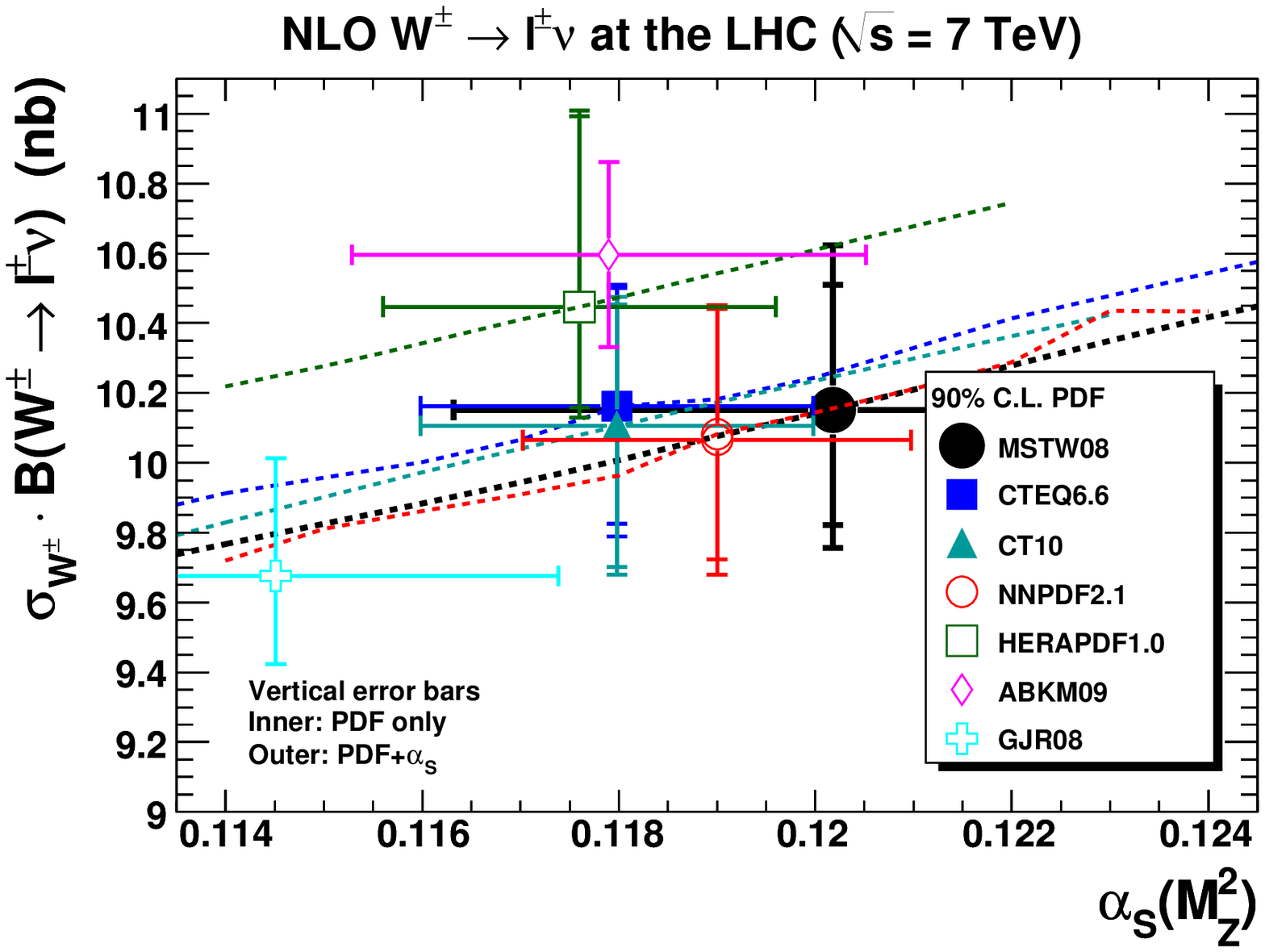}
  \end{minipage}
  \begin{minipage}{0.5\textwidth}
    (c)\\
    \includegraphics[width=\textwidth]{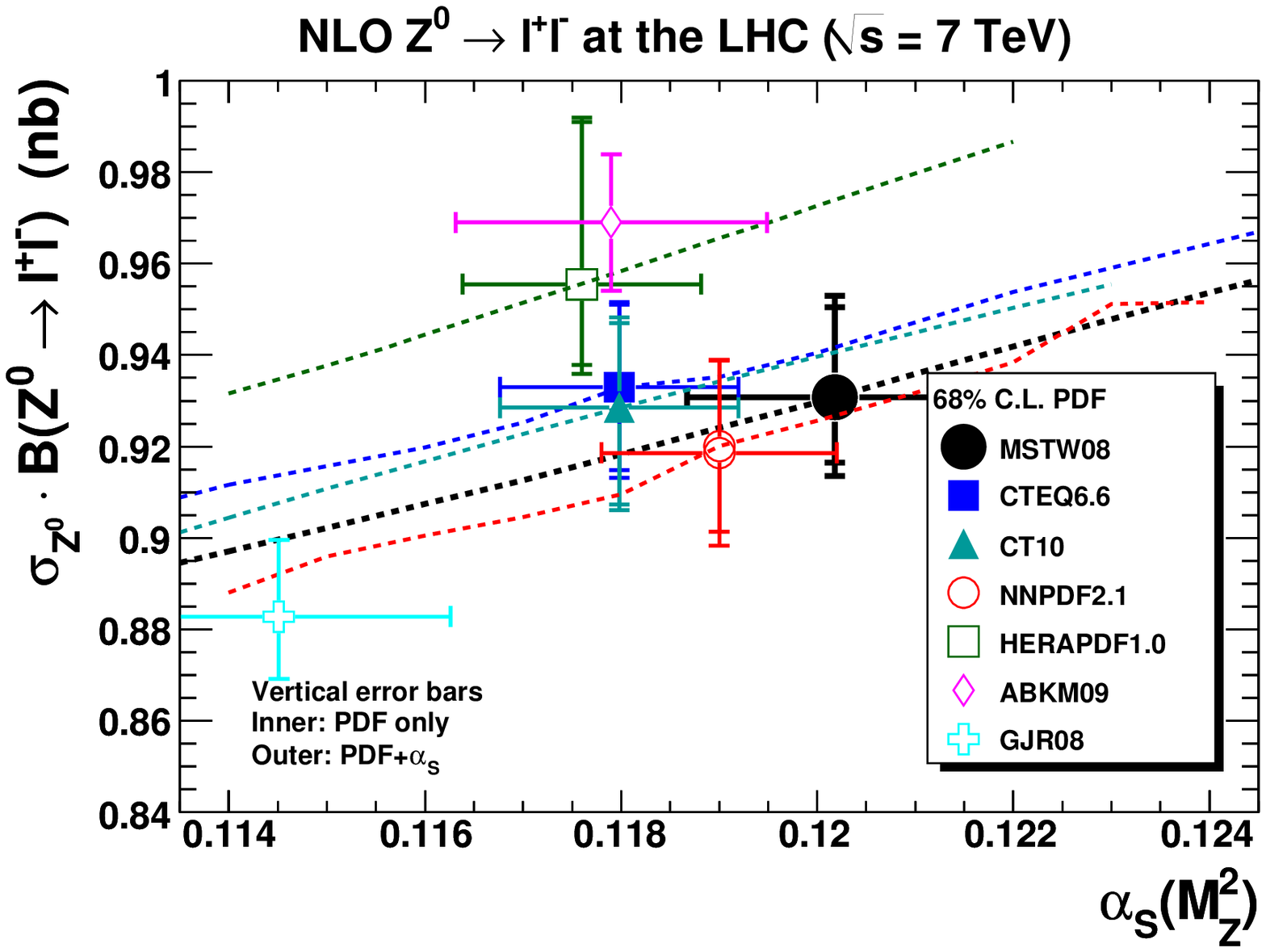}
  \end{minipage}%
  \begin{minipage}{0.5\textwidth}
    (d)\\
    \includegraphics[width=\textwidth]{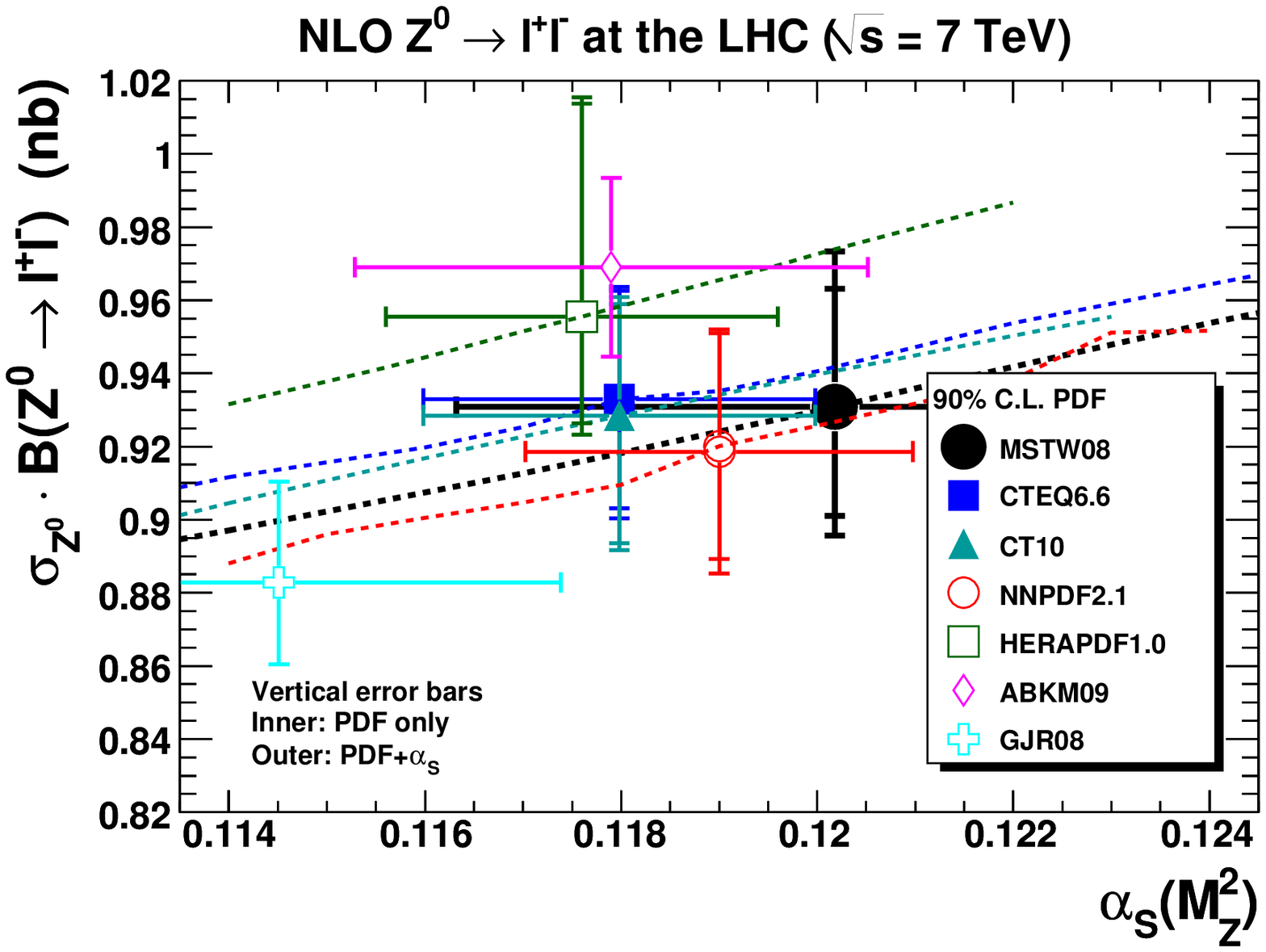}
  \end{minipage}
  \begin{minipage}{0.5\textwidth}
    (e)\\
    \includegraphics[width=\textwidth]{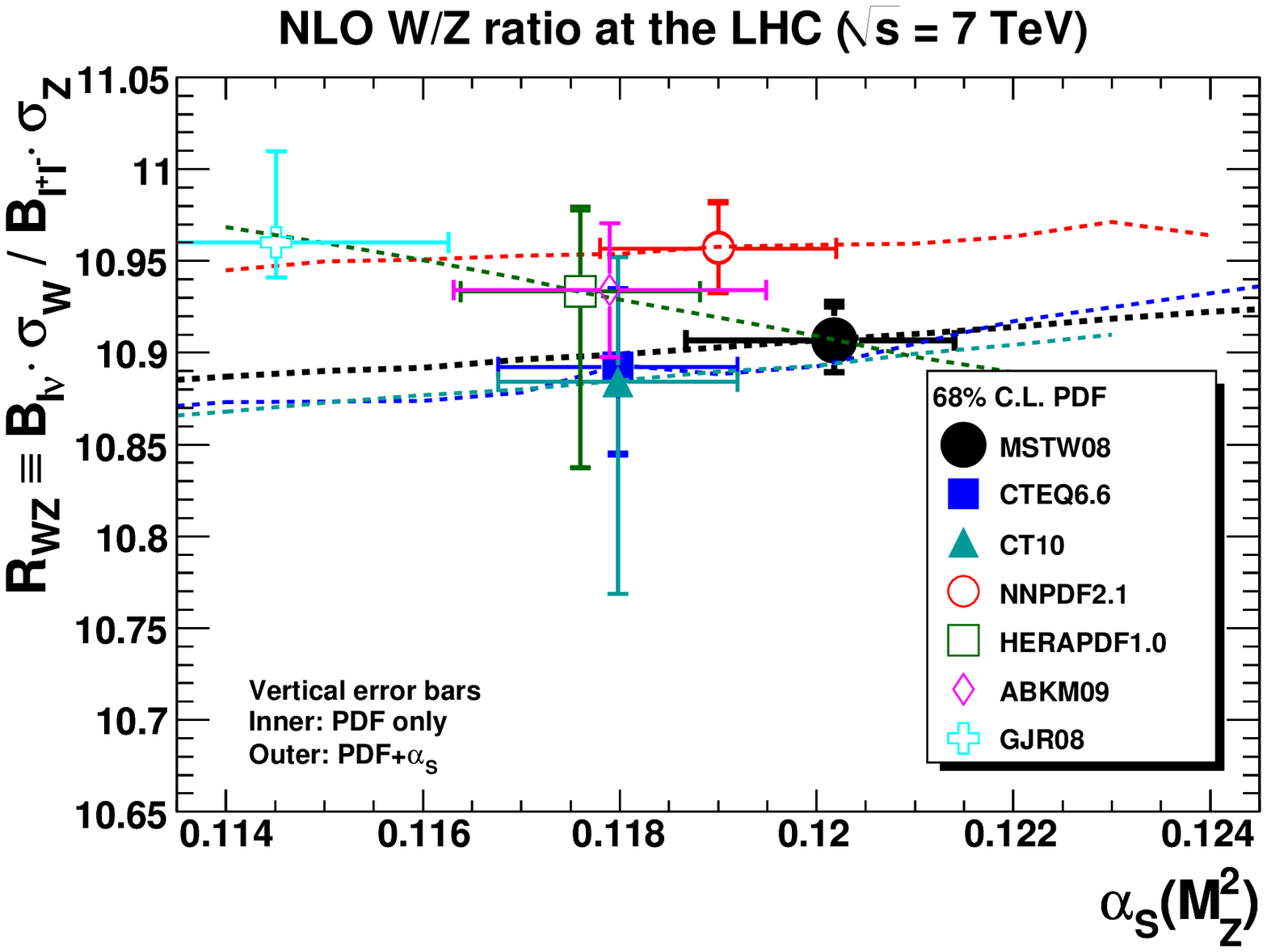}
  \end{minipage}%
  \begin{minipage}{0.5\textwidth}
    (f)\\
    \includegraphics[width=\textwidth]{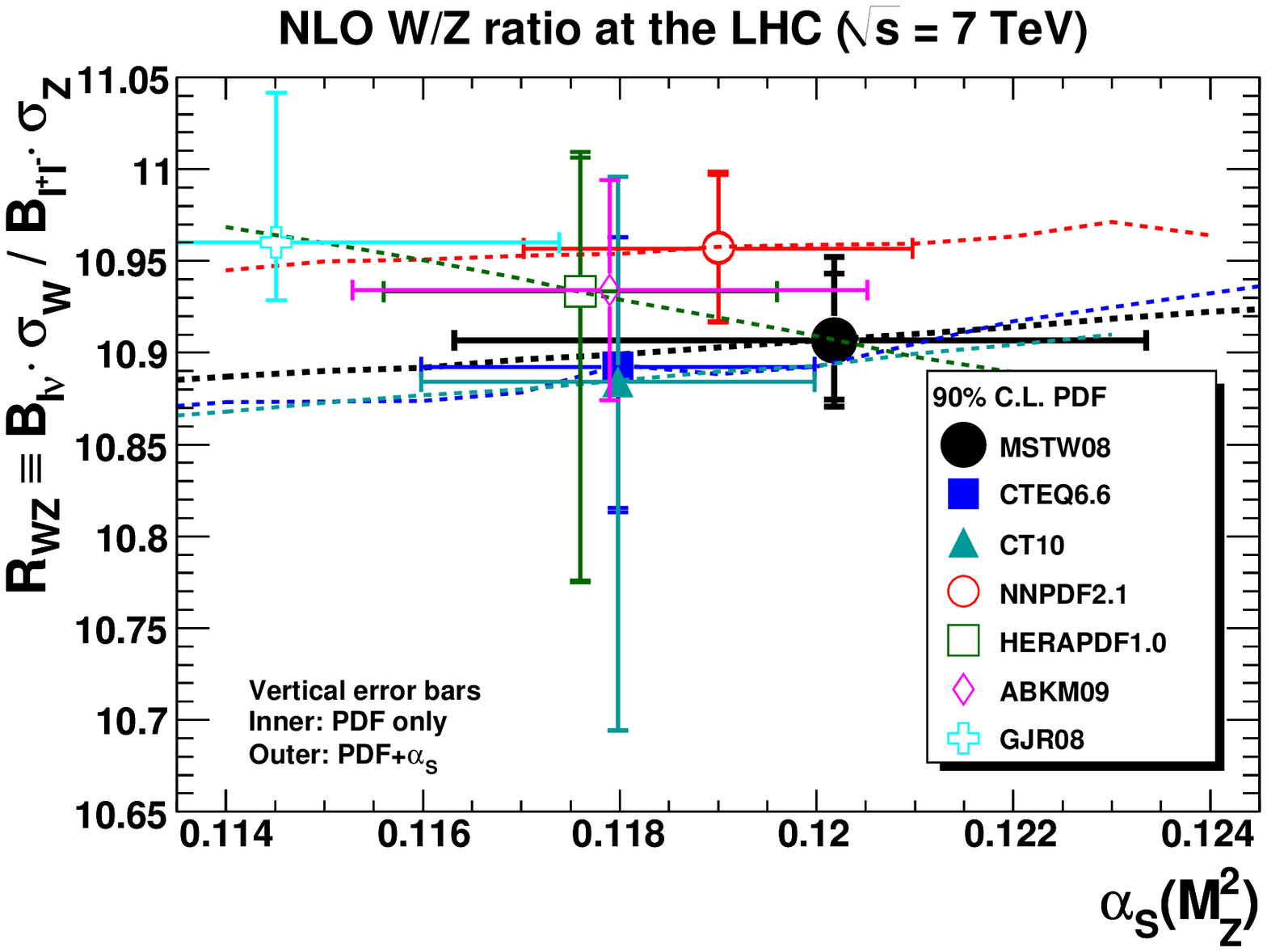}
  \end{minipage}
  \caption{$W^\pm$ and $Z^0$ total cross sections, plotted as a function of $\alpha_S(M_Z^2)$, at NLO.}
  \label{fig:wzvsasmznlo}
\end{figure}

\begin{figure}
  \centering
  \begin{minipage}{0.5\textwidth}
    (a)\\
    \includegraphics[width=\textwidth]{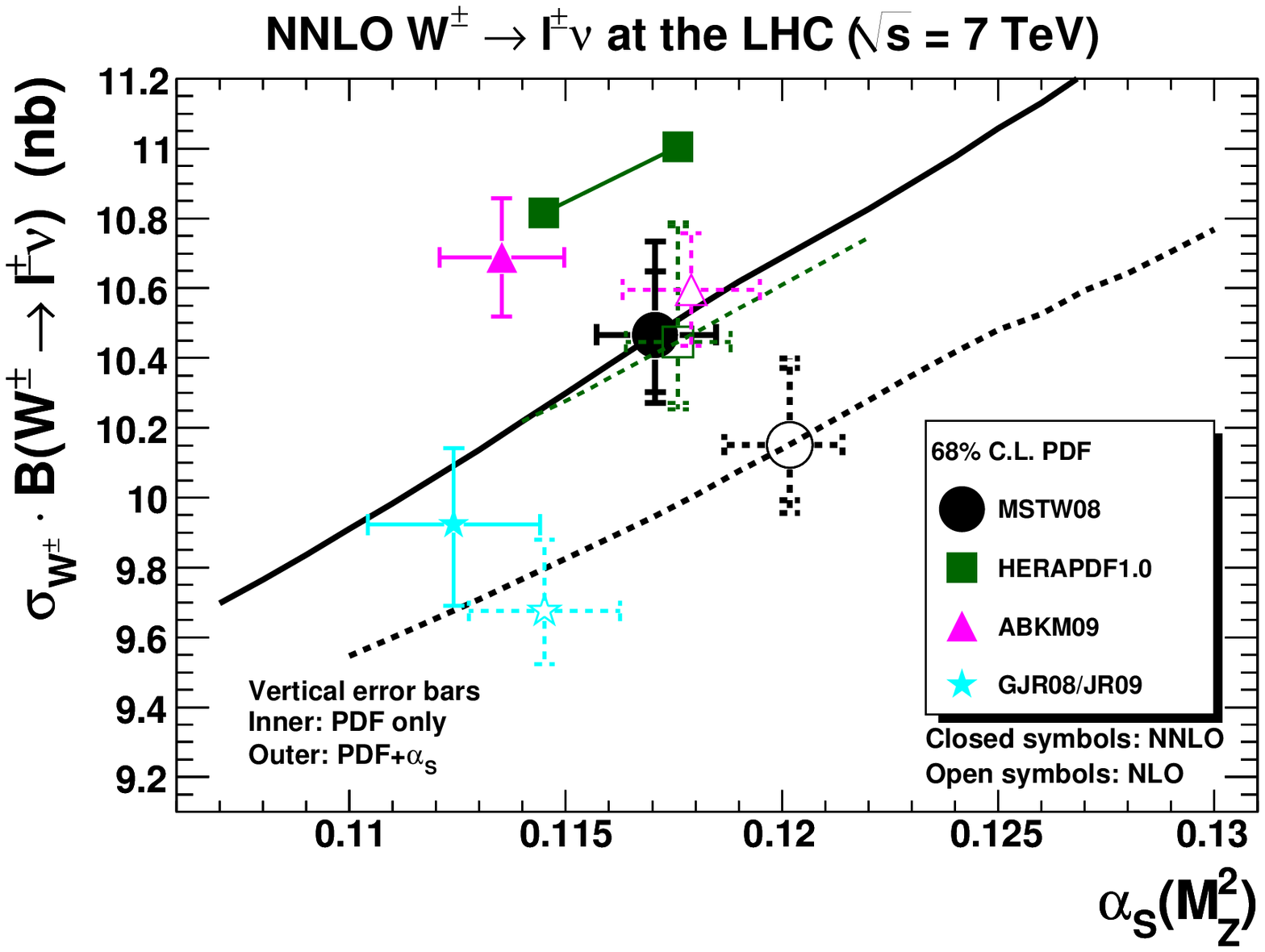}
  \end{minipage}%
  \begin{minipage}{0.5\textwidth}
    (b)\\
    \includegraphics[width=\textwidth]{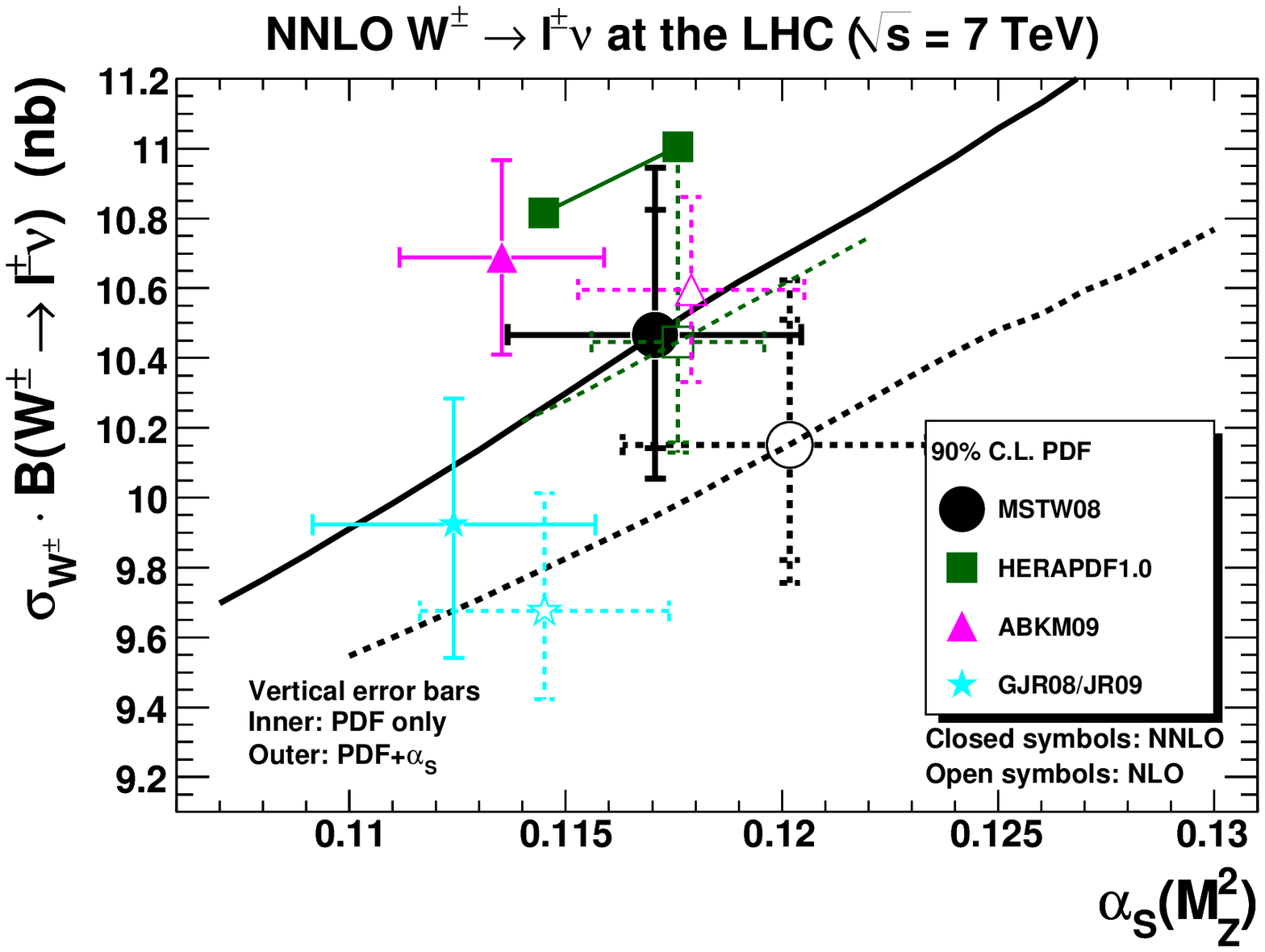}
  \end{minipage}
  \begin{minipage}{0.5\textwidth}
    (c)\\
    \includegraphics[width=\textwidth]{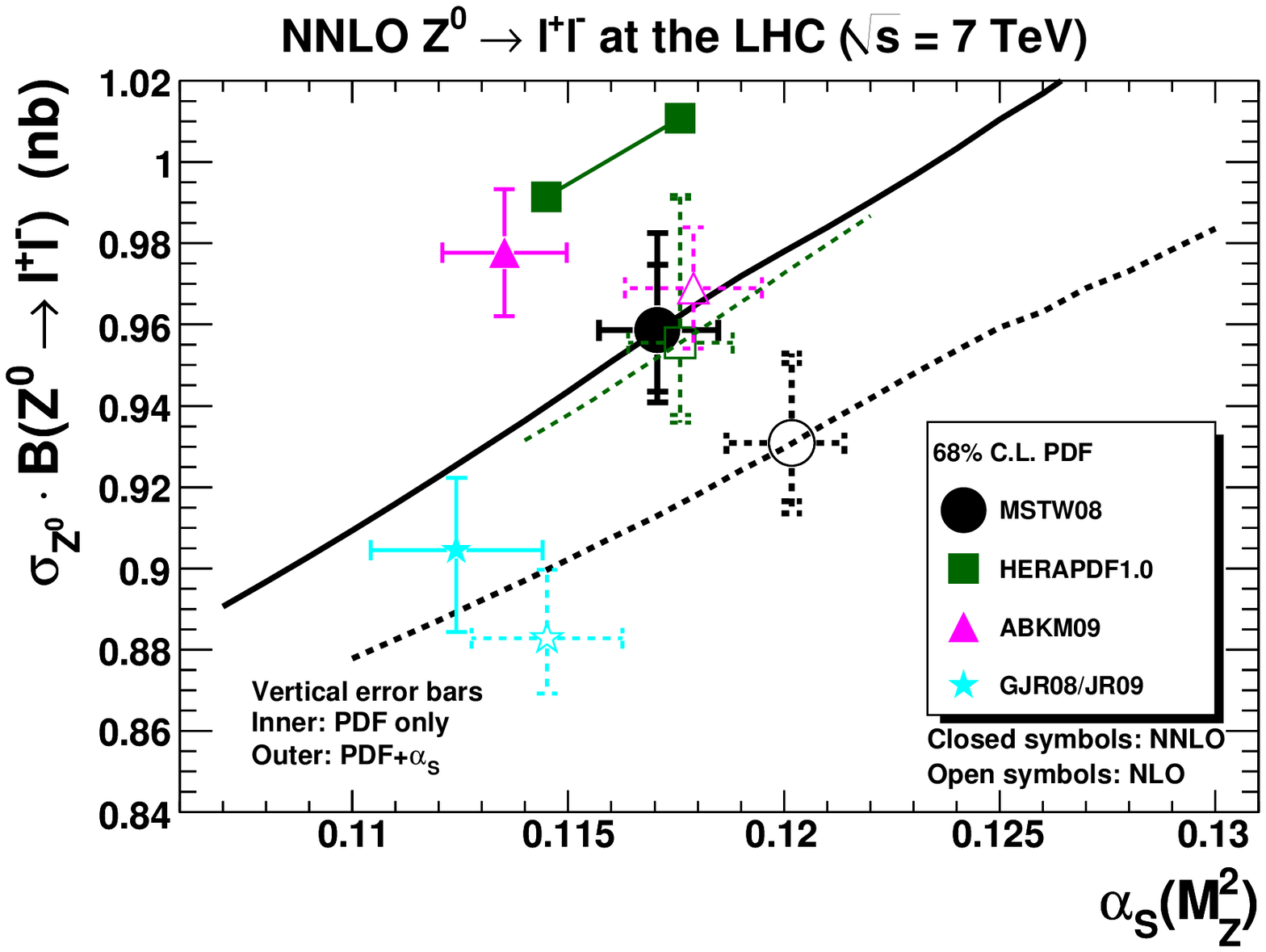}
  \end{minipage}%
  \begin{minipage}{0.5\textwidth}
    (d)\\
    \includegraphics[width=\textwidth]{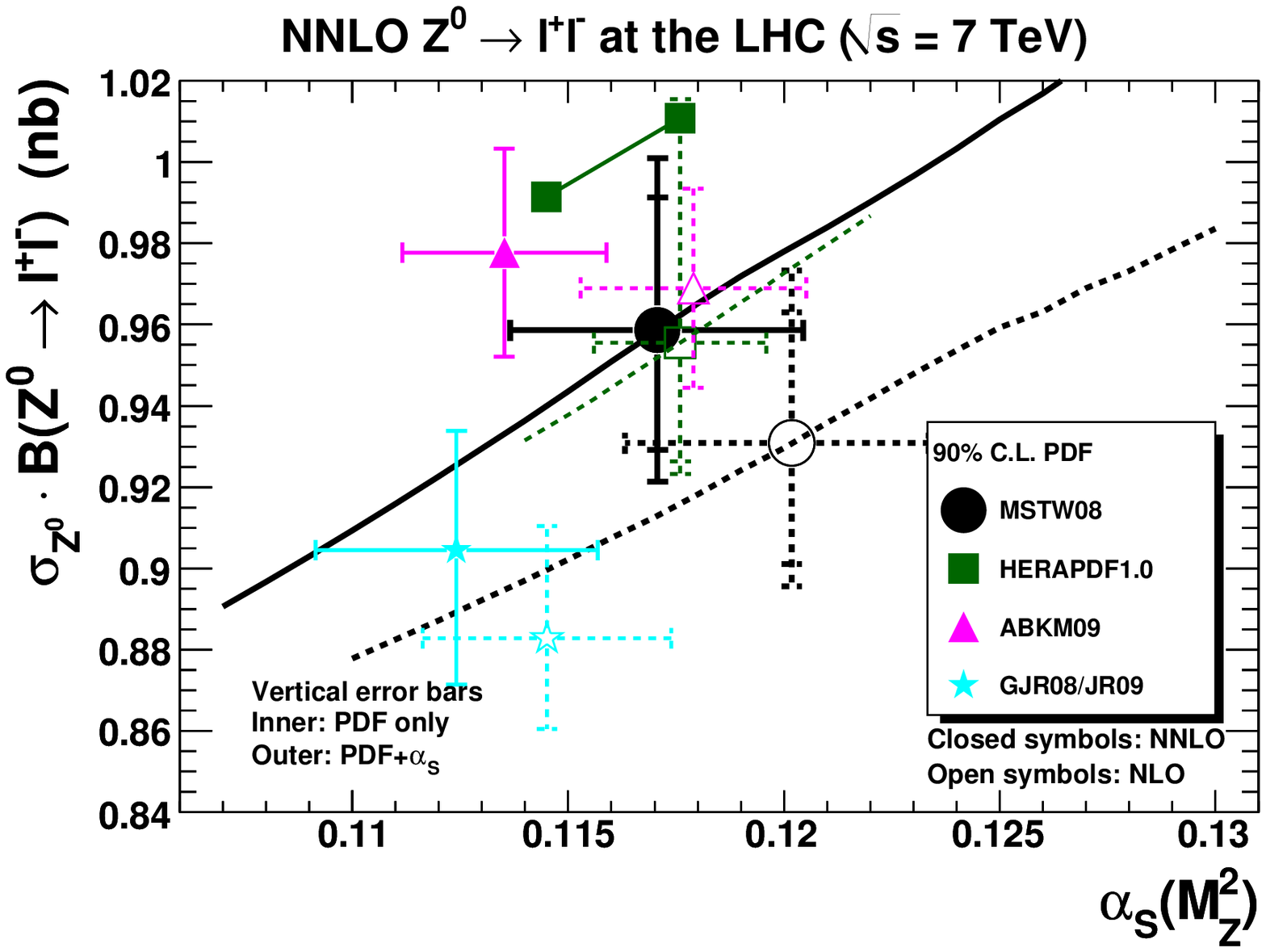}
  \end{minipage}
  \begin{minipage}{0.5\textwidth}
    (e)\\
    \includegraphics[width=\textwidth]{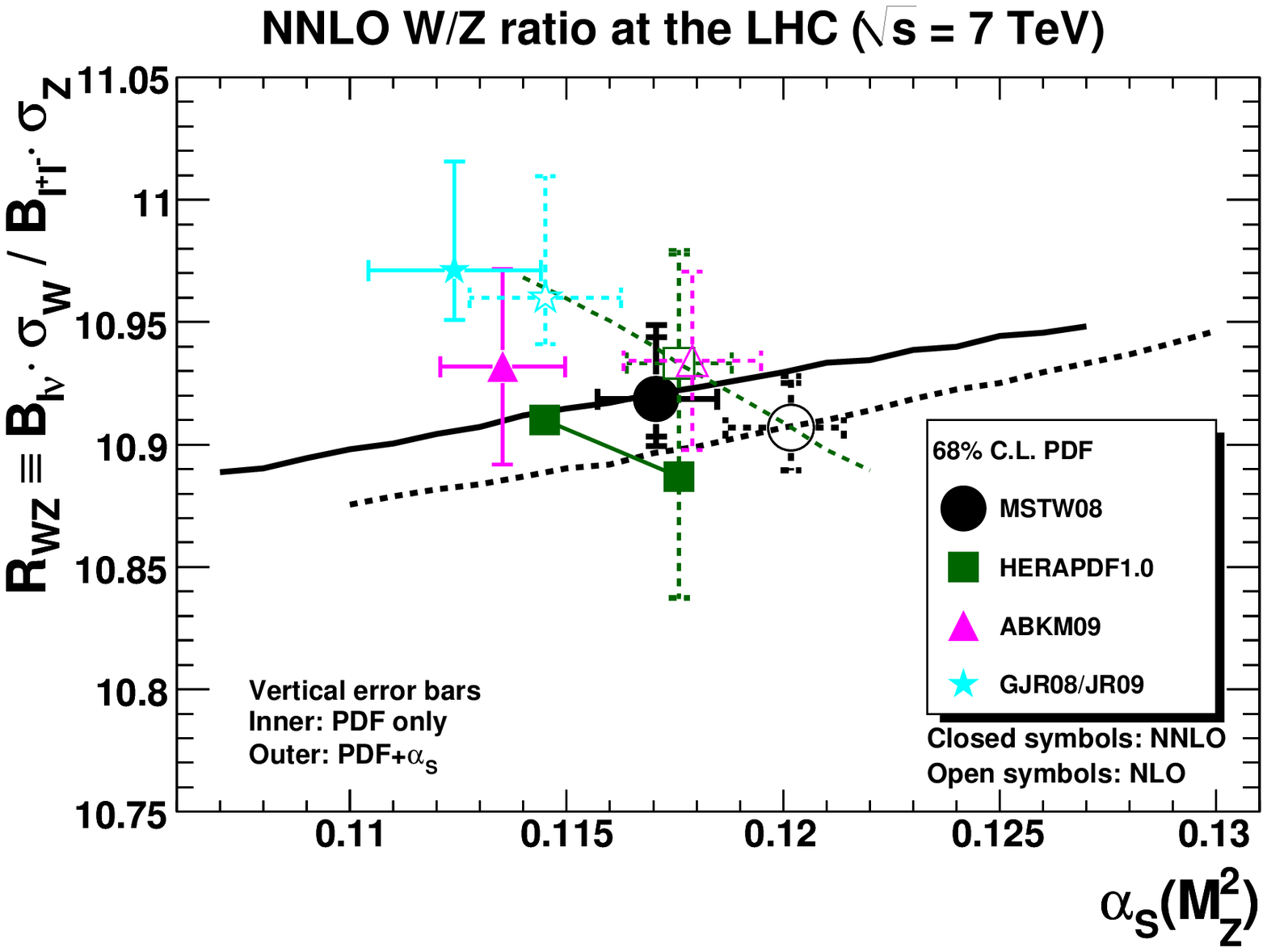}
  \end{minipage}%
  \begin{minipage}{0.5\textwidth}
    (f)\\
    \includegraphics[width=\textwidth]{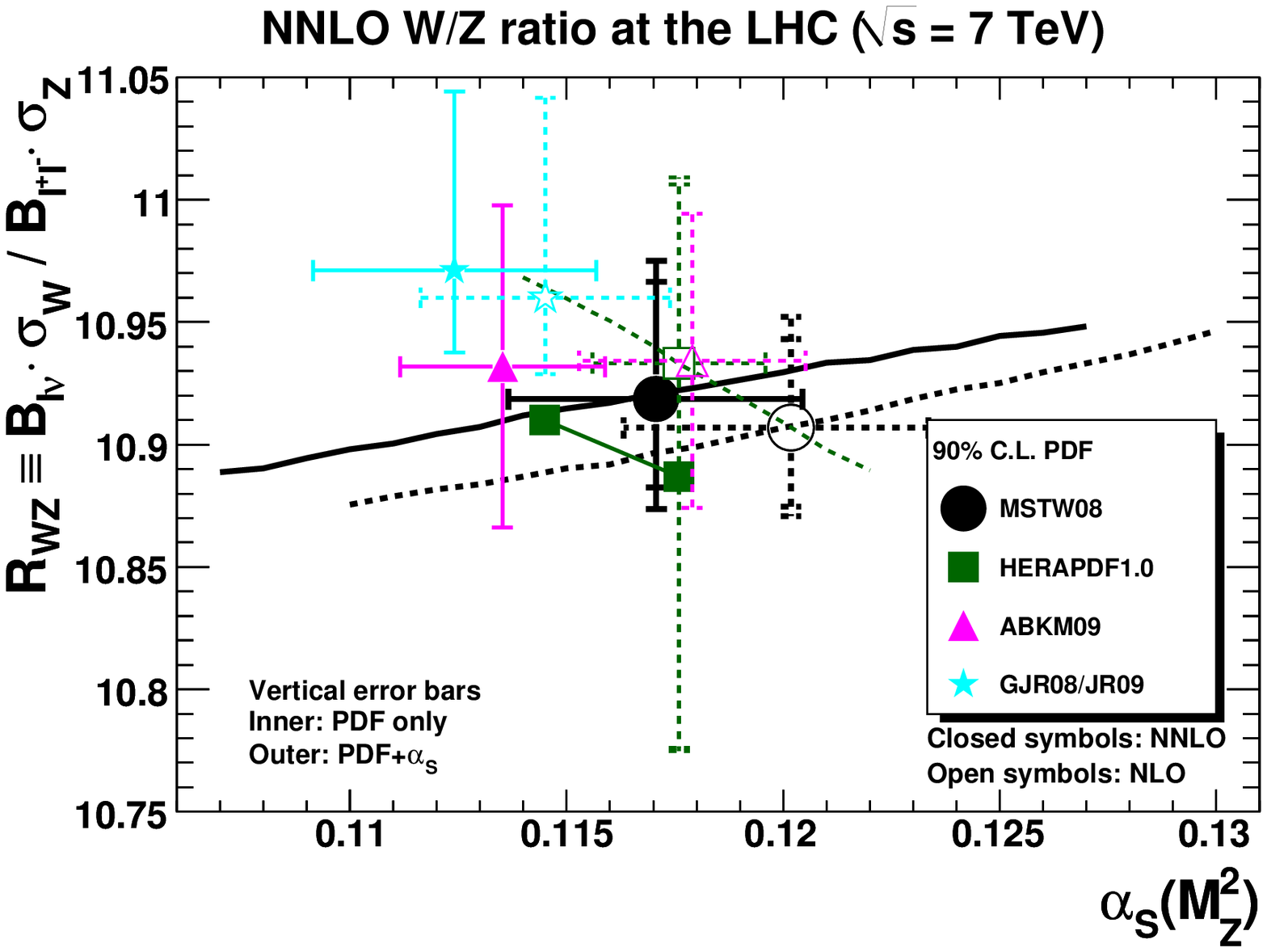}
  \end{minipage}
  \caption{$W^\pm$ and $Z^0$ total cross sections, plotted as a function of $\alpha_S(M_Z^2)$, at NNLO.}
  \label{fig:wzvsasmznnlo}
\end{figure}

\begin{figure}
  \centering
  \begin{minipage}{0.5\textwidth}
    (a)\\
    \includegraphics[width=\textwidth]{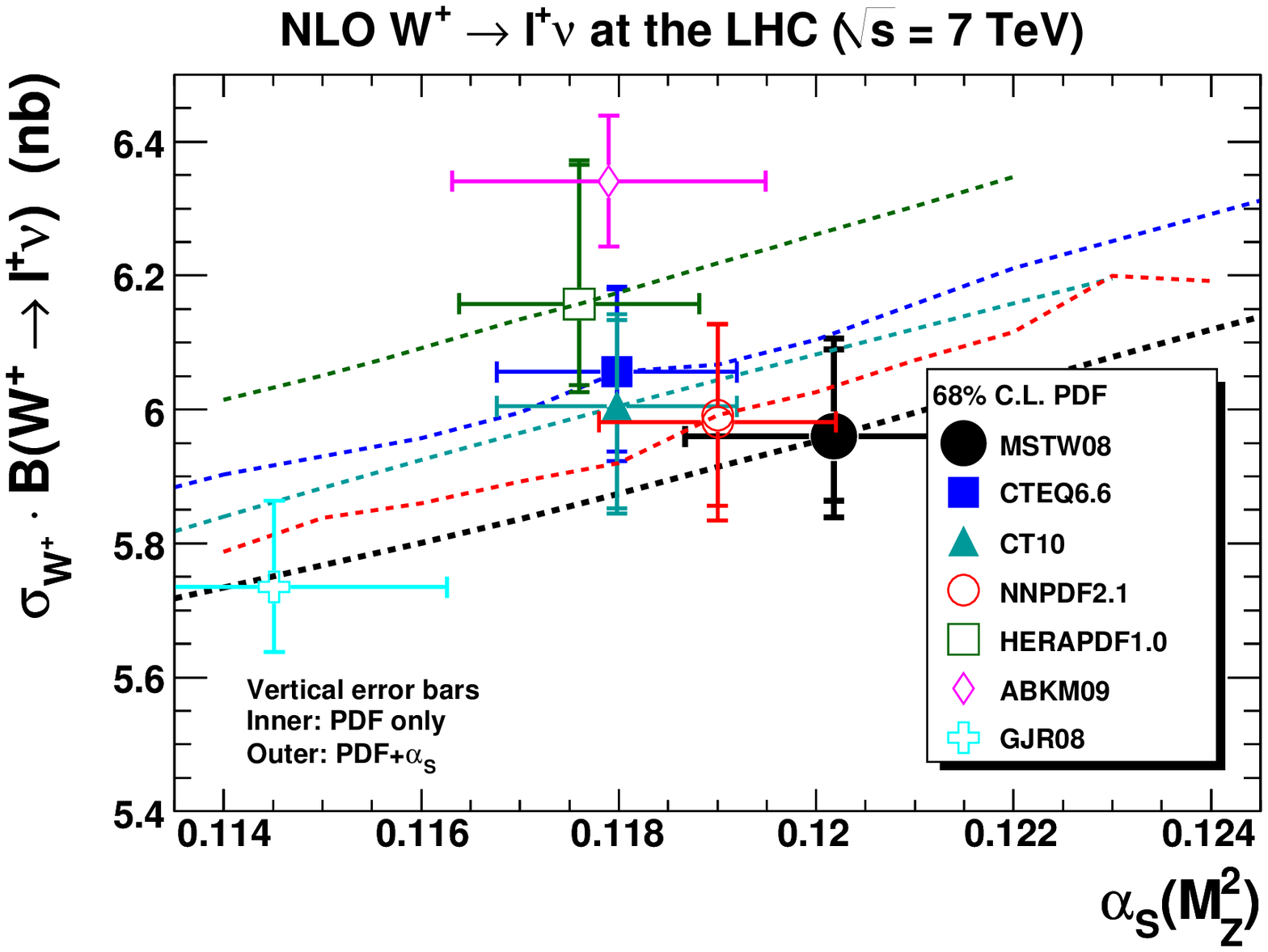}
  \end{minipage}%
  \begin{minipage}{0.5\textwidth}
    (b)\\
    \includegraphics[width=\textwidth]{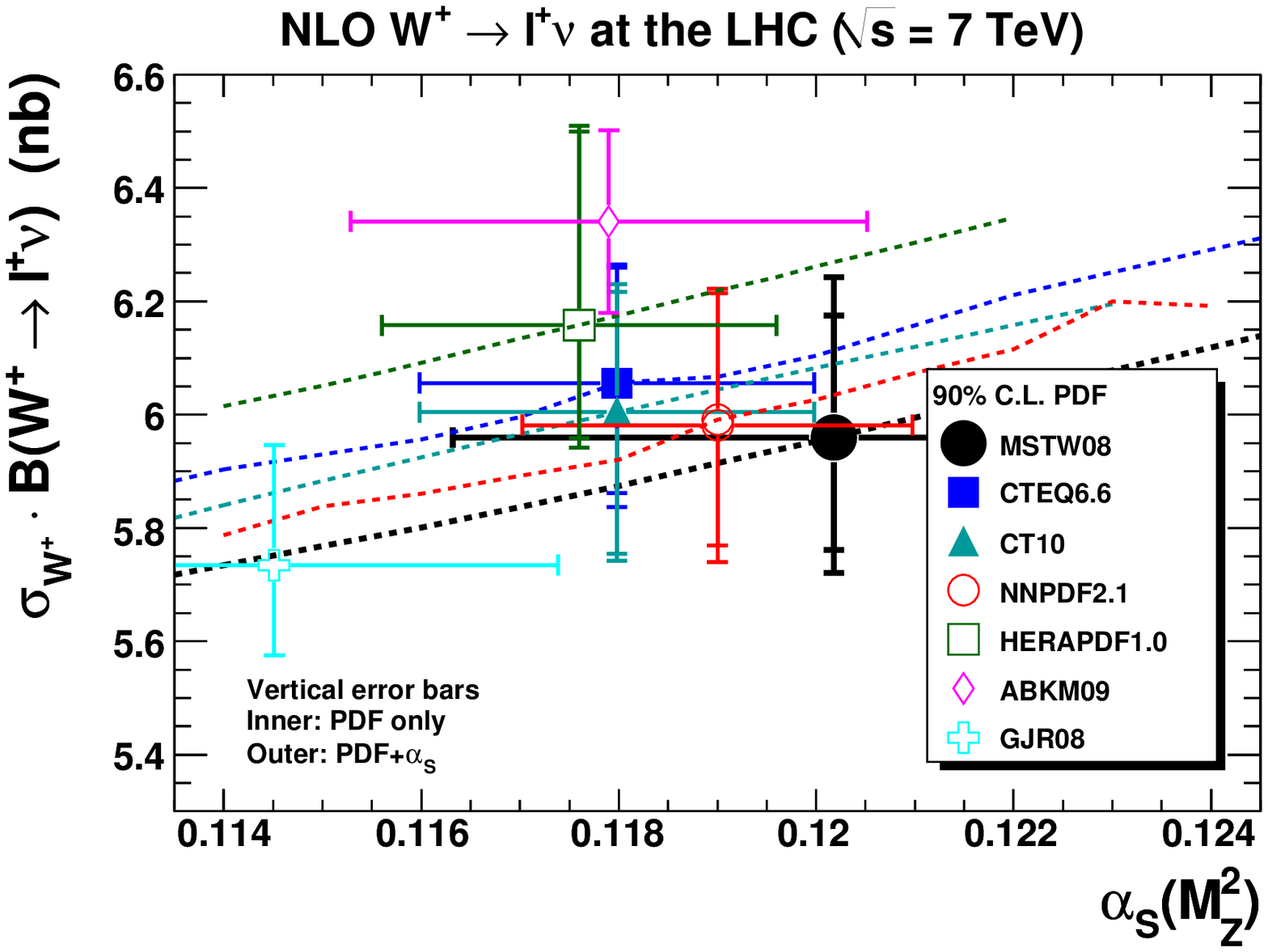}
  \end{minipage}
  \begin{minipage}{0.5\textwidth}
    (c)\\
    \includegraphics[width=\textwidth]{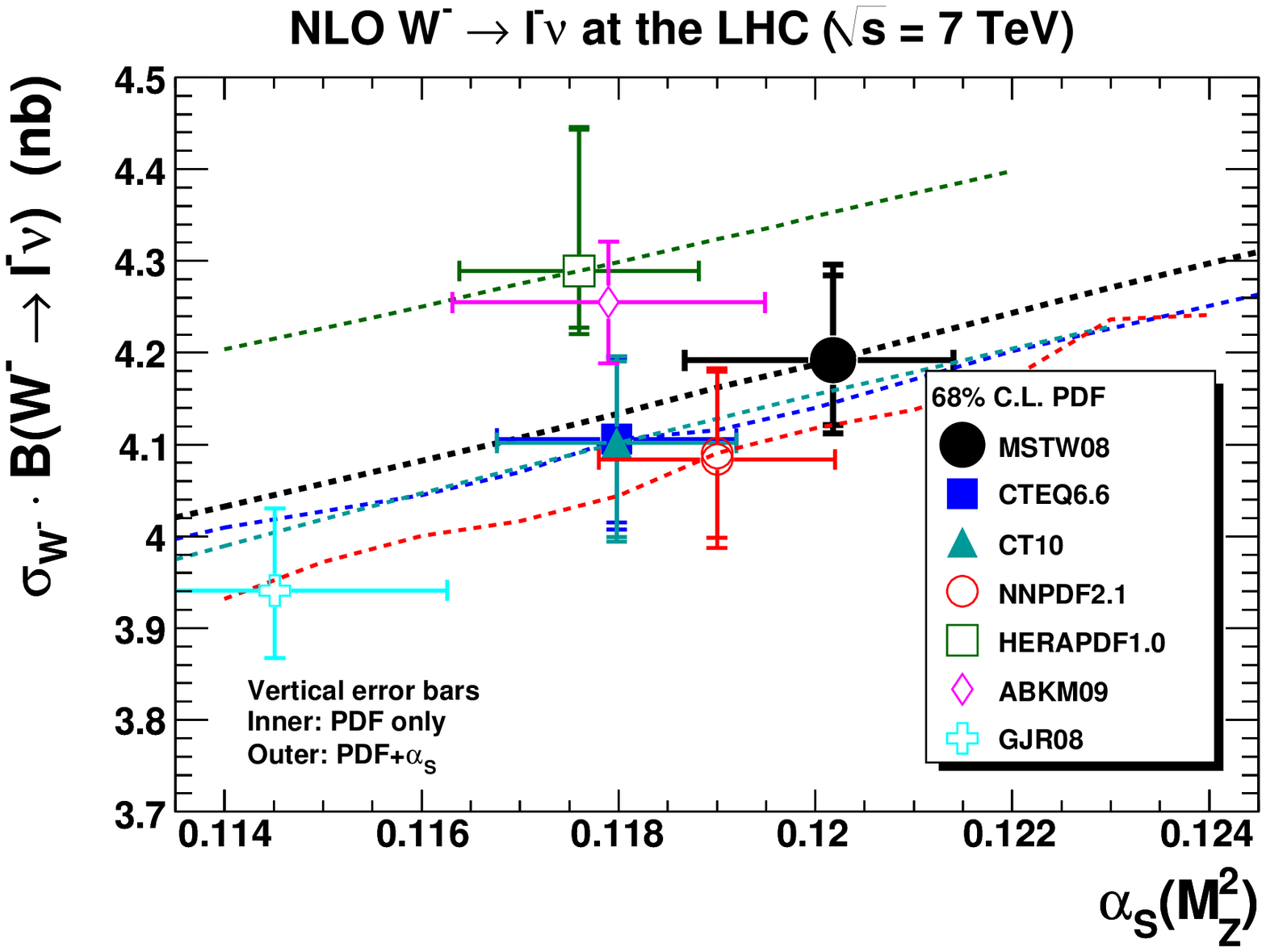}
  \end{minipage}%
  \begin{minipage}{0.5\textwidth}
    (d)\\
    \includegraphics[width=\textwidth]{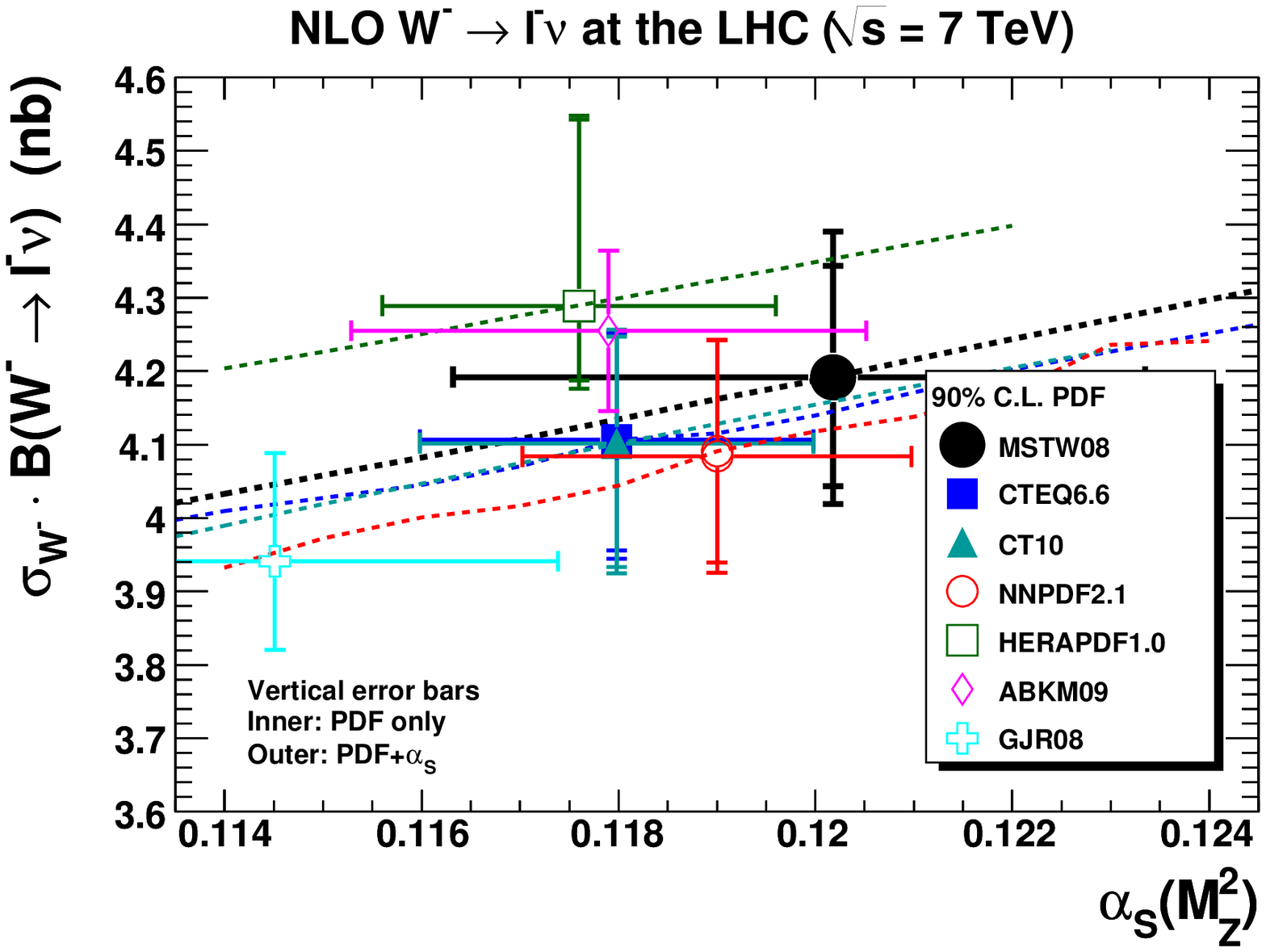}
  \end{minipage}
  \begin{minipage}{0.5\textwidth}
    (e)\\
    \includegraphics[width=\textwidth]{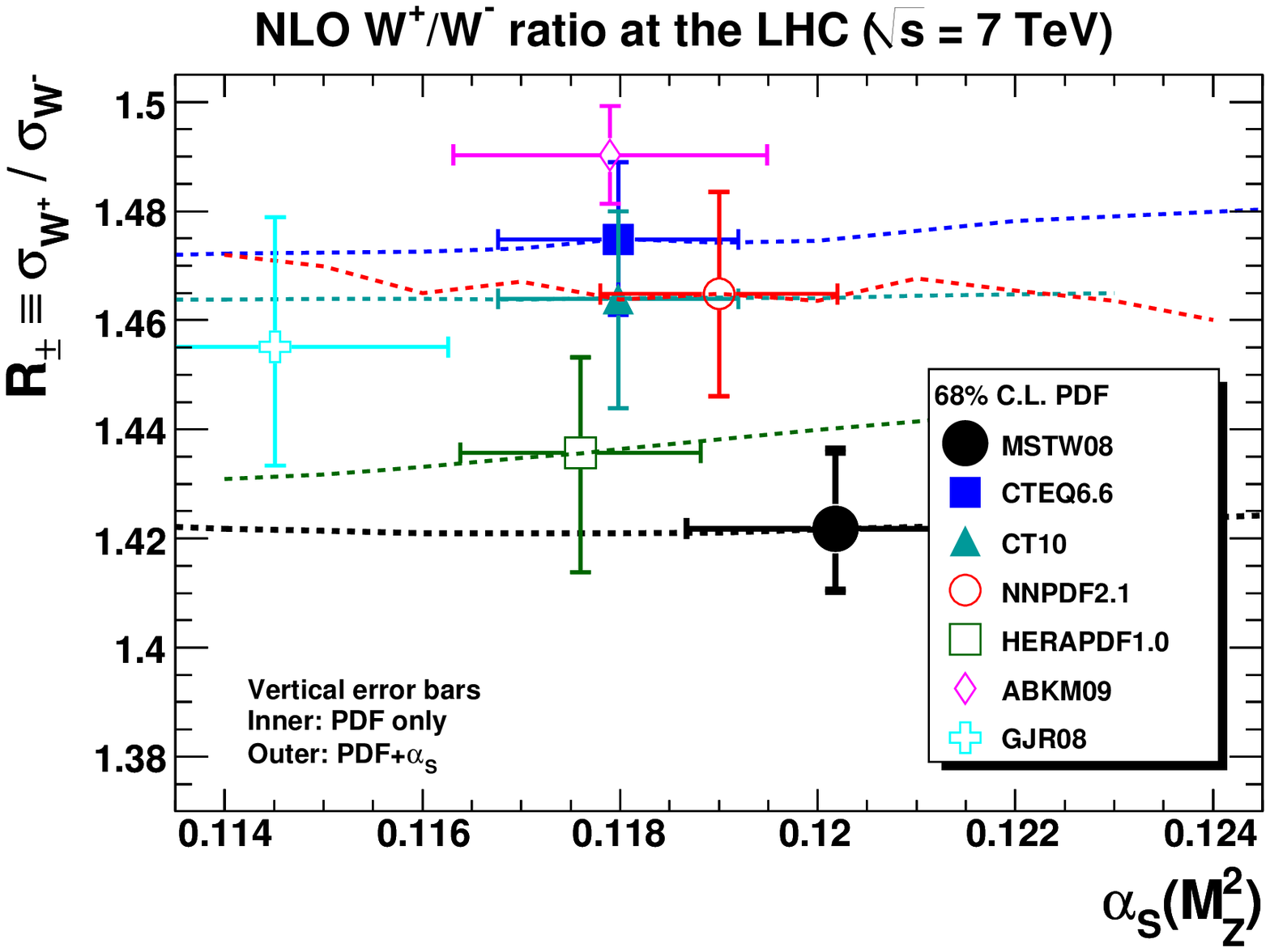}
  \end{minipage}%
  \begin{minipage}{0.5\textwidth}
    (f)\\
    \includegraphics[width=\textwidth]{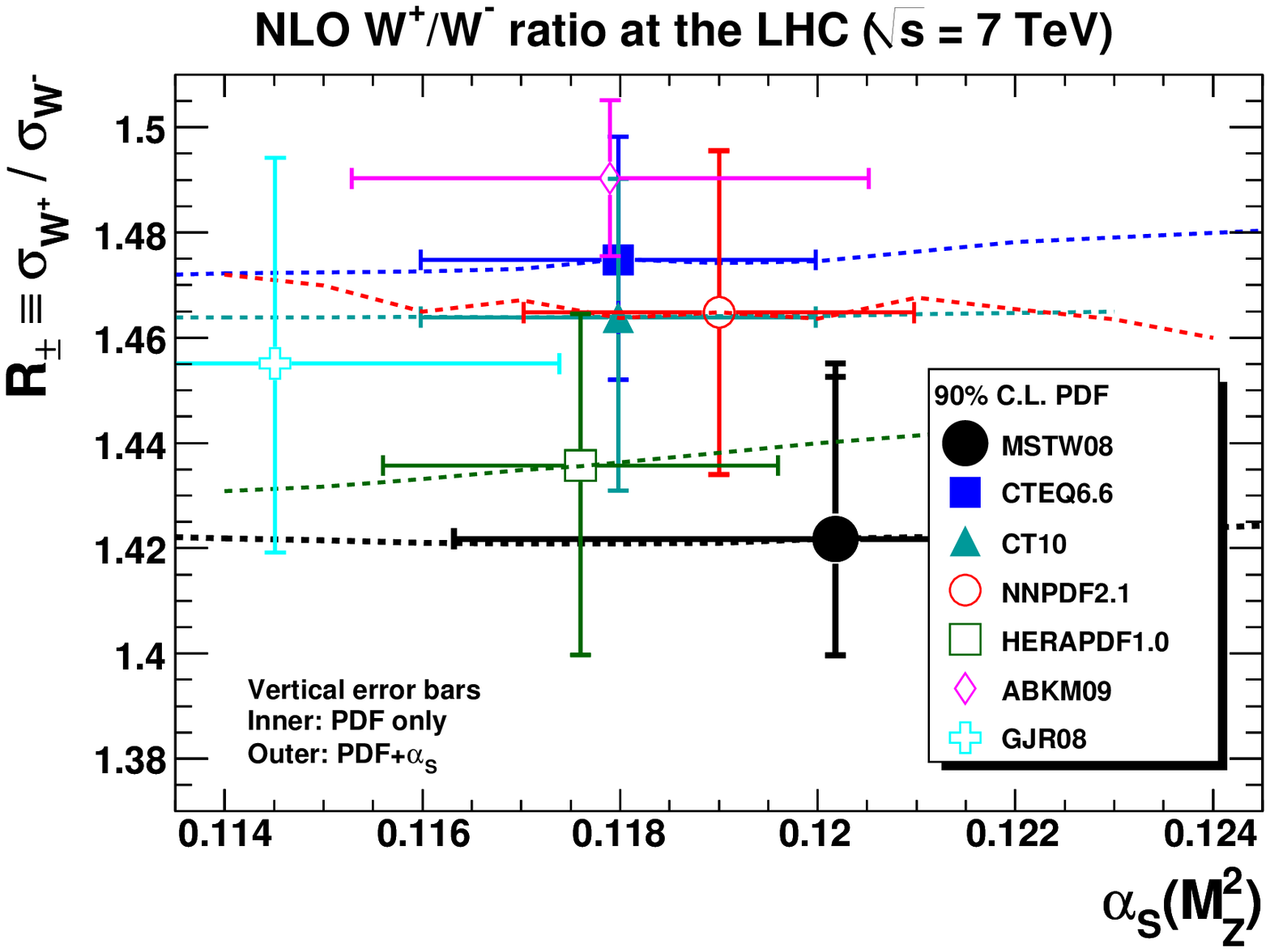}
  \end{minipage}
  \caption{$W^+$ and $W^-$ total cross sections, plotted as a function of $\alpha_S(M_Z^2)$, at NLO.}
  \label{fig:wpwmvsasmznlo}
\end{figure}

\begin{figure}
  \centering
  \begin{minipage}{0.5\textwidth}
    (a)\\
    \includegraphics[width=\textwidth]{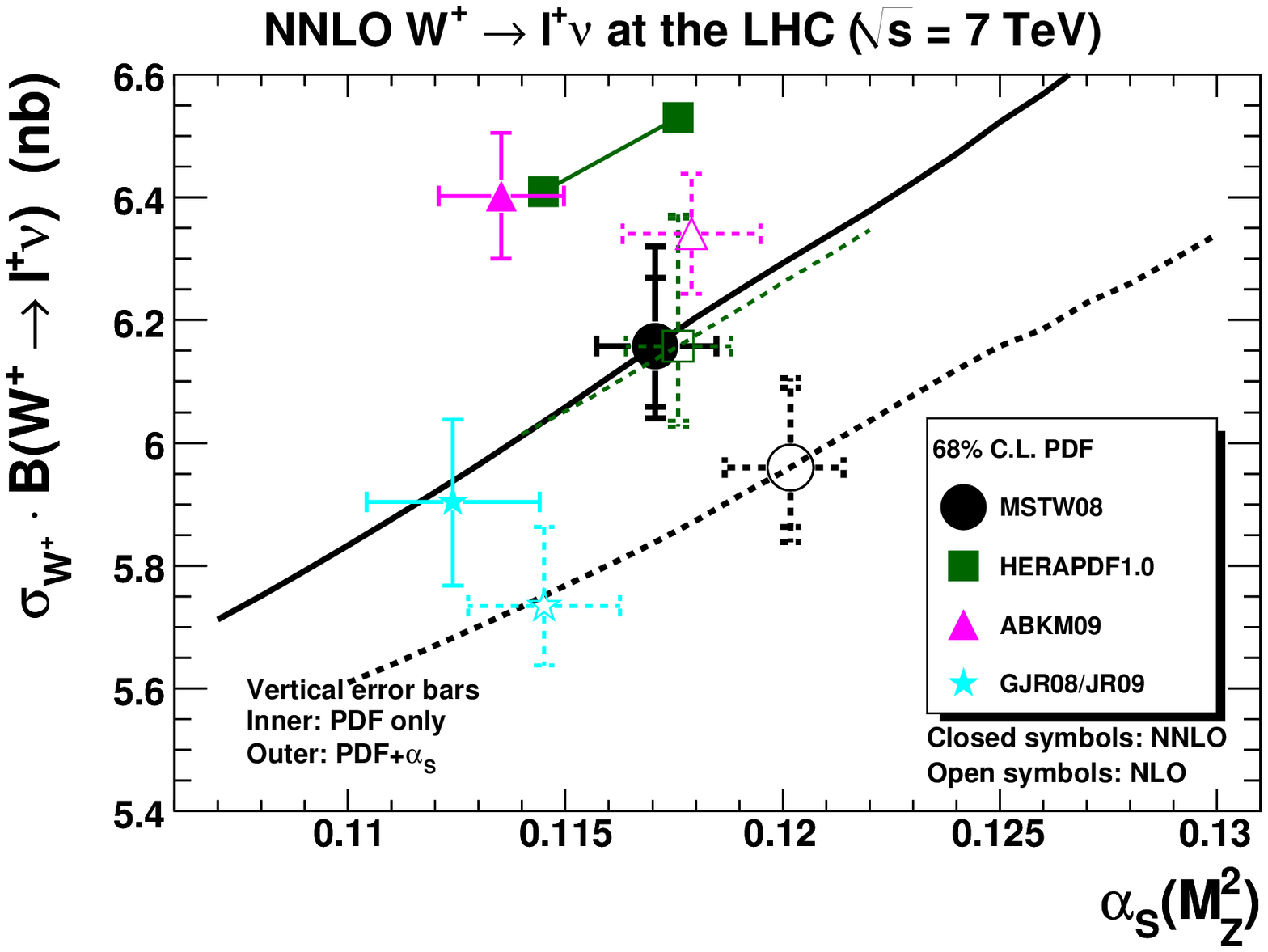}
  \end{minipage}%
  \begin{minipage}{0.5\textwidth}
    (b)\\
    \includegraphics[width=\textwidth]{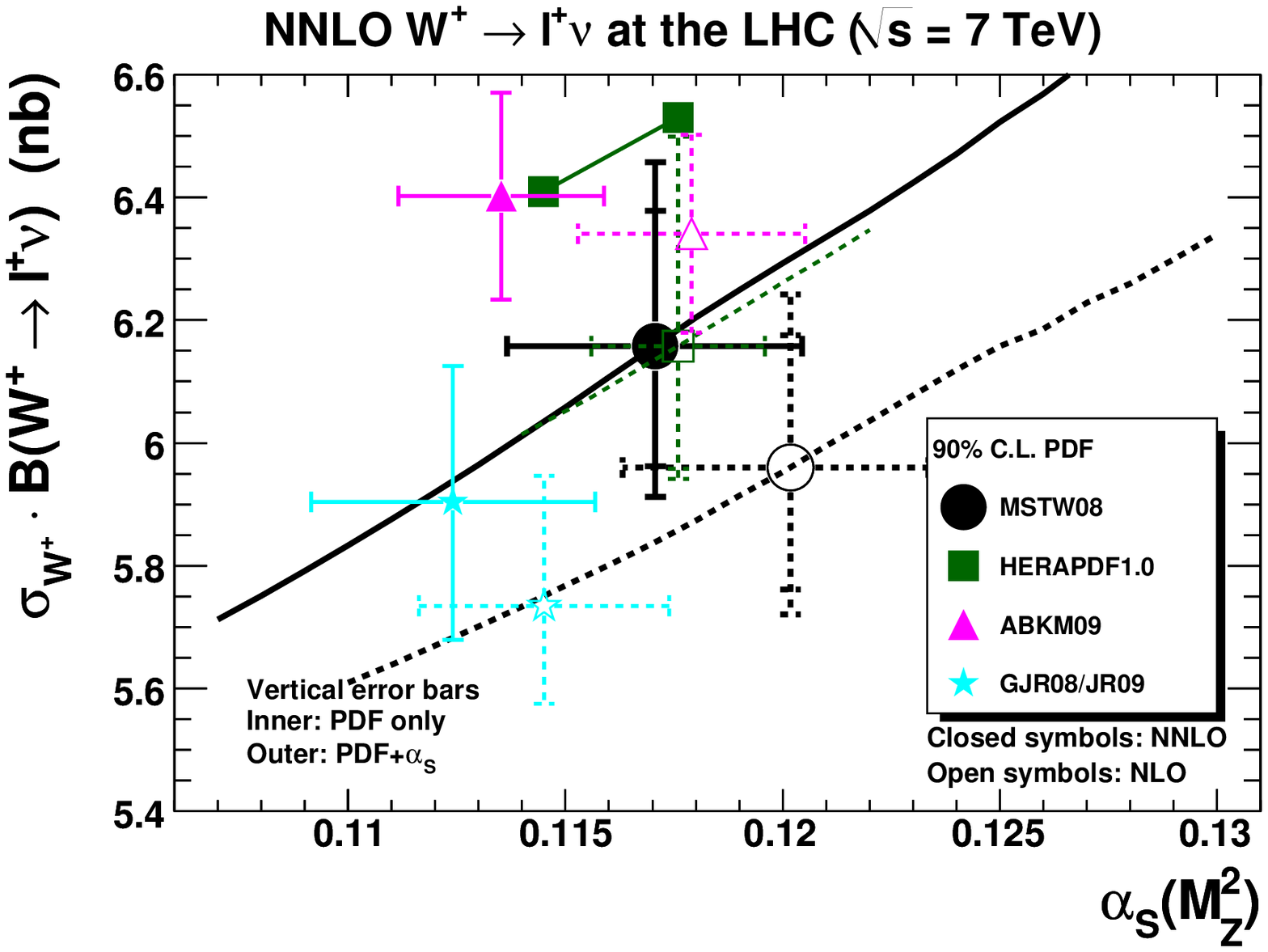}
  \end{minipage}
  \begin{minipage}{0.5\textwidth}
    (c)\\
    \includegraphics[width=\textwidth]{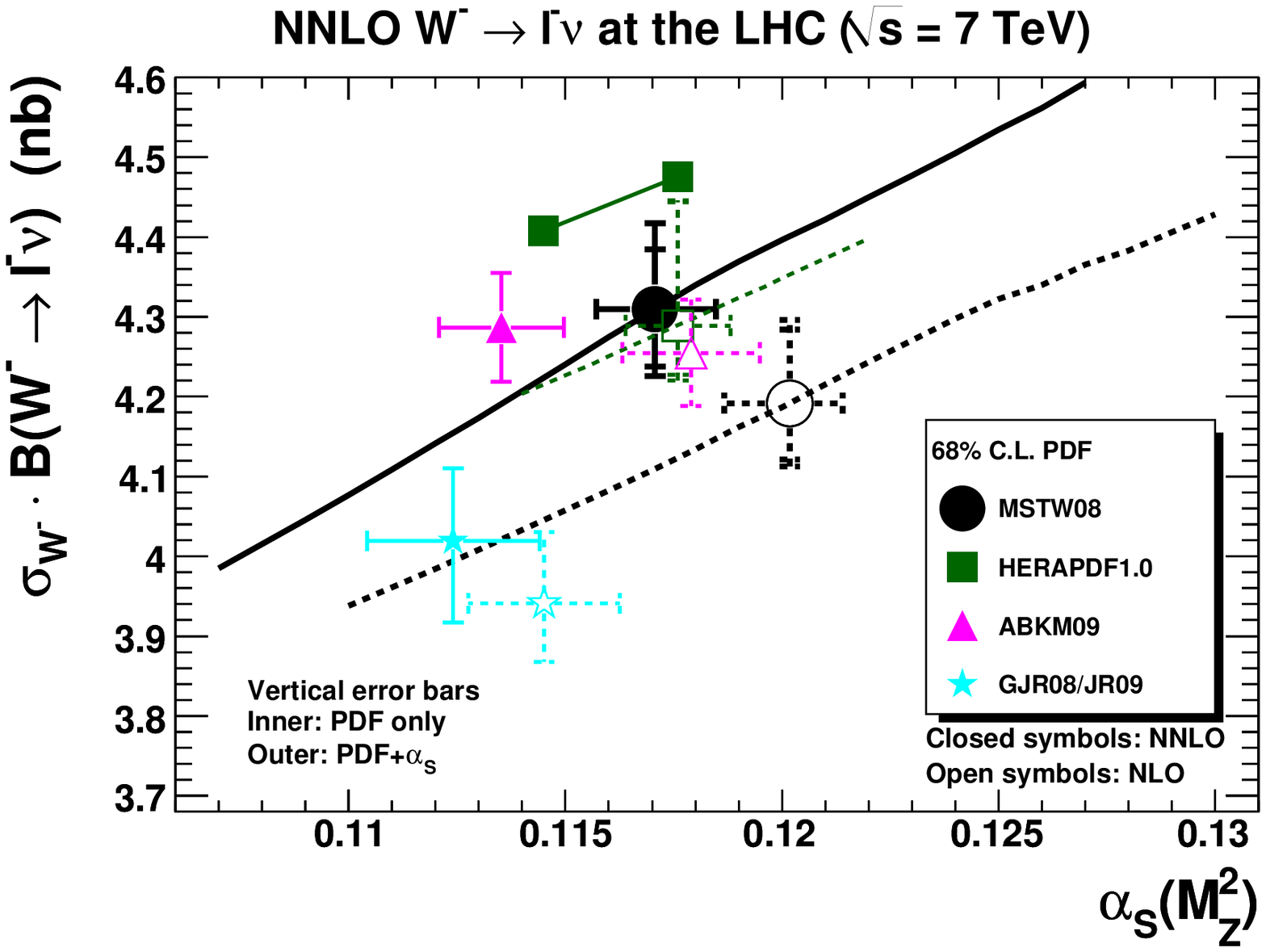}
  \end{minipage}%
  \begin{minipage}{0.5\textwidth}
    (d)\\
    \includegraphics[width=\textwidth]{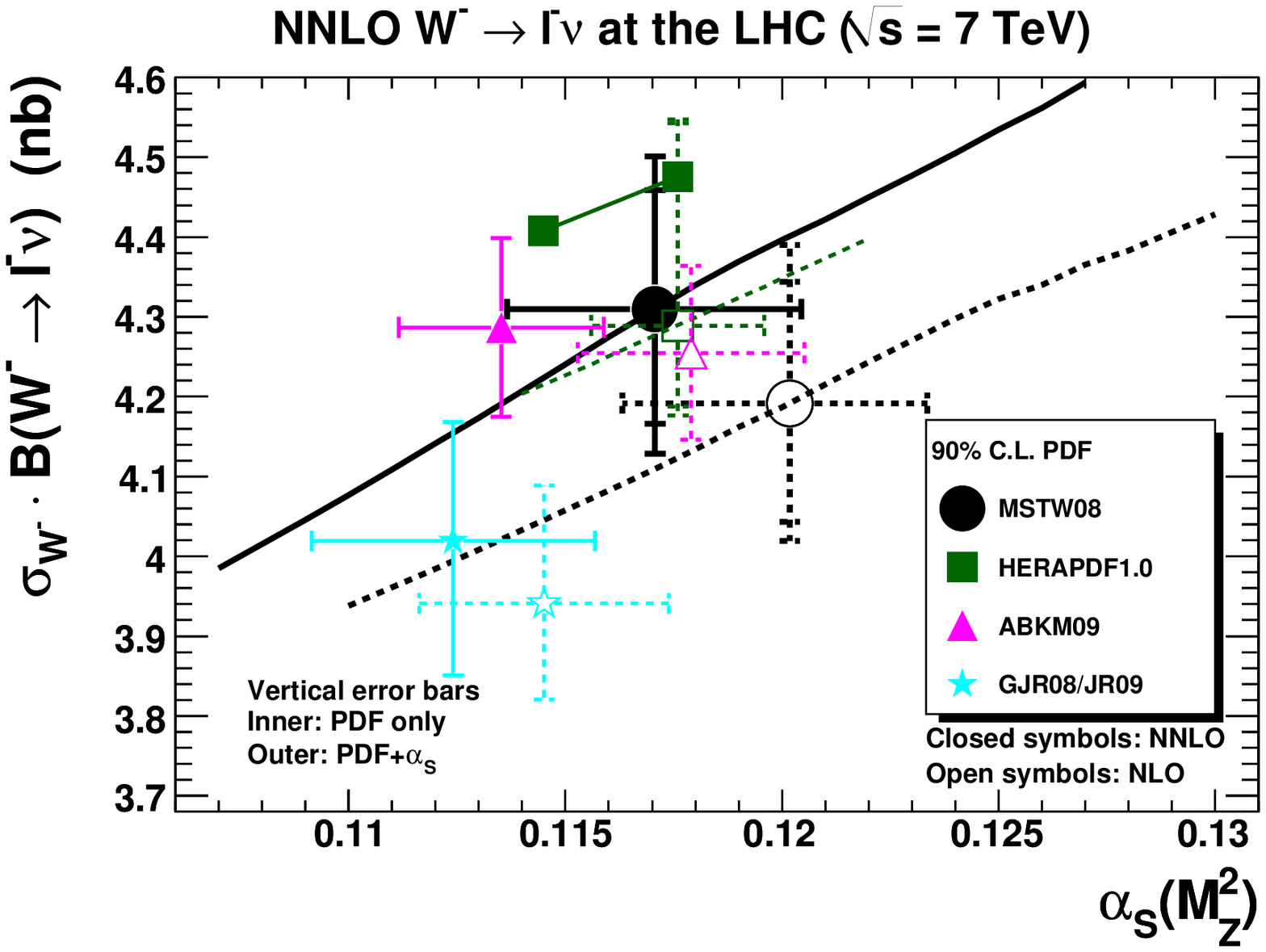}
  \end{minipage}
  \begin{minipage}{0.5\textwidth}
    (e)\\
    \includegraphics[width=\textwidth]{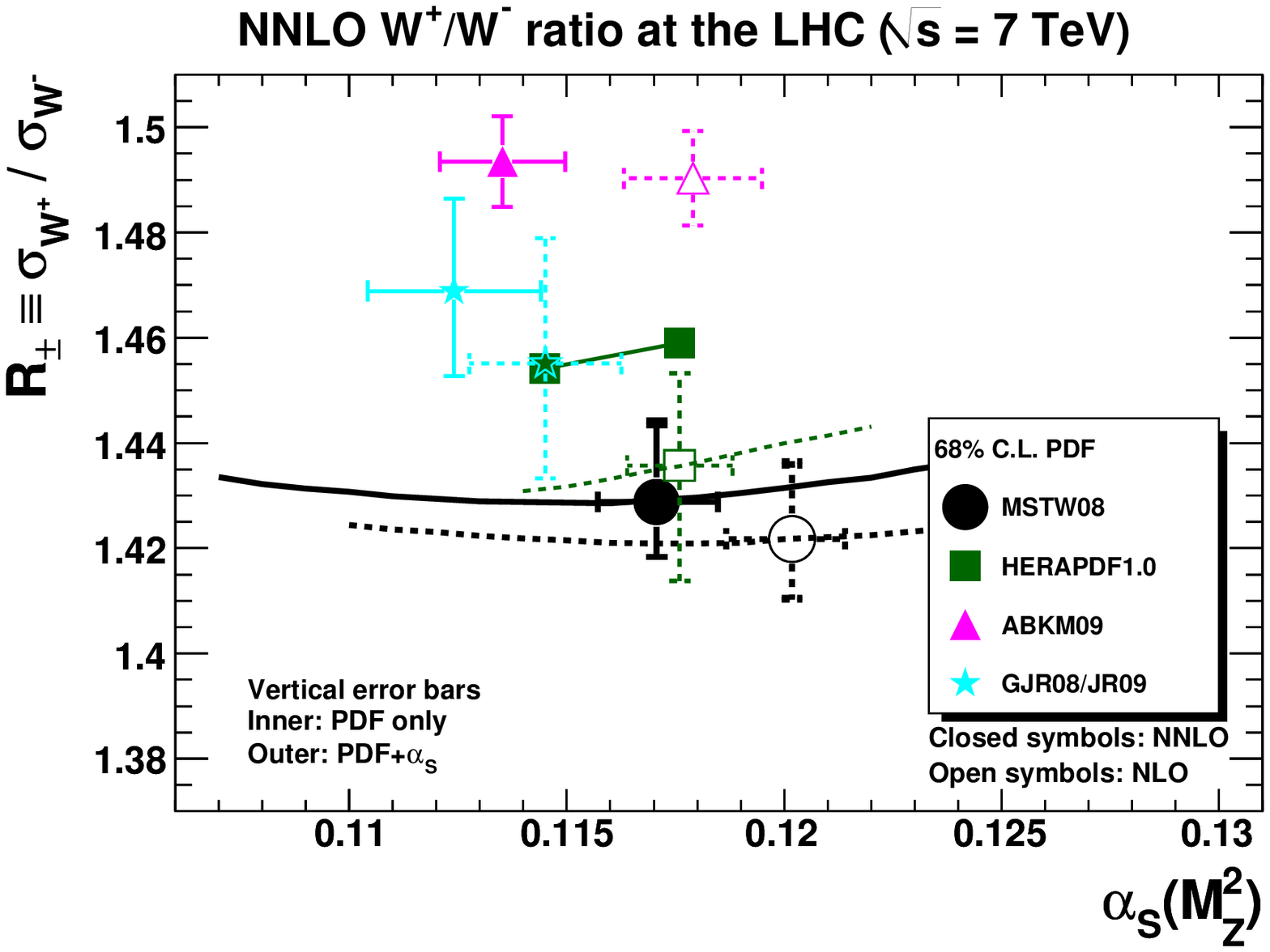}
  \end{minipage}%
  \begin{minipage}{0.5\textwidth}
    (f)\\
    \includegraphics[width=\textwidth]{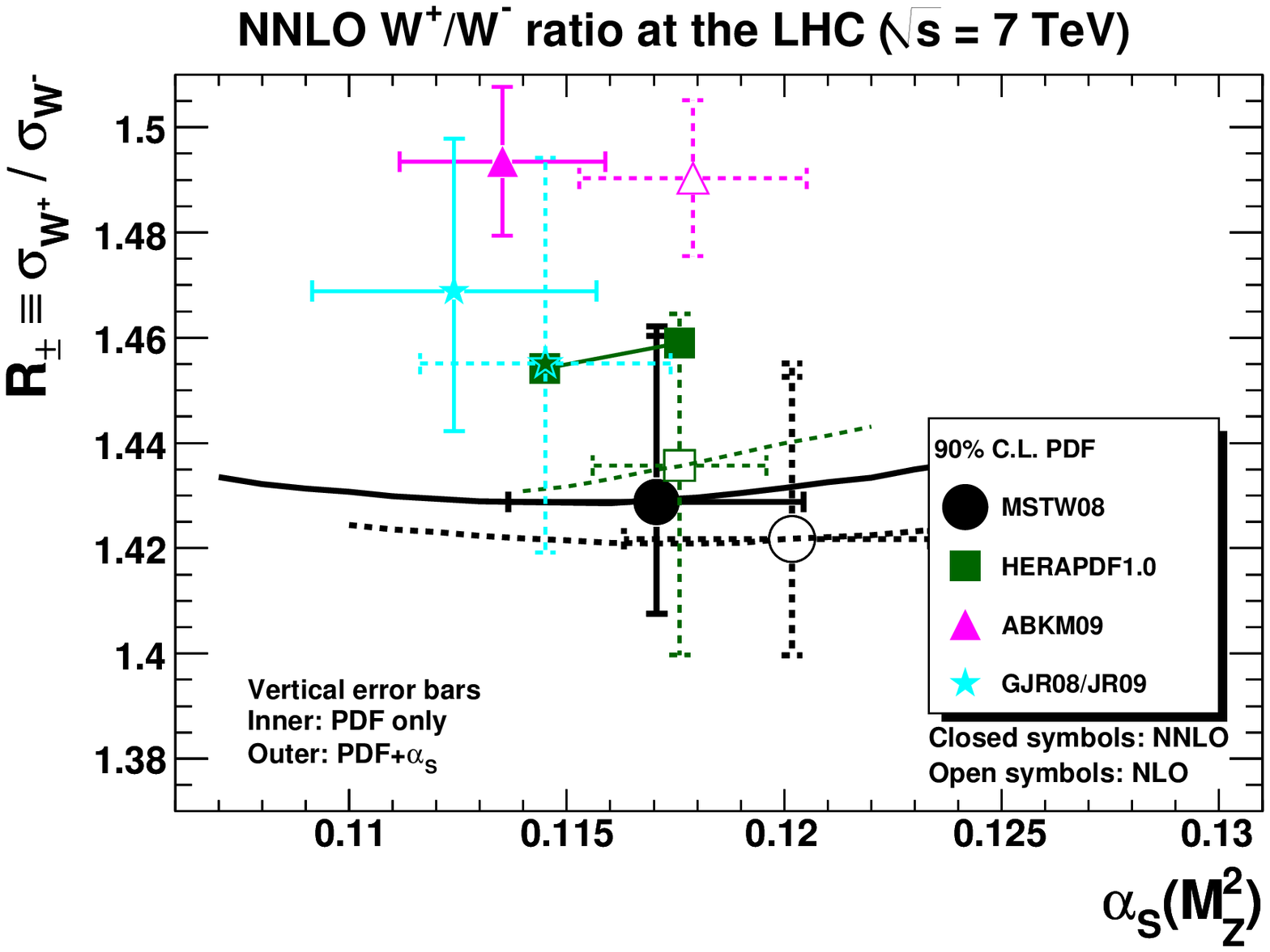}
  \end{minipage}
  \caption{$W^+$ and $W^-$ total cross sections, plotted as a function of $\alpha_S(M_Z^2)$, at NNLO.}
  \label{fig:wpwmvsasmznnlo}
\end{figure}

We calculate $W^\pm$ and $Z^0$ cross sections, multiplied by the appropriate leptonic branching ratio, in the zero-width approximation with no virtual photon ($\gamma^*$) included in the $Z^0$ cross sections, as in section 15 of ref.~\cite{Martin:2009iq}.  We use the PDG 2008~\cite{Amsler:2008zzb} electroweak parameters, with the CKM matrix taken from a global fit in the Standard Model with unitarity constraints imposed, i.e.~the central values of the CKM matrix in eq.~(11.27) of ref.~\cite{Amsler:2008zzb}.  The leptonic branching ratios are taken to be $B(W\to\ell\nu)=0.1080$ and $B(Z\to\ell\ell)=0.33658$.  The $W$ and $Z$ masses are taken to be $M_W=80.3980$~GeV and $M_Z=91.1876$~GeV, the effective weak-mixing angle $\sin^2(\theta_W)=0.23149$ (defining the vector coupling of the $Z$ boson to quarks), the Fermi coupling constant $G_F=1.16637\times10^{-5}$~GeV$^{-2}$ and the conversion constant 1~GeV$^{-2} = 389379.304$~nb.  The NLO total cross sections (with MSTW08 PDFs) calculated using the private code (originally from J.~Stirling) were checked against the \textsc{mcfm} package provided for the PDF4LHC benchmark exercise~\cite{Campbell:2010ff,terascale} and agreement was found to within the numerical integration error (of at most 0.1\%) for all processes apart from $Z^0$.  The discrepancy in $Z^0$ production was traced to an ambiguity in the benchmark specification, where the provided \textsc{mcfm} version~\cite{terascale} used $\sin^2(\theta_W)=1-(M_W/M_Z)^2=0.22265$, whereas we used the effective angle $\sin^2(\theta_W)=0.23149$~\cite{Amsler:2008zzb} which absorbs electroweak radiative corrections, leading to a $Z^0$ total cross section 1.4\% lower than the provided \textsc{mcfm} version~\cite{terascale}.  The NNLO corrections were calculated in ref.~\cite{Hamberg:1990np} (see also ref.~\cite{Harlander:2002wh} for a correction).

We show the $W^\pm$ ($=W^++W^-$) and $Z^0$ total cross sections (and their ratio), plotted as a function of $\alpha_S(M_Z^2)$, in figures~\ref{fig:wzvsasmznlo} and \ref{fig:wzvsasmznnlo}, with (a,c,e)~68\% C.L.~uncertainties on the left and (b,d,f)~90\% C.L.~uncertainties on the right.  We show the separate $W^+$ and $W^-$ total cross sections (and their ratio), plotted as a function of $\alpha_S(M_Z^2)$, in figures~\ref{fig:wpwmvsasmznlo} and \ref{fig:wpwmvsasmznnlo}.

The common format of all the plots in figures~\ref{fig:wzvsasmznlo}, \ref{fig:wzvsasmznnlo}, \ref{fig:wpwmvsasmznlo} and \ref{fig:wpwmvsasmznnlo} is that the markers are centred on the default $\alpha_S(M_Z^2)$ value and the corresponding predicted cross section of each PDF fitting group.  The horizontal error bars span the $\alpha_S(M_Z^2)$ uncertainty, the inner vertical error bars span the ``PDF only'' uncertainty where possible (i.e.~not for ABKM09 or GJR08/JR09), and the outer vertical error bars span the ``PDF+$\alpha_S$'' uncertainty.  The effect of the additional $\alpha_S$ uncertainty is small for $W$ and $Z$ production.  The dashed lines at NLO or the solid lines at NNLO interpolate the cross-section predictions calculated with the alternative PDF sets with different $\alpha_S(M_Z^2)$ values provided by each group, represented by the smaller symbols in figure~\ref{fig:asmzvalues}.  The NNLO plots in figures~\ref{fig:wzvsasmznnlo} and \ref{fig:wpwmvsasmznnlo} also show the NLO predictions (open symbols and dashed lines) together with the corresponding NNLO predictions (closed symbols and solid lines) to explicitly demonstrate how the size of the NNLO corrections depends on both the $\alpha_S(M_Z^2)$ choice and the PDF choice.  The NNLO corrections are reduced by taking different $\alpha_S(M_Z^2)$ values at different perturbative orders.

The \emph{global} fits are in good agreement for $W^\pm$ and $Z^0$ production, as was apparent from the $q\bar{q}$ luminosity plots in section~\ref{sec:luminosities}, and there are small uncertainties in predictions for the $W^\pm/Z^0$ ratio.  The $W^\pm/Z^0$ ratio is insensitive to NNLO corrections (and $\alpha_S$).  There is slightly more spread in the separate $W^+$ and $W^-$ total cross sections.  The $W^+/W^-$ ratio is sensitive to the $u/d$ ratio, but insensitive to NNLO corrections (and $\alpha_S$).  We will discuss $W$ and $Z$ production in more detail in section~\ref{sec:wandz}.

\subsection{\texorpdfstring{$gg\to H$}{gg->H} production}

\begin{figure}
  \centering
  \begin{minipage}{0.5\textwidth}
    (a)\\
    \includegraphics[width=\textwidth]{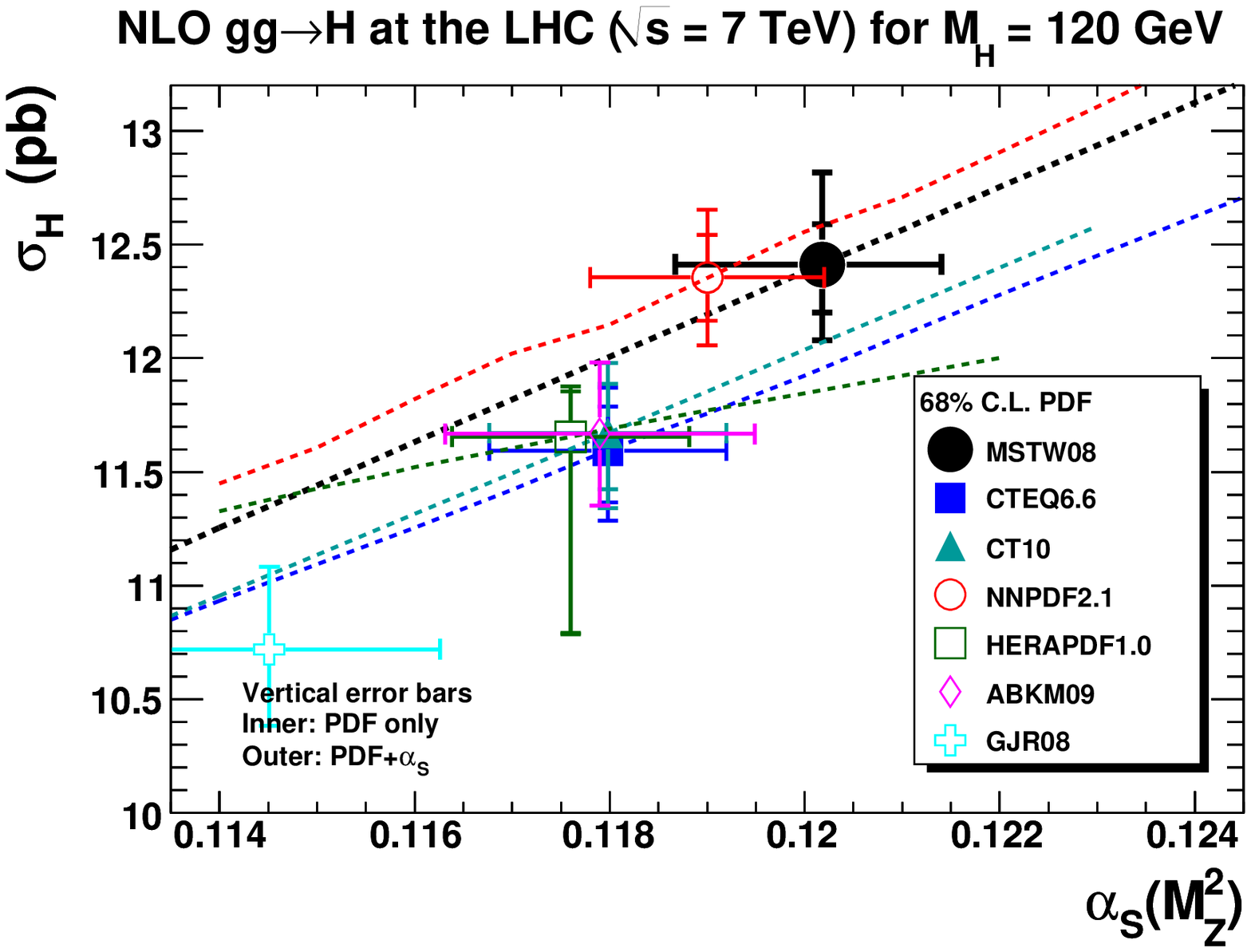}
  \end{minipage}%
  \begin{minipage}{0.5\textwidth}
    (b)\\
    \includegraphics[width=\textwidth]{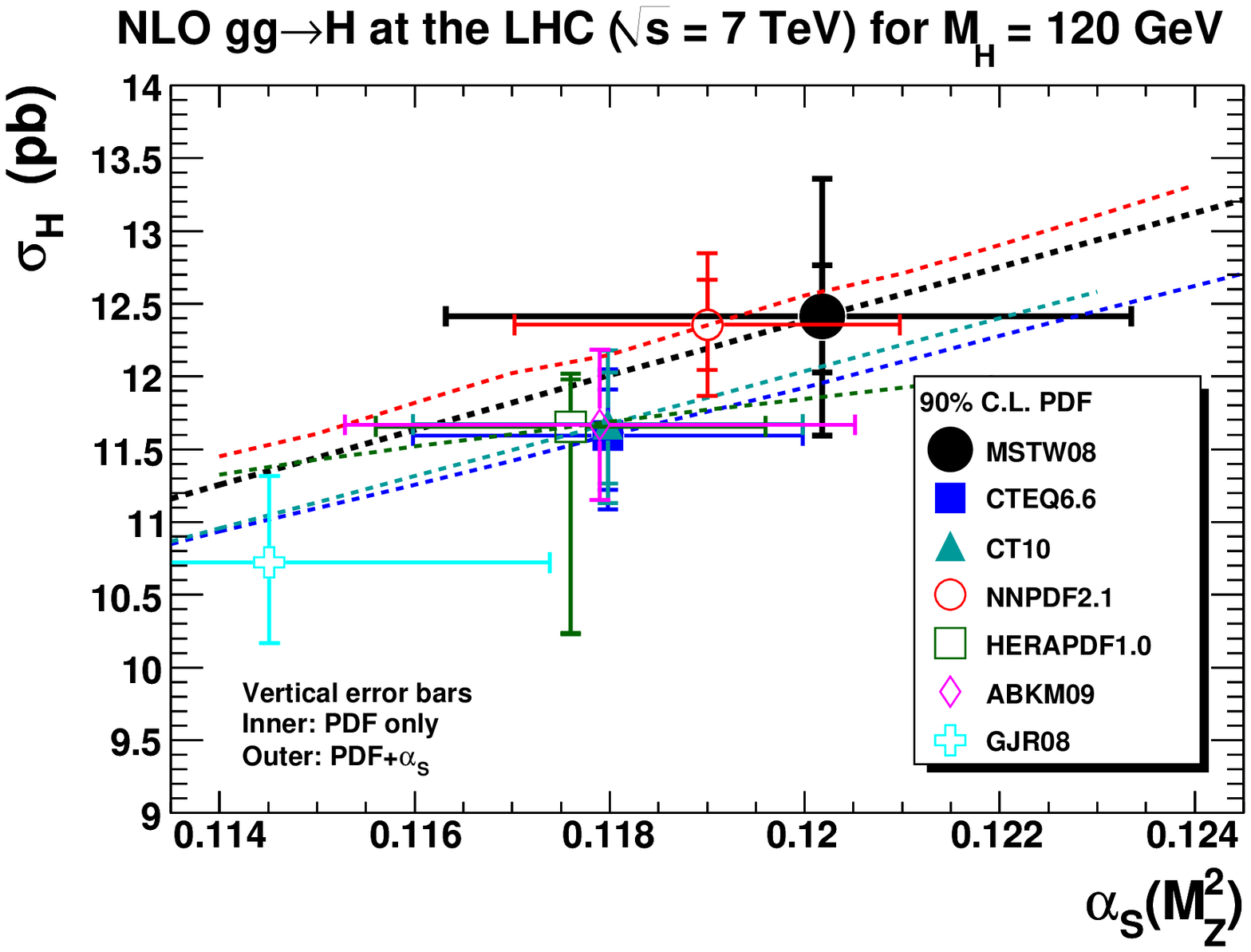}
  \end{minipage}
  \begin{minipage}{0.5\textwidth}
    (c)\\
    \includegraphics[width=\textwidth]{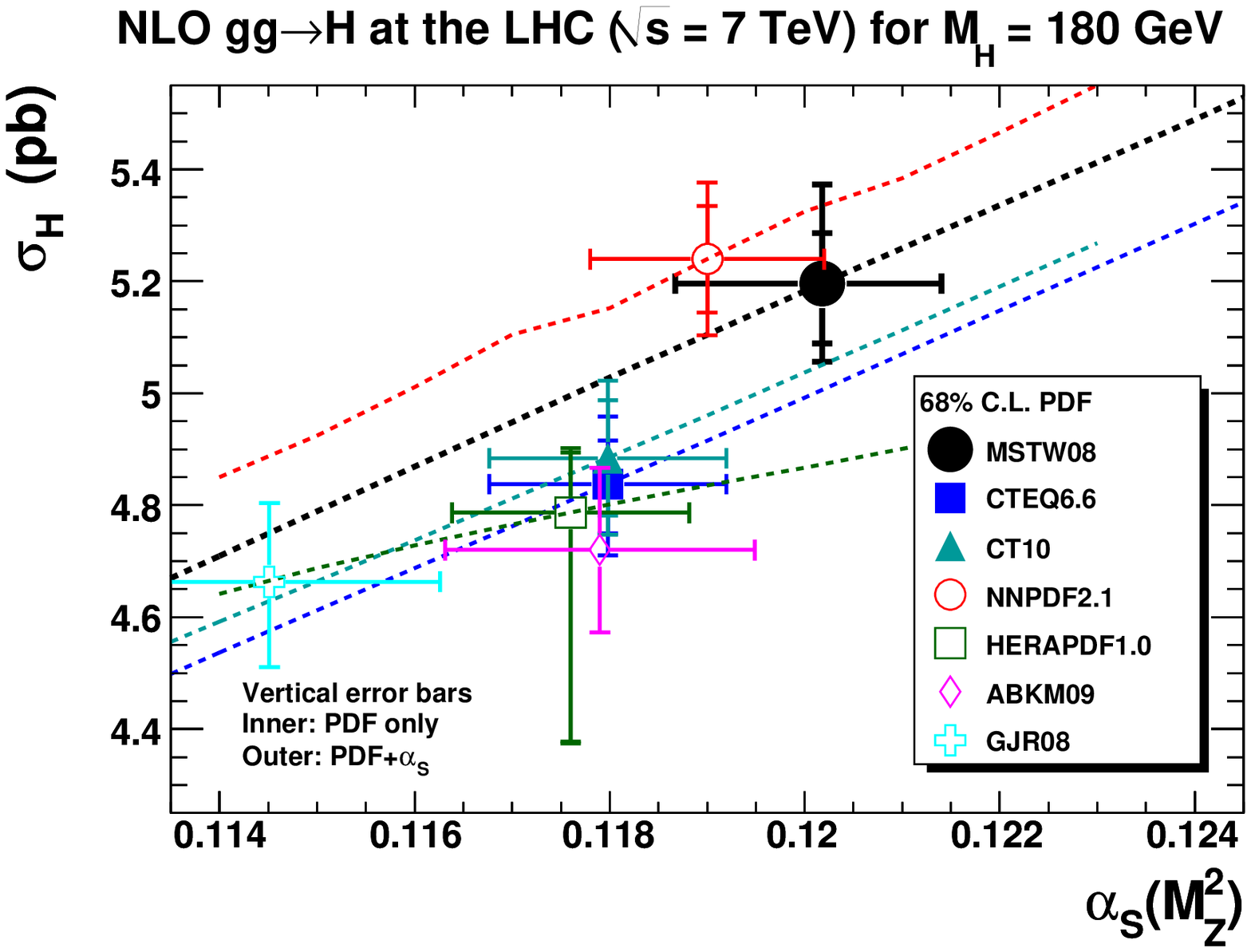}
  \end{minipage}%
  \begin{minipage}{0.5\textwidth}
    (d)\\
    \includegraphics[width=\textwidth]{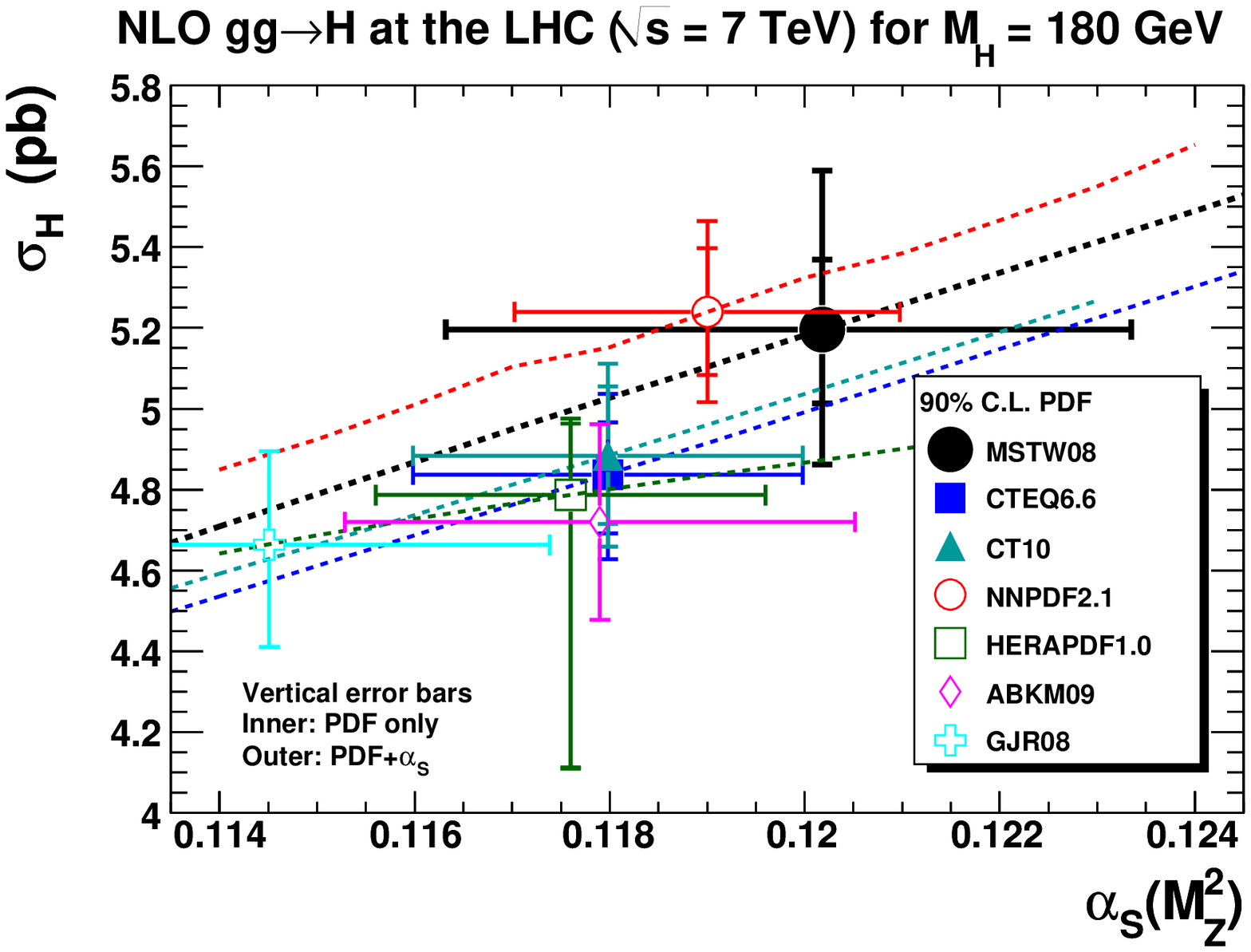}
  \end{minipage}
  \begin{minipage}{0.5\textwidth}
    (e)\\
    \includegraphics[width=\textwidth]{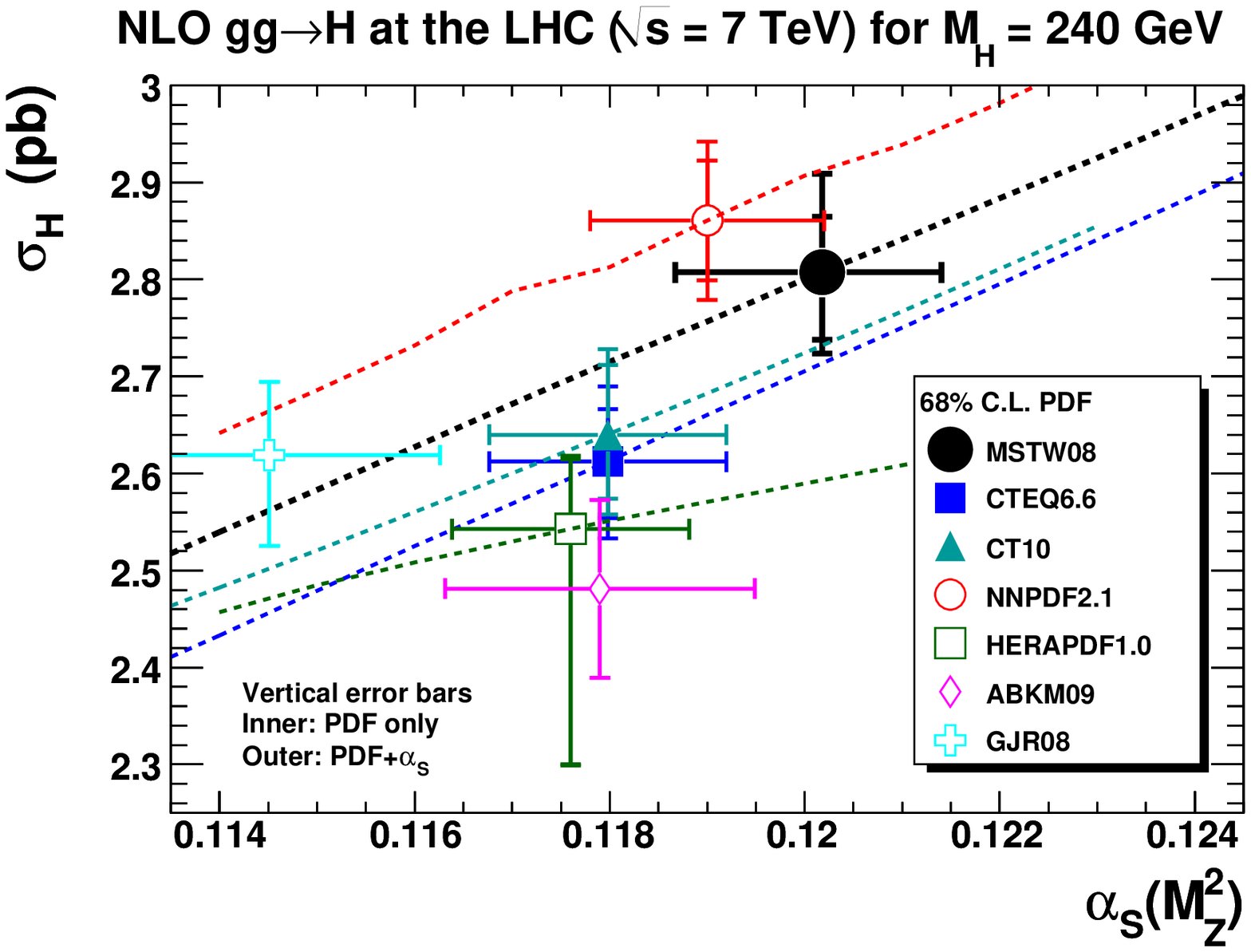}
  \end{minipage}%
  \begin{minipage}{0.5\textwidth}
    (f)\\
    \includegraphics[width=\textwidth]{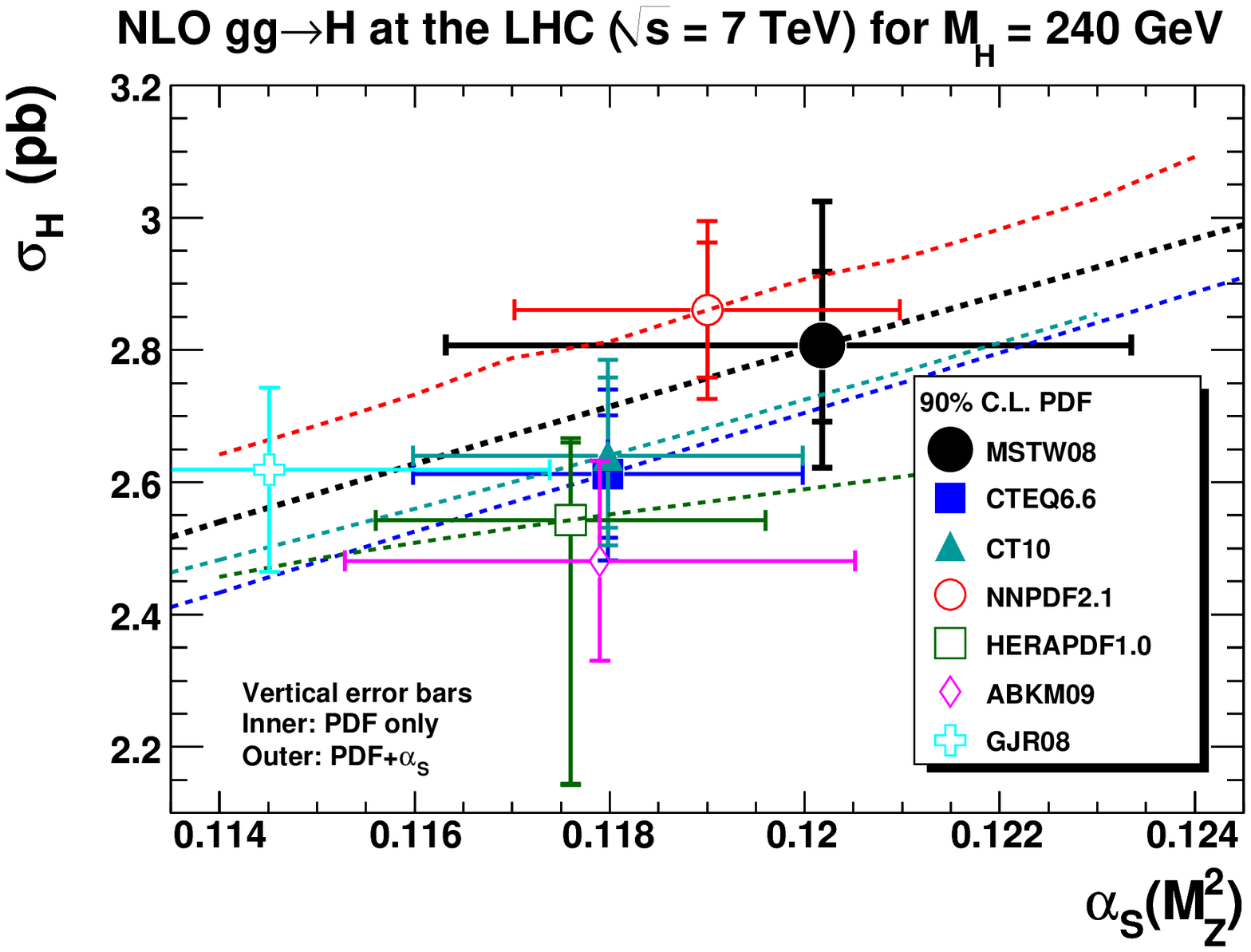}
  \end{minipage}
  \caption{$gg\to H$ total cross sections, plotted as a function of $\alpha_S(M_Z^2)$, at NLO.}
  \label{fig:gghvsasmznlo}
\end{figure}
\begin{figure}
  \centering
  \begin{minipage}{0.5\textwidth}
    (a)\\
    \includegraphics[width=\textwidth]{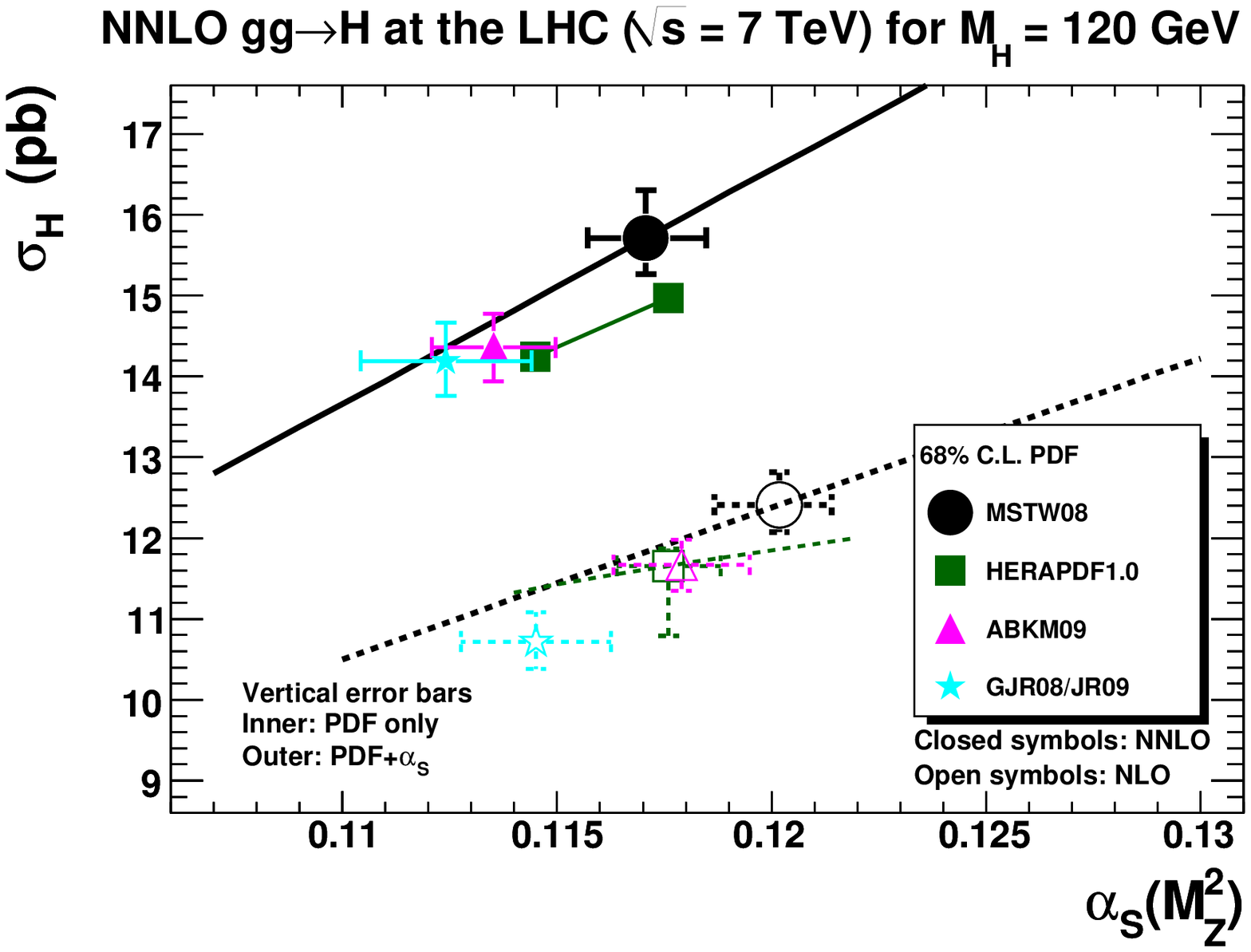}
  \end{minipage}%
  \begin{minipage}{0.5\textwidth}
    (b)\\
    \includegraphics[width=\textwidth]{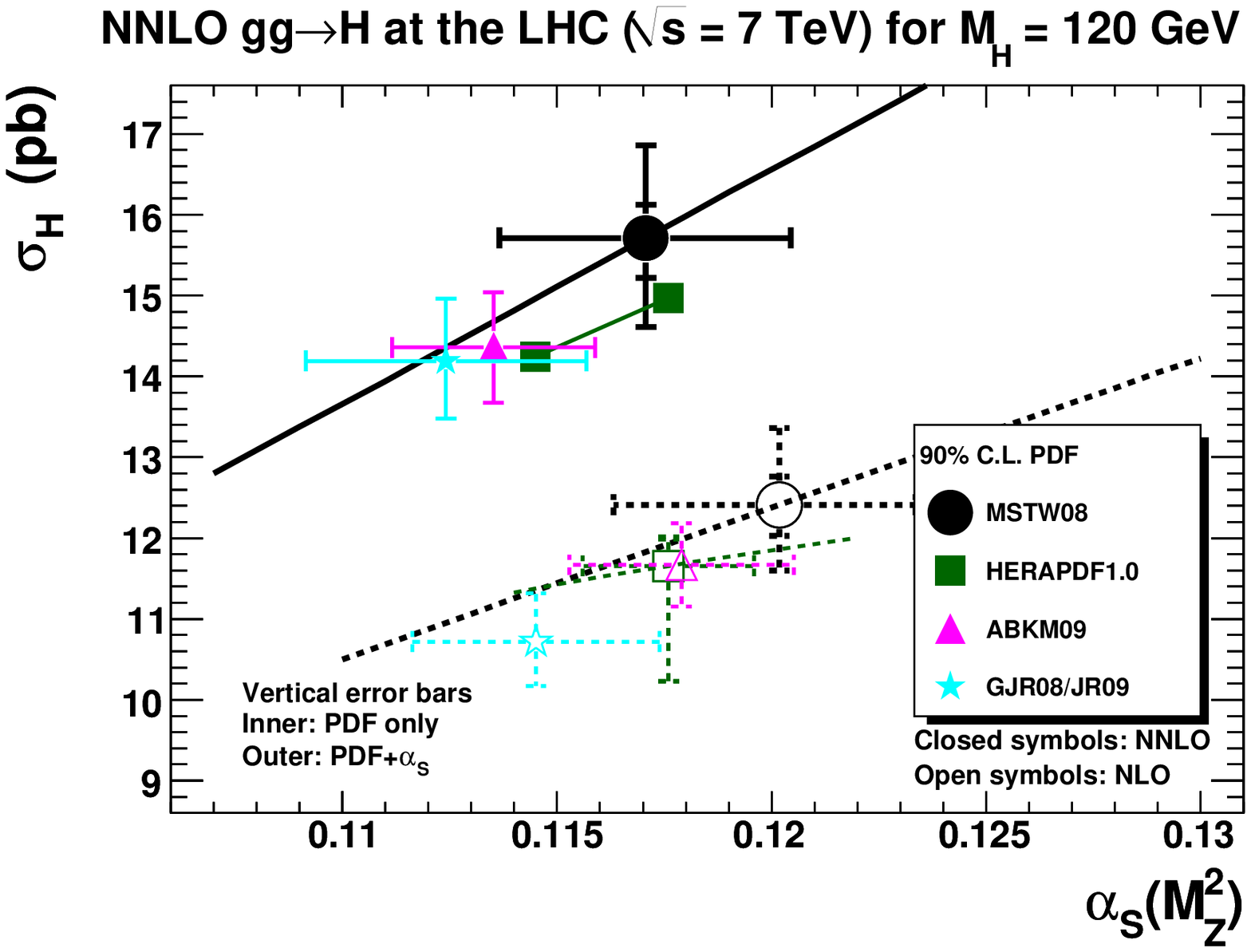}
  \end{minipage}
  \begin{minipage}{0.5\textwidth}
    (c)\\
    \includegraphics[width=\textwidth]{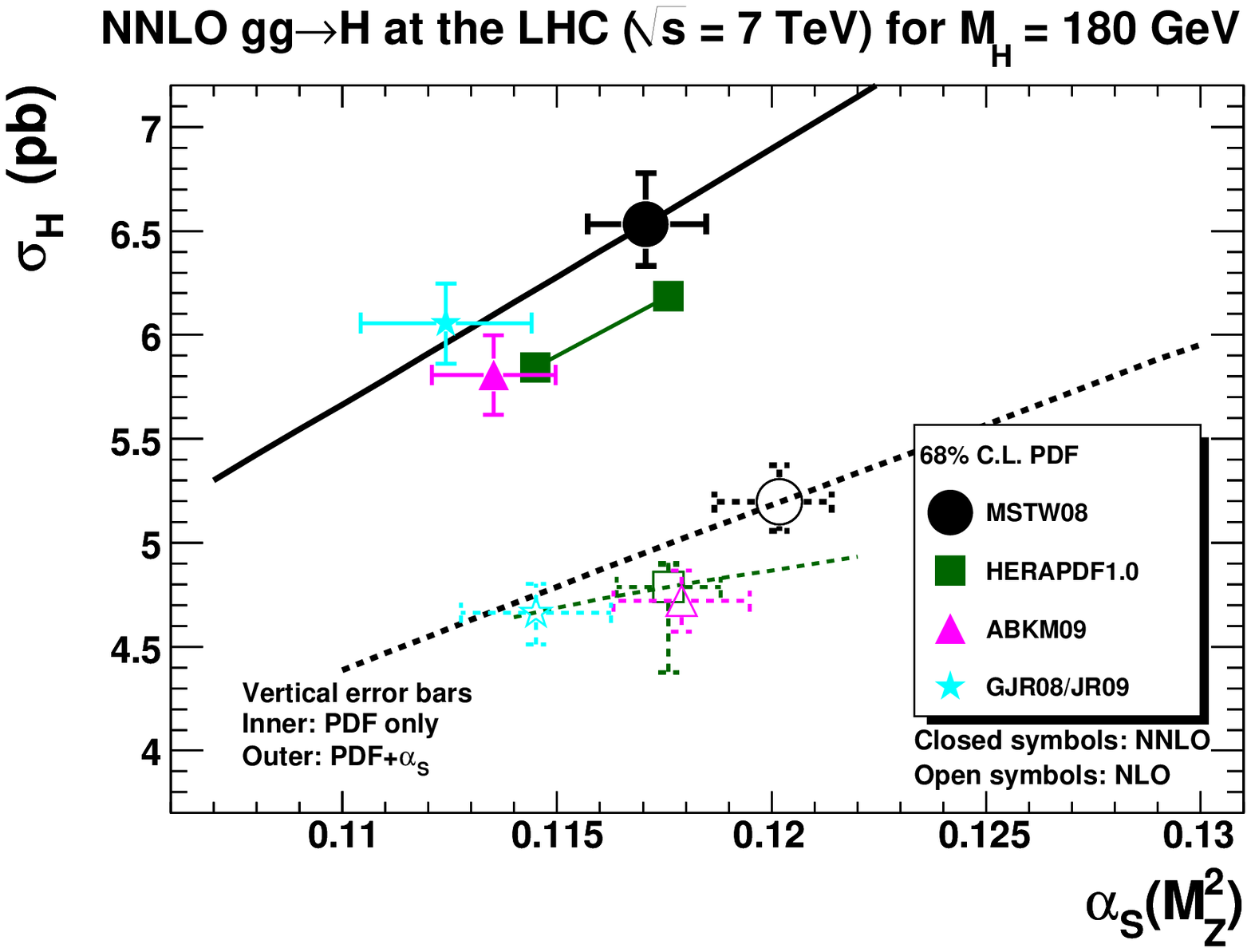}
  \end{minipage}%
  \begin{minipage}{0.5\textwidth}
    (d)\\
    \includegraphics[width=\textwidth]{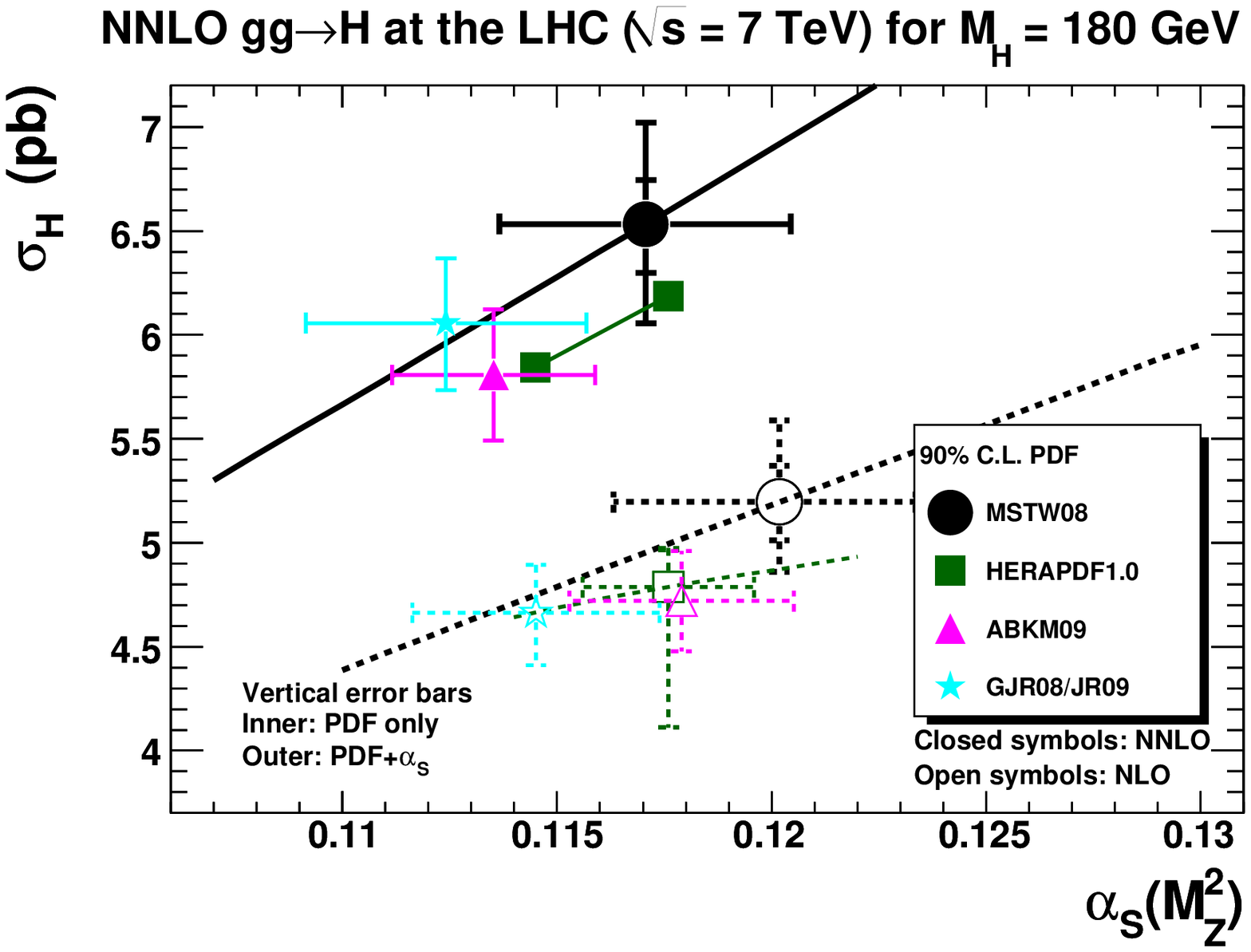}
  \end{minipage}
  \begin{minipage}{0.5\textwidth}
    (e)\\
    \includegraphics[width=\textwidth]{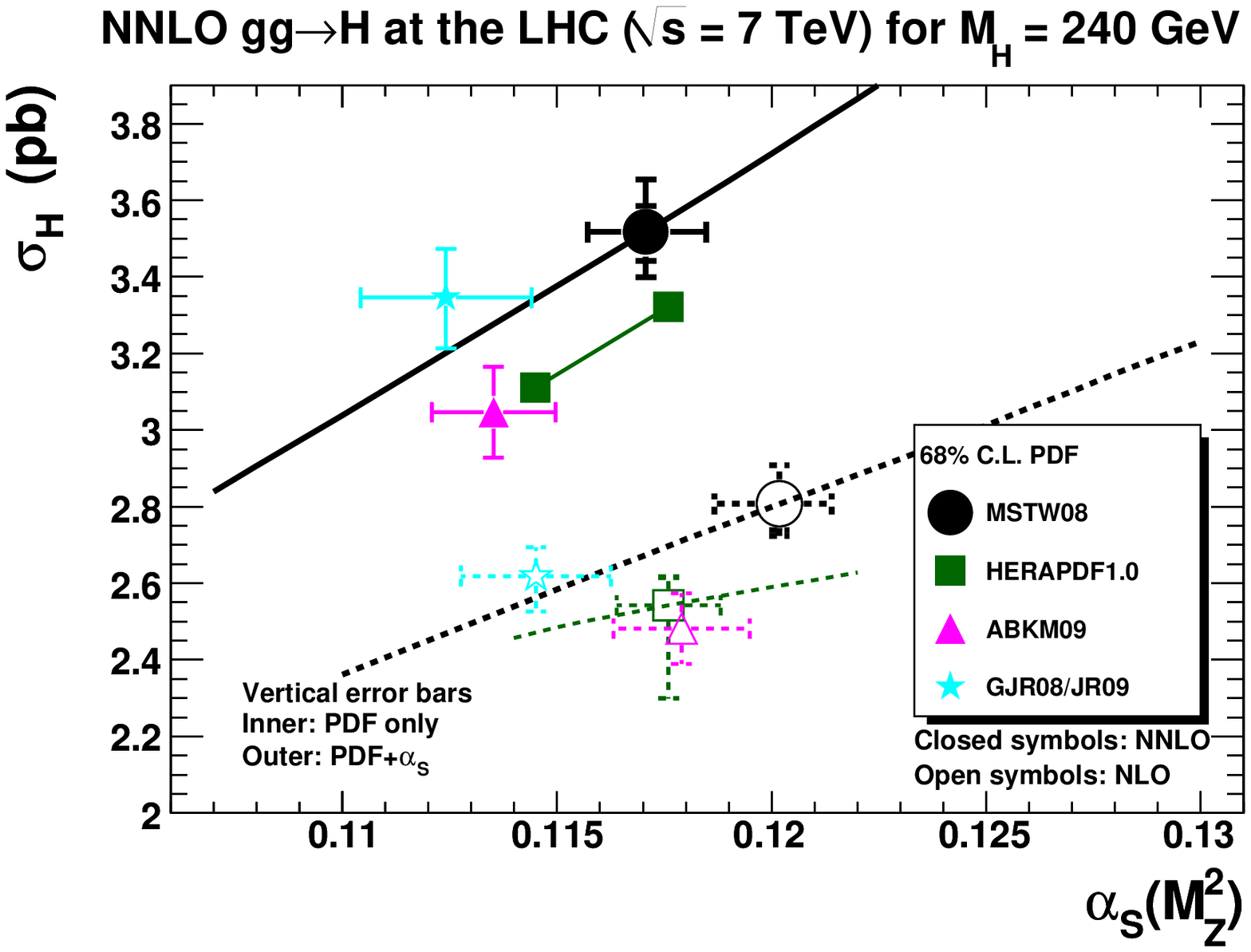}
  \end{minipage}%
  \begin{minipage}{0.5\textwidth}
    (f)\\
    \includegraphics[width=\textwidth]{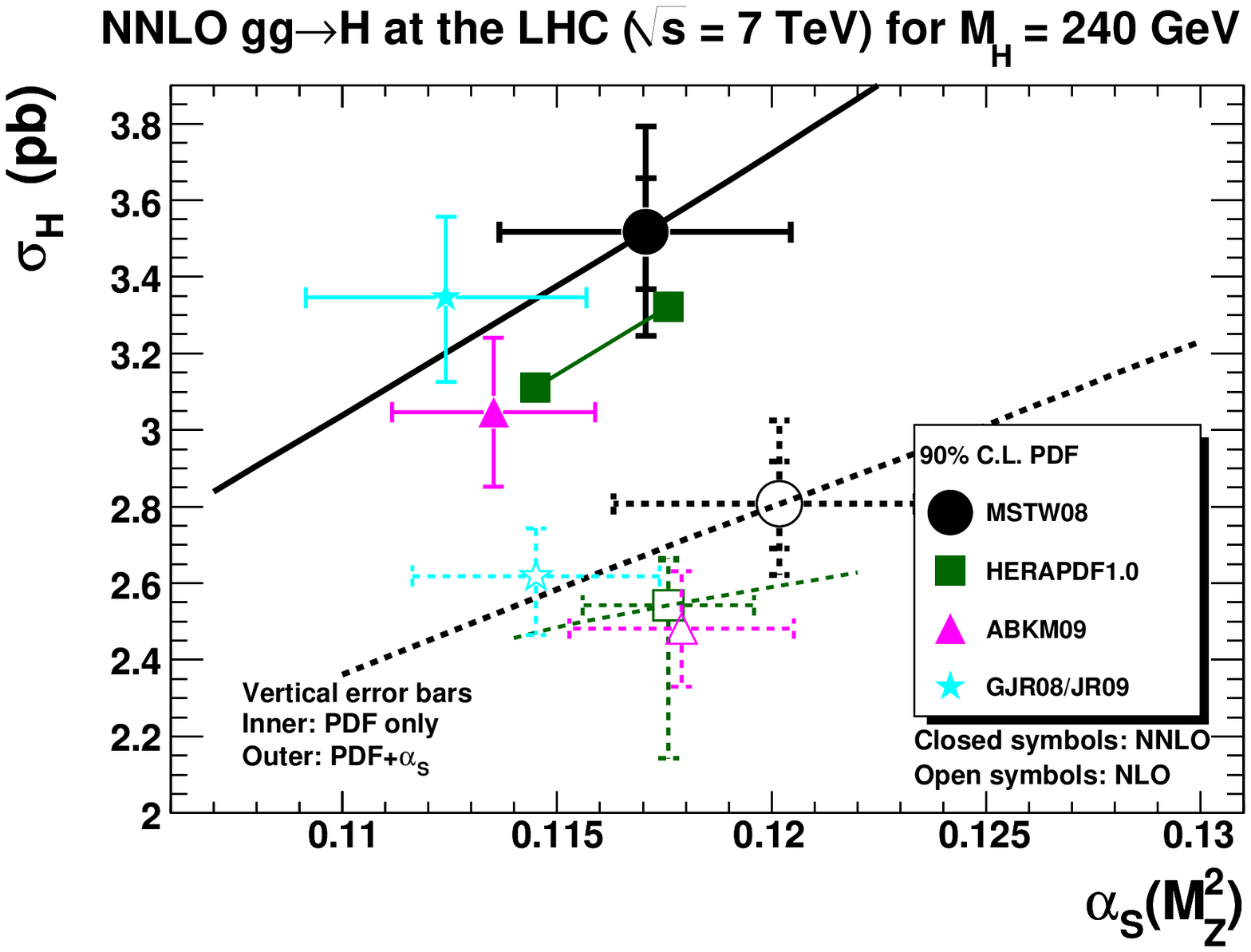}
  \end{minipage}
  \caption{$gg\to H$ total cross sections, plotted as a function of $\alpha_S(M_Z^2)$, at NNLO.}
  \label{fig:gghvsasmznnlo}
\end{figure}
We calculate the total cross section for Standard Model Higgs boson production (without decay) from gluon--gluon fusion ($gg\to H$) via a top-quark loop.  The $m_t$ dependence is retained only at leading-order (LO), with $m_t=171.3$~GeV (PDG 2009 best value), and the higher-order corrections are calculated in the limit of an infinite top-quark mass, with NNLO corrections from ref.~\cite{Harlander:2002wh}.  Again we use private code (originally by J.~Stirling), with the NLO corrections taken from ref.~\cite{Djouadi:1991tka} and agreement with \textsc{mcfm} was checked to within the numerical integration error (of at most 0.1\%).  The NNLO calculation is improved with respect to the numbers quoted in previous MSTW papers~\cite{Martin:2009bu,Thorne:2010pa,Thorne:2010kj,Martin:2010db} (which used $m_t=175$~GeV), where only terms up to $(1-z)^1$ (where $z\equiv M_H^2/\sqrt{s}$) were included in the expansion of the ``hard'' pieces of the NNLO partonic cross section around the kinematic point $z=1$ in powers of $(1-z)$~\cite{Harlander:2002wh}.  The improved calculation gives a 7~TeV LHC cross section for $M_H=120$~GeV which is 1.1\% larger than the expansion only up to $(1-z)^1$ and is in agreement with the exact NNLO calculation~\cite{Anastasiou:2002yz} to much less than 0.1\%.  We do not include the small bottom-quark loop contributions to the $gg\to H$ cross section.  We show the $gg\to H$ total cross sections, plotted as a function of $\alpha_S(M_Z^2)$, in figures~\ref{fig:gghvsasmznlo} and \ref{fig:gghvsasmznnlo}, for Higgs boson masses ($M_H$) of (a,b)~$120$~GeV, (c,d)~$180$~GeV and (e,f)~$240$~GeV, with (a,c,e)~68\% C.L.~uncertainties on the left and (b,d,f)~90\% C.L.~uncertainties on the right.  The $gg\to H$ cross sections start at $\mathcal{O}(\alpha_S^2)$ at LO, with anomalously large higher-order corrections, therefore they are directly sensitive to the value of $\alpha_S(M_Z^2)$.  Moreover, there is a known correlation between the value of $\alpha_S$ and the gluon distribution, which additionally affects the $gg\to H$ cross sections.  There is reasonable agreement between the \emph{global} fits at NLO, although CTEQ6.6/CT10 give somewhat lower cross sections than MSTW08/NNPDF2.1, even after accounting for the slightly lower default $\alpha_S(M_Z^2)$ value.  This behaviour reflects the differences in the $gg$ luminosity seen in figure~\ref{fig:rationlolumi}(c,d), i.e.~compared to MSTW08 and NNPDF2.1 the CTEQ6.6/CT10 $gg$ luminosity is larger at low $\hat{s}$ from the positive-definite input gluon constraint and larger at high $\hat{s}$ from fitting Tevatron Run I jet data, leading to a smaller $gg$ luminosity at the relevant intermediate $\sqrt{\hat{s}}\sim M_H$ values from the momentum sum rule.  Note from figure~\ref{fig:gghvsasmznnlo} that the NNLO corrections are huge, and therefore essential, and are only slightly reduced by taking different $\alpha_S(M_Z^2)$ values at different perturbative orders.  At NNLO, there is a significant discrepancy between the MSTW08 prediction and the ABKM09 prediction, even accounting for the very different $\alpha_S(M_Z^2)$ values.  The dependence of the $gg\to H$ cross sections at the LHC (and Tevatron) on the choice of PDFs (and $\alpha_S$) will be discussed in much more detail in ref.~\cite{higgs}, where we argue that the Tevatron jet data provide an important discriminator between the different PDF sets and strongly favour MSTW08 over ABKM09.

\subsection{\texorpdfstring{$t\bar{t}$}{ttbar} production}

We calculate $t\bar{t}$ production (without decay) for a top-quark pole mass $m_t=171.3$~GeV (PDG 2009 best value).  At NLO we use private code (originally by J.~Stirling) with NLO corrections from ref.~\cite{Nason:1987xz}, again checked to agree with \textsc{mcfm} to within the numerical integration error (of around 0.1\%).  The NNLO calculation of the total cross section for $t\bar{t}$ production is still in progress (see, for example, ref.~\cite{Bonciani:2010ue} for the status), although various approximations based on threshold resummation are available, in particular in the \textsc{hathor} public code~\cite{Aliev:2010zk}.  We will use the \textsc{hathor}~\cite{Aliev:2010zk} code with the default settings for an approximate ``NNLO'' calculation, although we will make no attempt to quantify the theoretical uncertainty (other than from PDFs and $\alpha_S$); see also the discussion in ref.~\cite{Salam:2011bj}.  There are very slight differences at NLO because \textsc{hathor}~\cite{Aliev:2010zk} uses a parameterisation~\cite{Langenfeld:2009wd} of the exact analytic result~\cite{Czakon:2008ii} instead of the original numerical result~\cite{Nason:1987xz}, giving cross sections around 0.03\% larger (i.e.~an insignificant amount) when both results are evaluated with high integration precision.

\begin{figure}
  \centering
  \begin{minipage}{0.5\textwidth}
    (a)\\
    \includegraphics[width=\textwidth]{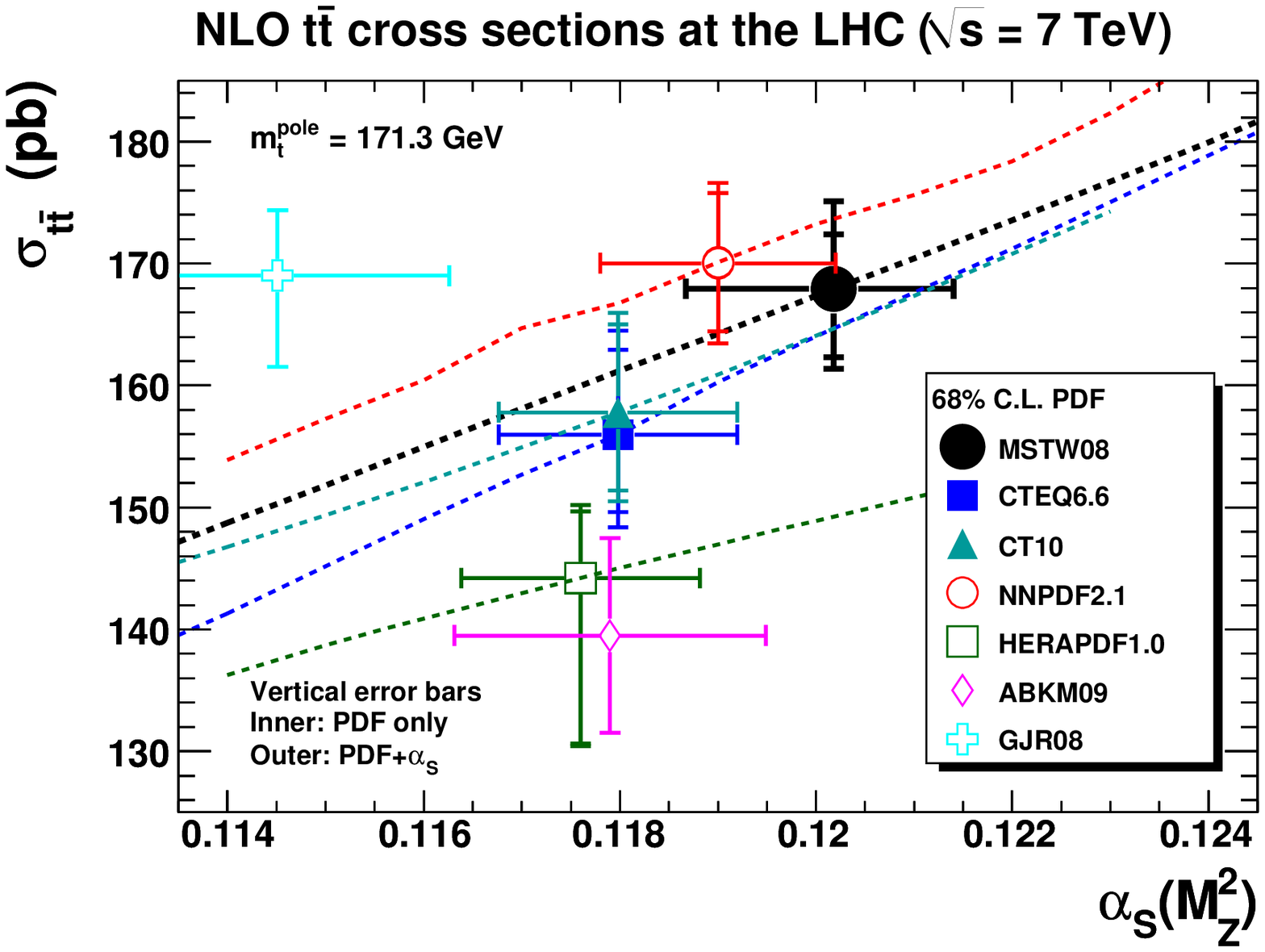}
  \end{minipage}%
  \begin{minipage}{0.5\textwidth}
    (b)\\
    \includegraphics[width=\textwidth]{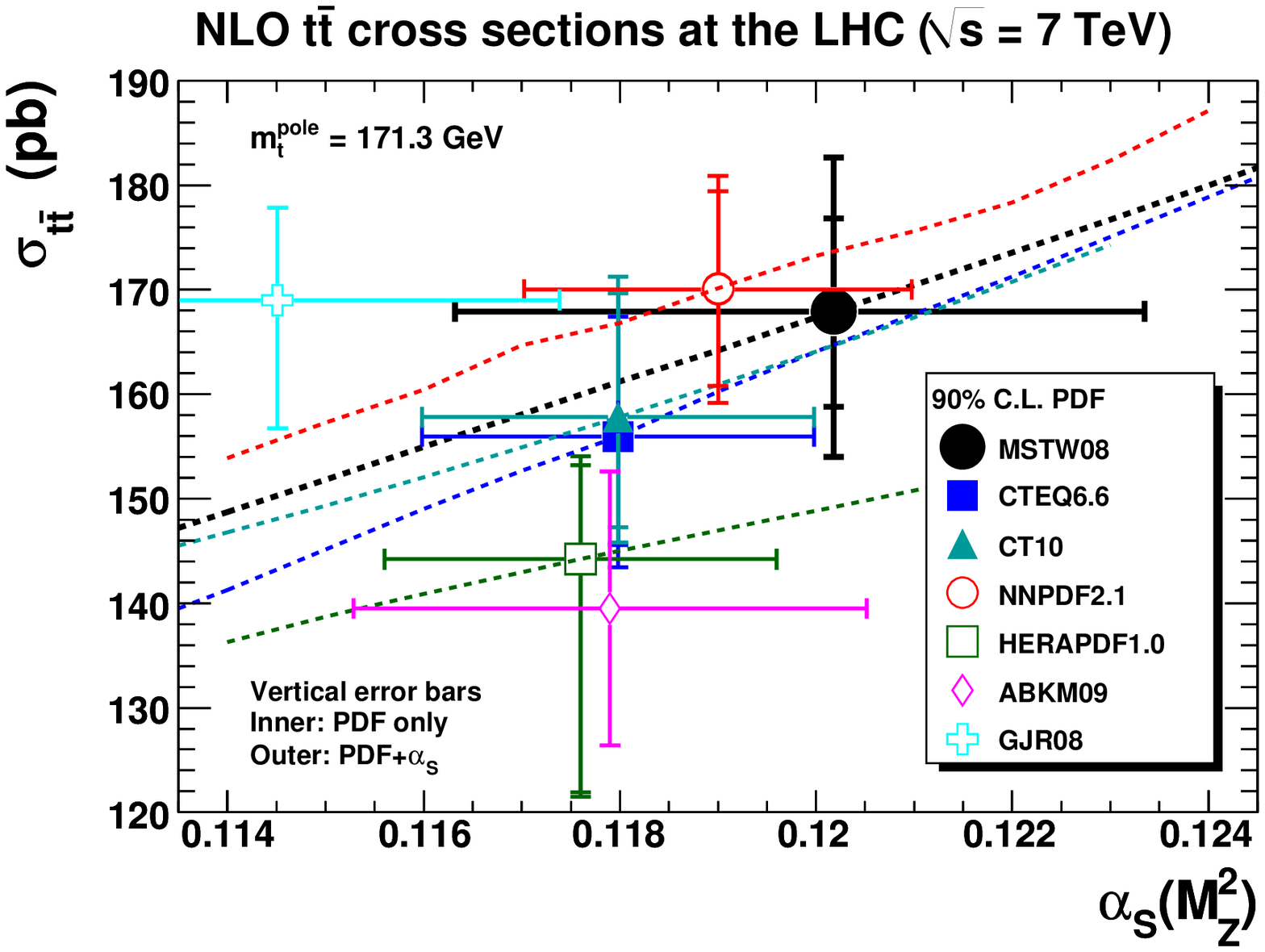}
  \end{minipage}
  \caption{$t\bar{t}$ total cross sections, plotted as a function of $\alpha_S(M_Z^2)$, at NLO, for $m_t=171.3$~GeV.}
  \label{fig:ttbarvsasmznlo}
\end{figure}
\begin{figure}
  \centering
  \begin{minipage}{0.5\textwidth}
    (a)\\
    \includegraphics[width=\textwidth]{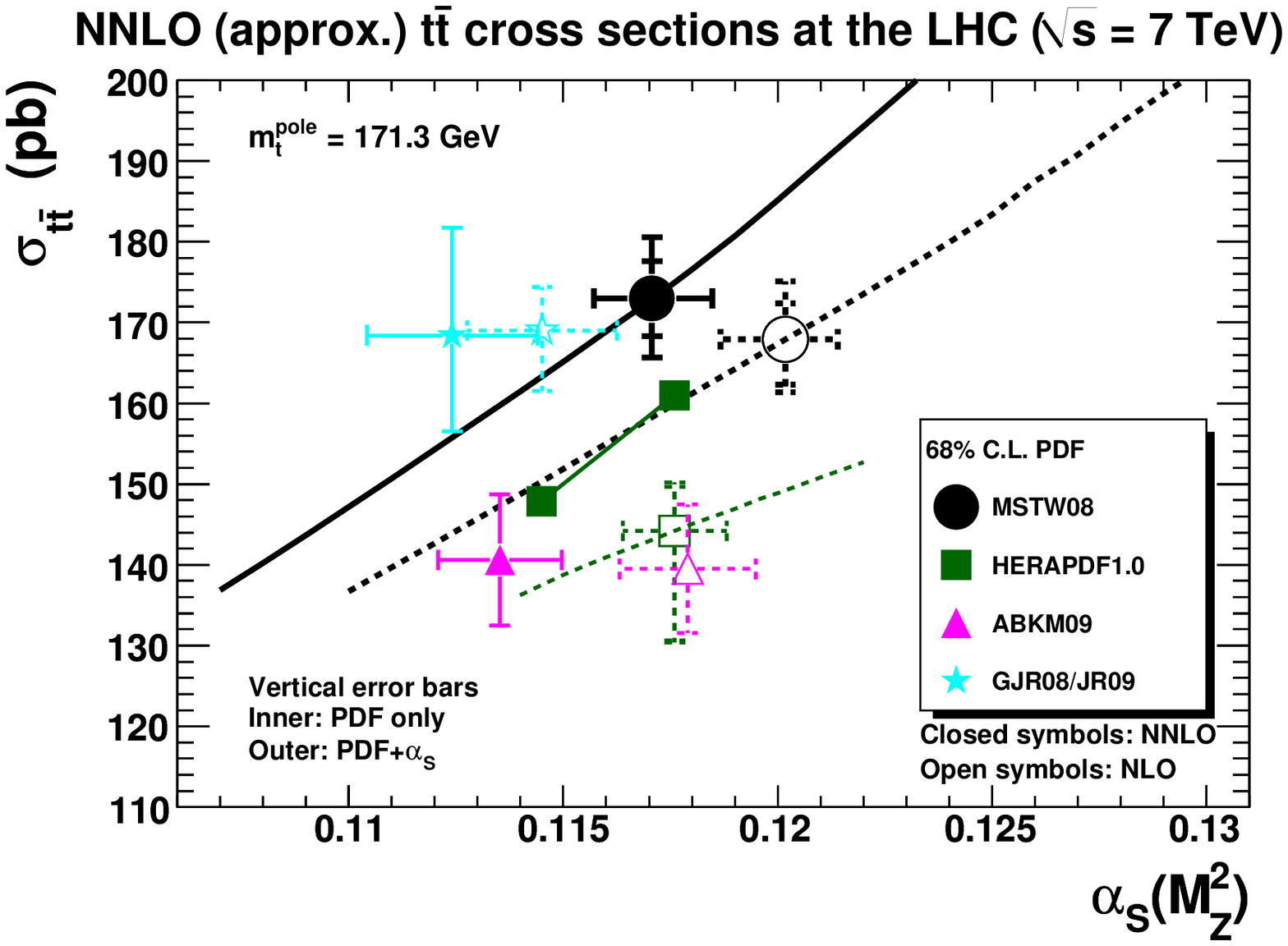}
  \end{minipage}%
  \begin{minipage}{0.5\textwidth}
    (b)\\
    \includegraphics[width=\textwidth]{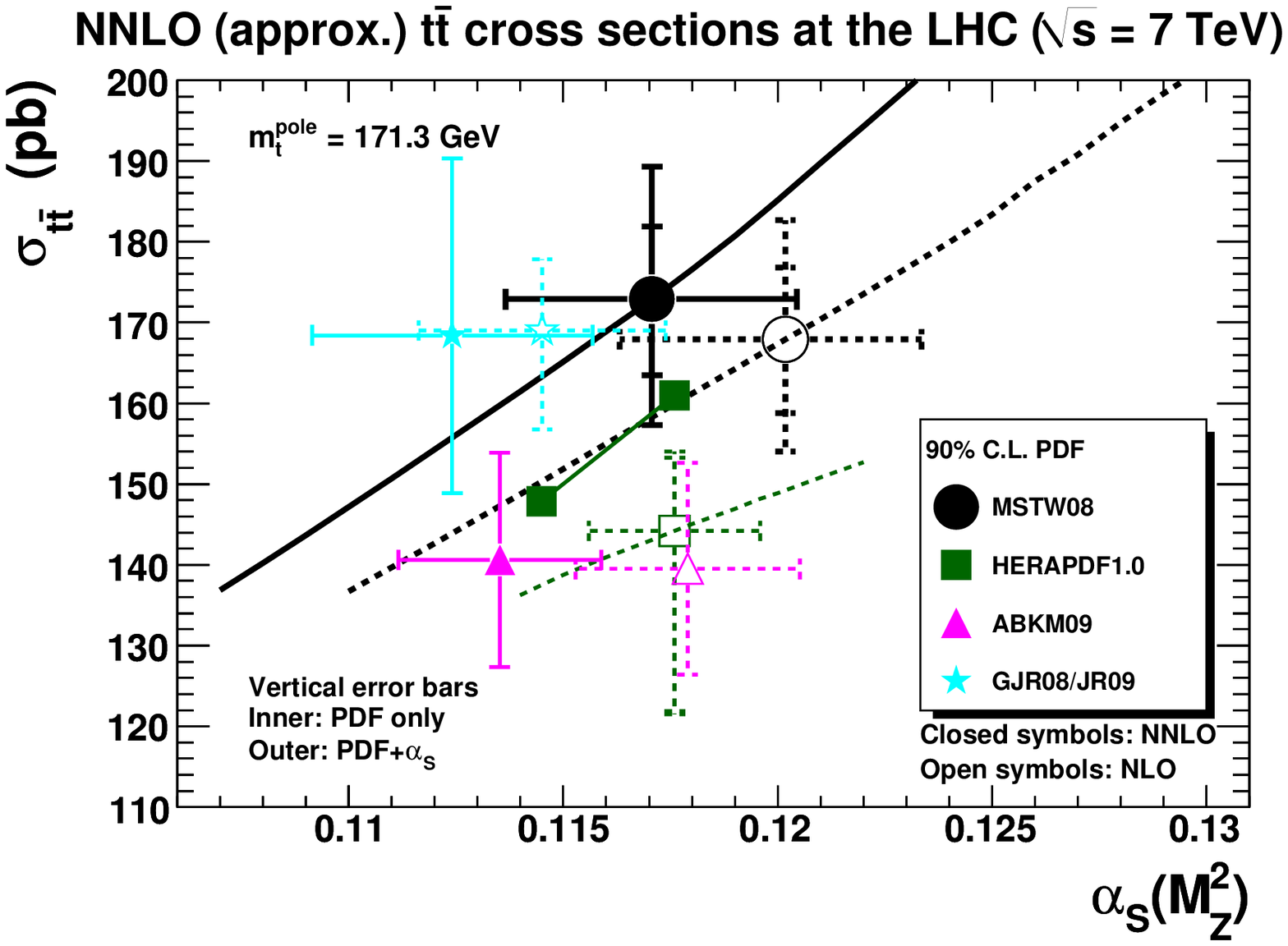}
  \end{minipage}
  \caption{$t\bar{t}$ total cross sections, plotted as a function of $\alpha_S(M_Z^2)$, at NNLO, for $m_t=171.3$~GeV.}
  \label{fig:ttbarvsasmznnlo}
\end{figure}
We show the $t\bar{t}$ total cross sections, plotted as a function of $\alpha_S(M_Z^2)$, in figures~\ref{fig:ttbarvsasmznlo} and \ref{fig:ttbarvsasmznnlo}, with (a)~68\% C.L.~uncertainties on the left and (b)~90\% C.L.~uncertainties on the right.  More than 80\% of the NLO $t\bar{t}$ cross section comes from the $gg$ channel for the LHC with $\sqrt{s} = 7$~TeV, rising to almost 90\% at $\sqrt{s} = 14$~TeV, compared to less than 15\% at the Tevatron ($\sqrt{s} = 1.96$~TeV).  The significant difference in the initial parton composition for $t\bar{t}$ production is due partly to the lower Tevatron energy ($pp$ collisions at $\sqrt{s}=1.96$~TeV would give around 50\% of the $t\bar{t}$ cross section from the $gg$ channel), but mainly due to the valence--valence nature of the $q\bar{q}\to t\bar{t}$ channel in $p\bar{p}$ collisions.  The partonic subprocess is $\mathcal{O}(\alpha_S^2)$ at LO.  There is therefore a strong dependence on both the gluon distribution (at $x\sim 2m_t/\sqrt{s}=0.05$) and $\alpha_S$.  The approximate NNLO corrections seem to be reasonably small, especially when taking different $\alpha_S(M_Z^2)$ values at different perturbative orders, but there are currently many ``NNLO'' choices and only one possibility is shown in figure~\ref{fig:ttbarvsasmznnlo}.  A more complete study of the theoretical uncertainties in the approximate NNLO calculation is clearly important, but it is beyond the scope of this paper.  Additionally, the predicted $t\bar{t}$ cross section has a fairly strong dependence on the assumed top-quark mass $m_t$, such that comparison of the measured cross section with theory predictions even allows an extraction of $m_t$~\cite{Langenfeld:2009wd}.  As some indication of the $m_t$ dependence, moving from $m_t=171.3$~GeV (PDG 2009 best value) to $m_t=173.3$~GeV~\cite{CDF:1900yx} decreases the $t\bar{t}$ cross section at the 7~TeV LHC by about 10~pb (or 6\%) at both NLO and NNLO with MSTW08 PDFs.  Moreover, the measured $t\bar{t}$ cross section, particularly at the Tevatron, is commonly used to constrain new physics contributions, therefore it is questionable whether it should be used directly as a PDF constraint.  Rather, we would hope that the gluon distribution (and $\alpha_S$) would be sufficiently constrained by other data sets, such as Tevatron jet production, that the $t\bar{t}$ cross section is a \emph{prediction} rather than a direct PDF \emph{constraint}.  Bearing these caveats in mind, the preliminary ATLAS combined measurement with 35~pb$^{-1}$ of data is $180\pm18$~pb~\cite{ATLAS:ttbar} and the preliminary CMS combined measurement with 36~pb$^{-1}$ of data is $158\pm19$~pb~\cite{CMS:ttbar}, so the approximate NNLO prediction using MSTW08 shown in figure~\ref{fig:ttbarvsasmznnlo} is consistent with both ATLAS and CMS, while the central value using ABKM09 is slightly less than 1-$\sigma$ below CMS and more than 2-$\sigma$ below ATLAS.  The discrepancy for ABKM09 would increase further if using a more up-to-date value of $m_t=173.3$~GeV~\cite{CDF:1900yx} rather than $m_t=171.3$~GeV (PDG 2009 best value), which would reduce all theory predictions by around 6\%.  We repeat that in another paper~\cite{higgs} we show that the Tevatron data on jet production also show a definite preference for MSTW08 over ABKM09.

\section{Discussion of PDF4LHC \texorpdfstring{Interim Recommendations}{recipe}} \label{sec:pdf4lhc}

The PDF4LHC \emph{Interim Recommendations}~\cite{Botje:2011sn} were originally formulated by the PDF4LHC Steering Committee in June 2010~\cite{PDF4LHC} in response to a request from the \emph{LHC Higgs Cross Section Working Group}~\cite{LHCHiggs:2010}.  The NLO and NNLO prescriptions for PDF+$\alpha_S$ uncertainties at 68\% C.L.~are summarised below~\cite{Botje:2011sn}.
\begin{description}
\item[NLO :] ``For the calculation of uncertainties at the LHC, use the envelope provided by the central values and PDF+$\alpha_S$ errors from the MSTW08, CTEQ6.6 and NNPDF2.0 PDFs.''  ``As a central value, use the midpoint of this envelope.''
\item[NNLO :] ``As a central value, use the MSTW08 prediction.''  Rescale the MSTW08 NNLO uncertainty by ``the factor obtained by dividing the full uncertainty obtained from the envelope of MSTW, CTEQ and NNPDF results at NLO by the MSTW uncertainty at NLO''.  This rescaling factor was found to be $\sim 2$ for $gg\to H$ at the LHC.
\end{description}
These recommendations were motivated by the fact that the three global fits (MSTW, CTEQ and NNPDF) are generally in reasonable agreement, as seen in this paper, but the deviations between predictions are sometimes as large as their uncertainties for certain observables, not always for well-understood reasons.  Other fits to more limited data sets, particularly those not including Tevatron jet data, generally have larger deviations, again as clearly seen from the results of this paper.  At NNLO, only the MSTW group currently provides NNLO PDFs, so the rescaling factor was introduced to expand the MSTW uncertainty in a similar way as at NLO, given that (experimental) PDF uncertainties are largely independent of perturbative order, and the observed differences between the three global fits (MSTW, CTEQ and NNPDF) at NLO would be expected to be similar also at NNLO.

The PDF4LHC \emph{Interim Recommendations}~\cite{Botje:2011sn} summarised above are already somewhat dated in that CTEQ6.6~\cite{Nadolsky:2008zw} has now been replaced by CT10~\cite{Lai:2010vv} and NNPDF2.0~\cite{Ball:2010de} has been replaced by NNPDF2.1~\cite{Ball:2011mu}, in both cases correcting known defects.  In particular, CTEQ6.6 includes only Tevatron Run I jet data, which are known to prefer a larger high-$x$ gluon distribution than the more reliable Run II data~\cite{Martin:2009iq}, while NNPDF2.0 uses the inadequate ZM-VFNS for DIS structure functions leading to systematically low predictions for $W$ and $Z$ cross sections at the LHC~\cite{PDF4LHC:March2010}.  Clearly a minimal update to the prescription would retain the original form, but with these replacements, i.e.~CTEQ6.6 $\to$ CT10 and NNPDF2.0 $\to$ NNPDF2.1.

However, the key question is whether we need an official recommendation at all.  The PDF4LHC recommendations~\cite{PDF4LHC,Botje:2011sn} were formulated by only the Steering Committee and should not be taken to represent a consensus decision of the PDF fitting community, as might be assumed.  Moreover, it is difficult to see that any sensible consensus can ever be reached, given that many proponents of a given prescription have a vested interest that their own particular PDF set is used as widely as possible, thereby encouraging the use of ``envelope'' prescriptions to satisfy the majority of those involved in the decision-making process.  The importance of PDF uncertainties is \emph{process-dependent}: it is overkill to use a complicated prescription universally, for example, when other theoretical uncertainties such as scale dependence are overwhelmingly dominant.  For some processes, uncertainties due to $\alpha_S$ are not important (for example, in the $W^\pm\to\ell^\pm\nu$ charge asymmetry), and therefore the additional effort required to compute a ``PDF+$\alpha_S$'' uncertainty rather than a ``PDF only'' uncertainty is not worthwhile.  For other processes, it may be necessary to include additional PDF-related uncertainties beyond the usual PDF+$\alpha_S$ prescription.  This is the case, for example, for processes initiated by bottom quarks such as $t$-channel single-top production, $qb\to q^\prime t$, or (supersymmetric) Higgs boson production from bottom-quark fusion, $b\bar{b}\to \Phi$, where the additional uncertainty due to the choice of the bottom-quark mass $m_b$ can be included for the MSTW08 PDFs according to the prescription presented in ref.~\cite{Martin:2010db}; see also the NNPDF2.1 studies~\cite{Ball:2011mu}.

\begin{figure}
  \centering
  \begin{minipage}{0.5\textwidth}
    (a)\\
    \includegraphics[width=\textwidth]{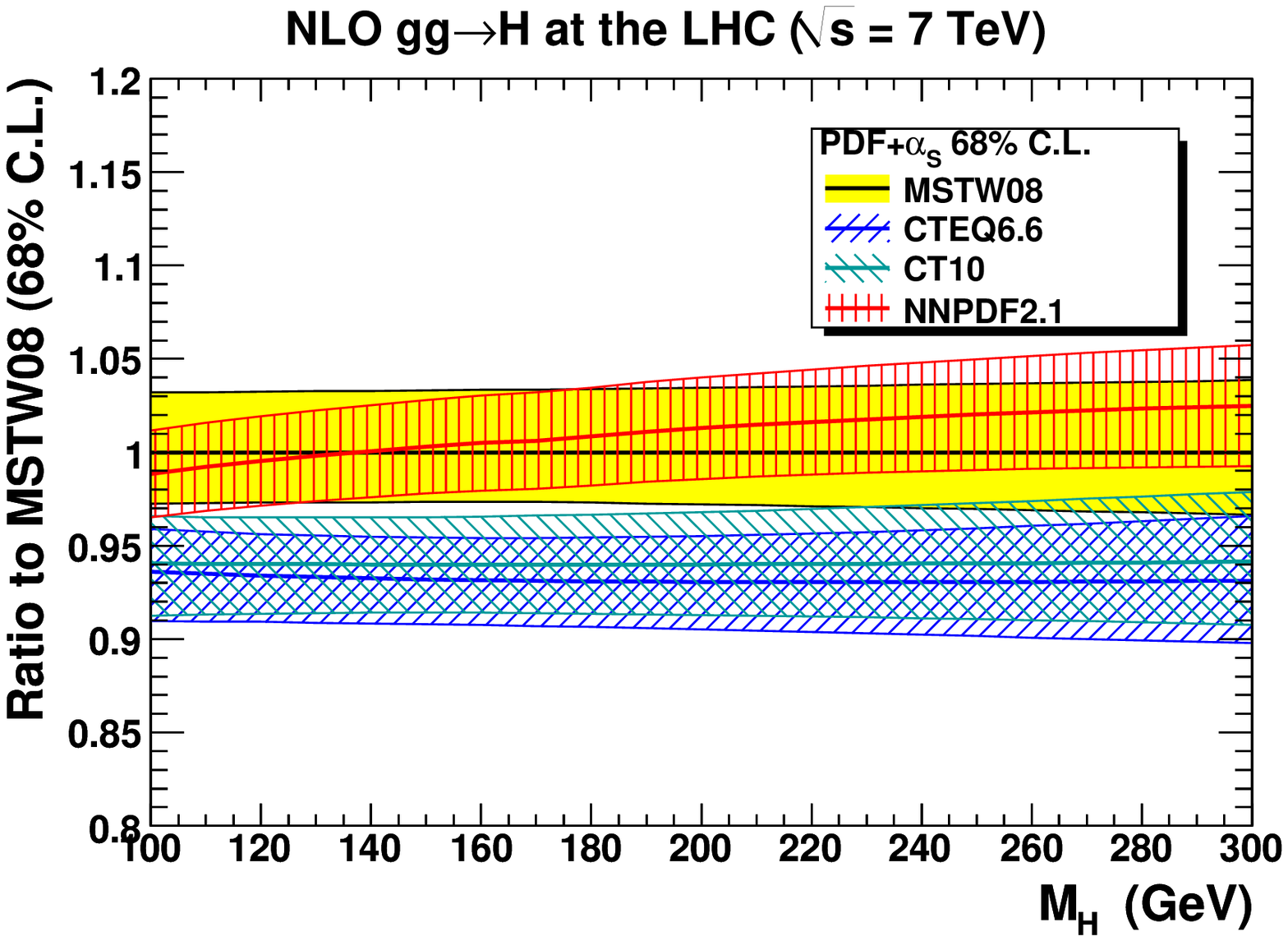}
  \end{minipage}%
  \begin{minipage}{0.5\textwidth}
    (b)\\
    \includegraphics[width=\textwidth]{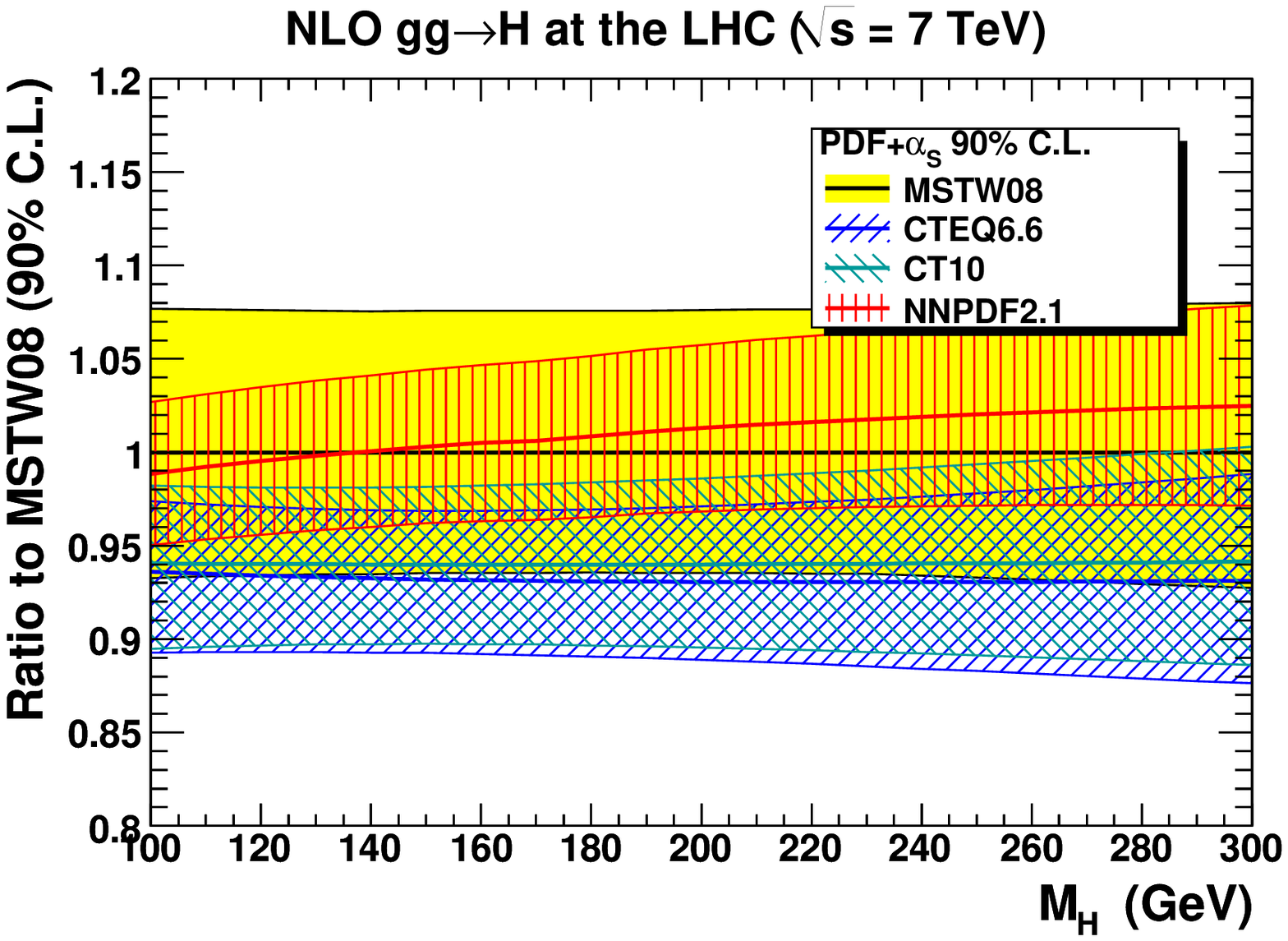}
  \end{minipage}
  \begin{minipage}{0.5\textwidth}
    (c)\\
    \includegraphics[width=\textwidth]{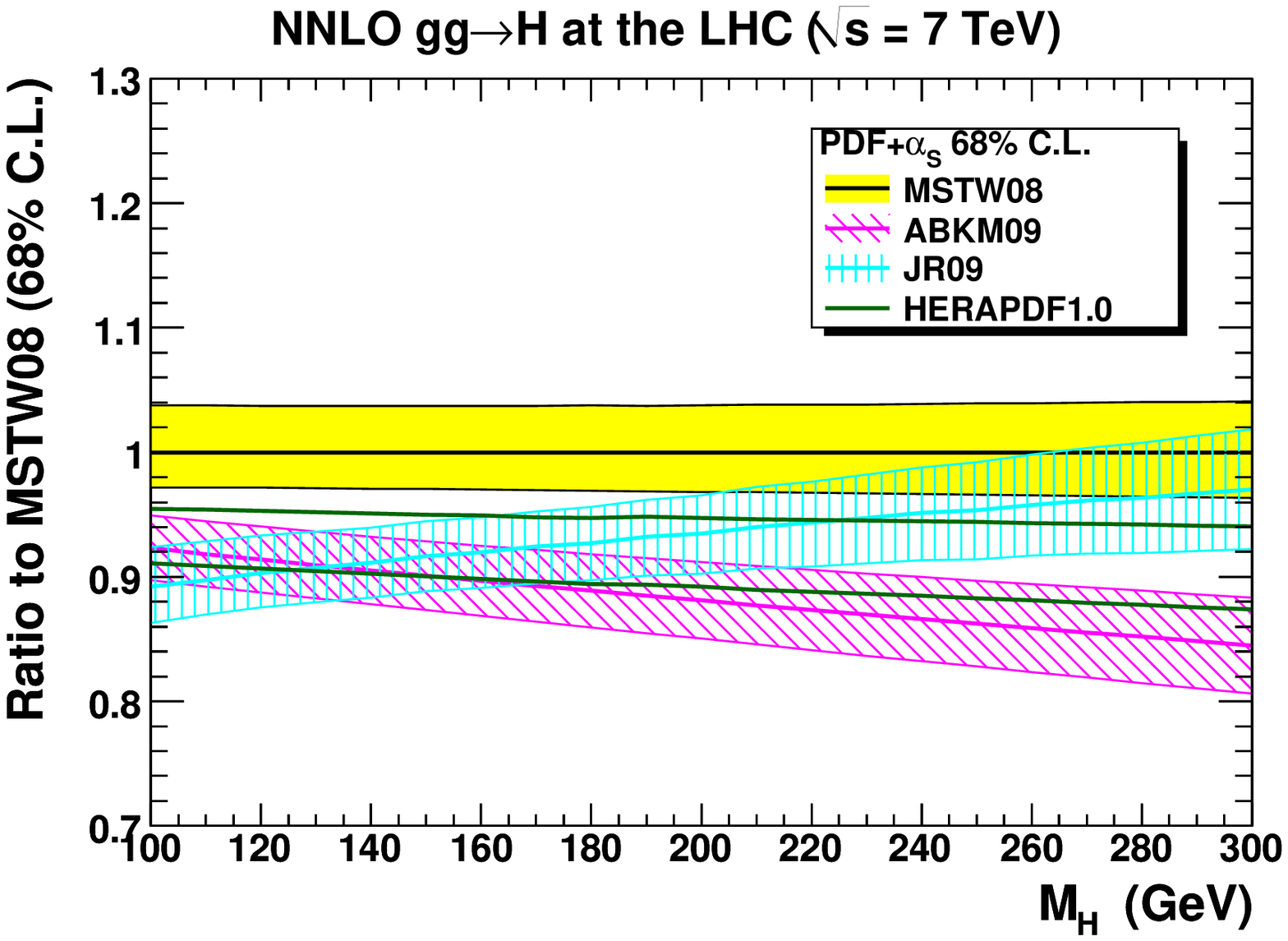}
  \end{minipage}%
  \begin{minipage}{0.5\textwidth}
    (d)\\
    \includegraphics[width=\textwidth]{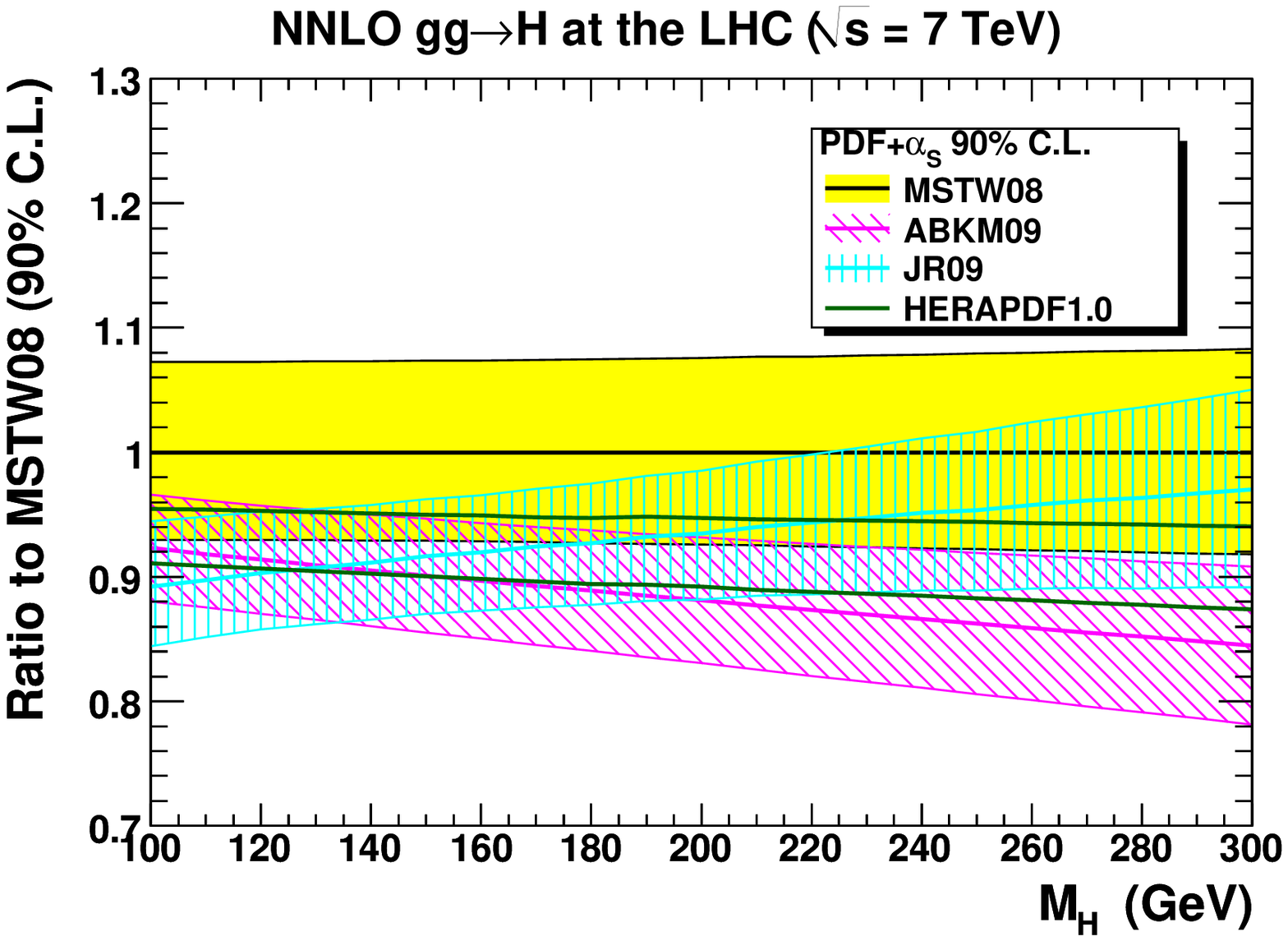}
  \end{minipage}
  \caption{Ratio to the MSTW08 prediction for $gg\to H$ with PDF+$\alpha_S$ uncertainties for (a)~NLO at 68\%~C.L., (b)~NLO at 90\%~C.L., (c)~NNLO at 68\%~C.L., (d)~NNLO at 90\%~C.L.}
  \label{fig:ggHvsMH}
\end{figure}

If there are large differences in predictions obtained with PDFs from different groups, these should be understood and results should be shown with multiple predictions, rather than simply taking the envelope.  On the other hand, if there are only small differences between the predictions and their uncertainties, the envelope method will give a similar result to the uncertainty of any one group, so it is redundant.  Moreover, the frequent nature of PDF releases means that any official recommendation will soon be outdated, as already seen with the use of CTEQ6.6 and NNPDF2.0.  Indeed the updated CT10 and NNPDF2.1 sets generally give predictions somewhat closer to MSTW08 than their predecessors.  One possible suggestion for a simpler recommendation is to use only MSTW08 uncertainties at 90\% C.L.~rather than the ``envelope of 68\%'' currently proposed.  This is a common prescription used, for example, in the 2010 Tevatron Higgs exclusion analysis~\cite{CDF:2010ar}, but not the most recent one~\cite{CDF:2011gs} where the PDF4LHC recommendations have been adopted.  It can be applied consistently at NLO and NNLO and gives similar results to the PDF4LHC recipe; see figure~\ref{fig:ggHvsMH} where we show the $gg\to H$ total cross section at the LHC versus the Higgs mass $M_H$, at both NLO and NNLO, and with PDF+$\alpha_S$ uncertainties at both 68\% and 90\% C.L.  This simpler recommendation could be reconsidered when CTEQ and NNPDF sets appear at NNLO, but these are expected to follow the same trend as at NLO, i.e.~to be in reasonably good agreement with MSTW when using 90\% C.L.~uncertainties.

Despite the criticism above, the PDF4LHC prescription is not unreasonable when a single conservative estimate of PDF uncertainties is necessary (although it would be worrying if the result depended strongly on the particular recipe).  This is the case, for example, in searches for new physics, in the calculation of acceptance corrections, and in the measurement of the $W$ boson mass at hadron colliders~\cite{Bozzi:2011ww}, and was the type of situation for which the PDF4LHC recommendations were formulated.  However, it is inappropriate for Standard Model measurements sensitive to PDFs, such as $W,Z$ and jet cross sections, where the PDF4LHC recipe has commonly been applied by CMS~\cite{Khachatryan:2010xn,CMS:WandZ,Chatrchyan:2011qt,Chatrchyan:2011me}, but not by ATLAS~\cite{Aad:2010yt,ATLAS:WandZ,ATLAS:jetsmeasured,ATLAS:jetscomparison}.  Showing only the envelope \emph{masks} rather than \emph{exposes} potential discrepancies between PDF sets, contrary to a primary motivation for these types of measurements as a potential PDF constraint.  Indeed the original recommendations noted that~\cite{Botje:2011sn}:
\begin{quote}
  ``\ldots it is useful to highlight the differences between two use cases: (1) cross sections which have not yet been measured (such as, for example, Higgs production) and (2) comparisons to existing cross sections.  For the latter, the most useful comparisons should be to the predictions using individual PDFs (and their uncertainty bands). Such cross sections have the potential, for example, to provide information useful for modification of those PDFs.''
\end{quote}

The PDF4LHC recipe was formulated only for case (1).  Due to the intrinsic complication such that it is difficult for non-experts to understand, let alone implement unambiguously in practice, the PDF4LHC recommmendations have already frequently been modified or misinterpreted, but still labelled as following the ``PDF4LHC'' recipe.  For example, theory comparisons with CMS data on dijet production~\cite{Chatrchyan:2011qt} and inclusive jet production~\cite{Chatrchyan:2011me} use CT10 rather than CTEQ6.6, construct the envelope and midpoint \emph{without} $\alpha_S$ uncertainties, then vary $\alpha_S(M_Z^2)$ by $\pm0.002$ using only CT10 (rather than the recommended $\pm0.0012$ for 1-$\sigma$).  Theory comparisons with the CMS data on isolated prompt photon production~\cite{Khachatryan:2010fm} obtained the uncertainty by constructing the envelope using \emph{only} the central values of CT10, MSTW08 and NNPDF2.0, which is arguably a worse prescription than using only the PDF uncertainties of any one group.  Envelope prescriptions for the uncertainty also have the disadvantage that PDF correlations between cross sections, such as those defined in section~\ref{sec:correlations}, cannot be easily calculated.

We conclude that rather than blindly follow the PDF4LHC recommmendations, or some ad-hoc modification, an appropriate treatment of PDF uncertainties should be chosen depending on the process under consideration and the PDF sets available at the time.  Of course, in some cases, but by no means all, constructing an envelope and midpoint may well be a sensible prescription, but this need not be applied universally without justification.  In many other cases, it is better to instead consistently evaluate the complete uncertainty using the prescription provided by one (or preferably, more than one) PDF fitting group.

\section{Discussion of \texorpdfstring{$W$ and $Z$}{W and Z} production\texorpdfstring{ at the LHC}{}} \label{sec:wandz}

In table~\ref{tab:WZuncertainties} we give percentage uncertainties at 68\% and 90\% C.L.~of the NNLO predictions for $W^\pm$ ($=W^++W^-$) and $Z^0$ total cross sections, and their ratio $R_{WZ}$, using MSTW 2008 PDFs~\cite{Martin:2009iq,Martin:2009bu,Martin:2010db}.  In table~\ref{tab:PMuncertainties} we show the same results for the $W^+$ and $W^-$ total cross sections, and their ratio $R_{\pm}$.  The asymmetric PDF~\cite{Martin:2009iq}, PDF+$\alpha_S$~\cite{Martin:2009bu} and PDF+$\alpha_S$+$m_{c,b}$~\cite{Martin:2010db} uncertainties are calculated using the prescriptions given in the respective papers.  The uncertainties due to the choice of heavy-quark masses ($m_c$ and $m_b$) are estimated, for fixed $\alpha_S(M_Z^2)$, by varying the charm-quark mass in the range $m_c=1.40\pm0.15$~GeV at 68\%~C.L.~and $m_c=1.40\pm0.25$~GeV at 90\% C.L., and the bottom-quark mass in the range $m_b=4.75\pm0.25$~GeV at 68\%~C.L.~and $m_b=4.75\pm0.50$~GeV at 90\% C.L., then adding the variation in quadrature with the PDF+$\alpha_S$ uncertainty.  The uncertainties coming from neglected higher-order corrections beyond NNLO are estimated at 68\% C.L.~by allowing the renormalisation and factorisation scales (with central values $\mu_R = \mu_F = M_{W,Z}$) to vary independently so that the ratios $\mu_R/M_{W,Z}$, $\mu_F/M_{W,Z}$ and $\mu_R/\mu_F$ all lie in the range between $1/2$ and $2$, then the 90\% C.L.~uncertainty is obtained by multiplying by 1.64485.  Finally, this uncertainty from scale variation is added in quadrature with the PDF+$\alpha_S$+$m_{c,b}$ uncertainty, but in all cases the final uncertainty is not much more than the PDF+$\alpha_S$ uncertainty used elsewhere in this paper.

\begin{table}
  \centering
  \begin{tabular}{|l|c|c|c|c|c|c|}
    \hline
    LHC, $\sqrt{s} = 7$ TeV & \multicolumn{2}{c|}{$B_{\ell\nu} \cdot \sigma_{W^\pm}$} & \multicolumn{2}{c|}{$B_{\ell^+\ell^-}\cdot\sigma_{Z^0}$} & \multicolumn{2}{c|}{$R_{WZ}$} \\ \hline
    Uncertainties (\%) & 68\% & 90\% & 68\% & 90\% & 68\% & 90\% \\ \hline
    PDF only & $^{+1.7}_{-1.6}$ & $^{+3.4}_{-3.1}$ & $^{+1.7}_{-1.6}$ & $^{+3.4}_{-3.1}$ & $^{+0.2}_{-0.1}$ & $^{+0.4}_{-0.3}$ \\
    PDF+$\alpha_S$ & $^{+2.6}_{-1.9}$ & $^{+4.6}_{-3.9}$ & $^{+2.5}_{-1.8}$ & $^{+4.4}_{-3.9}$ & $^{+0.3}_{-0.2}$ & $^{+0.5}_{-0.4}$ \\ \hline
    Charm mass, $m_c$ & $^{+1.1}_{-1.2}$ & $^{+1.8}_{-2.1}$ & $^{+1.4}_{-1.5}$ & $^{+2.3}_{-2.4}$ & $^{+0.2}_{-0.3}$ & $^{+0.4}_{-0.5}$ \\
    Bottom mass, $m_b$ & $^{+0.1}_{-0.1}$ & $^{+0.2}_{-0.3}$ & $^{+0.0}_{-0.0}$ & $^{+0.0}_{-0.0}$ & $^{+0.1}_{-0.1}$ & $^{+0.2}_{-0.2}$ \\
    PDF+$\alpha_S$+$m_{c,b}$ & $^{+2.8}_{-2.2}$ & $^{+4.9}_{-4.5}$ & $^{+2.9}_{-2.3}$ & $^{+5.0}_{-4.6}$ & $^{+0.4}_{-0.3}$ & $^{+0.7}_{-0.7}$ \\ \hline
    Scales ($\mu_R,\mu_F$) & $^{+0.5}_{-0.8}$ & $^{+0.9}_{-1.3}$ & $^{+0.5}_{-0.7}$ & $^{+0.8}_{-1.1}$ & $^{+0.1}_{-0.1}$ & $^{+0.1}_{-0.2}$ \\
    PDF+$\alpha_S$+$m_{c,b}$+$\mu_{R,F}$ & $^{+2.8}_{-2.4}$ & $^{+5.0}_{-4.6}$ & $^{+2.9}_{-2.4}$ & $^{+5.0}_{-4.7}$ & $^{+0.4}_{-0.4}$ & $^{+0.7}_{-0.7}$ \\
    \hline
  \end{tabular}
  \caption{Percentage uncertainties at 68\% and 90\% C.L.~of the NNLO predictions for $W^\pm$ ($=W^++W^-$) and $Z^0$ total cross sections, and their ratio $R_{WZ}$, using MSTW 2008 PDFs~\cite{Martin:2009iq,Martin:2009bu,Martin:2010db}.}
  \label{tab:WZuncertainties}
\end{table}

\begin{table}
  \centering
  \begin{tabular}{|l|c|c|c|c|c|c|}
    \hline
    LHC, $\sqrt{s} = 7$ TeV & \multicolumn{2}{c|}{$B_{\ell\nu} \cdot \sigma_{W^+}$} & \multicolumn{2}{c|}{$B_{\ell\nu}\cdot\sigma_{W^-}$} & \multicolumn{2}{c|}{$R_\pm$} \\ \hline
    Uncertainties (\%) & 68\% & 90\% & 68\% & 90\% & 68\% & 90\% \\ \hline
    PDF only & $^{+1.8}_{-1.6}$ & $^{+3.6}_{-3.2}$ & $^{+1.7}_{-1.7}$ & $^{+3.5}_{-3.3}$ & $^{+1.0}_{-0.7}$ & $^{+2.2}_{-1.5}$ \\
    PDF+$\alpha_S$ & $^{+2.6}_{-1.9}$ & $^{+4.9}_{-4.0}$ & $^{+2.5}_{-1.9}$ & $^{+4.4}_{-4.2}$ & $^{+1.1}_{-0.7}$ & $^{+2.3}_{-1.5}$ \\ \hline
    Charm mass, $m_c$ & $^{+1.2}_{-1.3}$ & $^{+1.9}_{-2.2}$ & $^{+1.0}_{-1.1}$ & $^{+1.5}_{-1.8}$ & $^{+0.2}_{-0.2}$ & $^{+0.4}_{-0.4}$ \\
    Bottom mass, $m_b$ & $^{+0.1}_{-0.1}$ & $^{+0.2}_{-0.2}$ & $^{+0.1}_{-0.1}$ & $^{+0.2}_{-0.3}$ & $^{+0.0}_{-0.0}$ & $^{+0.0}_{-0.0}$ \\
    PDF+$\alpha_S$+$m_{c,b}$ & $^{+2.9}_{-2.3}$ & $^{+5.2}_{-4.6}$ & $^{+2.7}_{-2.2}$ & $^{+4.7}_{-4.6}$ & $^{+1.1}_{-0.8}$ & $^{+2.4}_{-1.5}$ \\ \hline
    Scales ($\mu_R,\mu_F$) & $^{+0.5}_{-0.8}$ & $^{+0.9}_{-1.3}$ & $^{+0.6}_{-0.8}$ & $^{+1.0}_{-1.3}$ & $^{+0.1}_{-0.1}$ & $^{+0.2}_{-0.1}$ \\
    PDF+$\alpha_S$+$m_{c,b}$+$\mu_{R,F}$ & $^{+2.9}_{-2.5}$ & $^{+5.3}_{-4.8}$ & $^{+2.7}_{-2.3}$ & $^{+4.8}_{-4.8}$ & $^{+1.1}_{-0.8}$ & $^{+2.4}_{-1.6}$ \\
    \hline
  \end{tabular}
  \caption{Percentage uncertainties at 68\% and 90\% C.L.~of the NNLO predictions for $W^+$ and $W^-$ total cross sections, and their ratio $R_{\pm}$, using MSTW 2008 PDFs~\cite{Martin:2009iq,Martin:2009bu,Martin:2010db}.}
  \label{tab:PMuncertainties}
\end{table}

\begin{figure}
  \centering
  \begin{minipage}{0.5\textwidth}
    (a)\\
    \includegraphics[width=\textwidth]{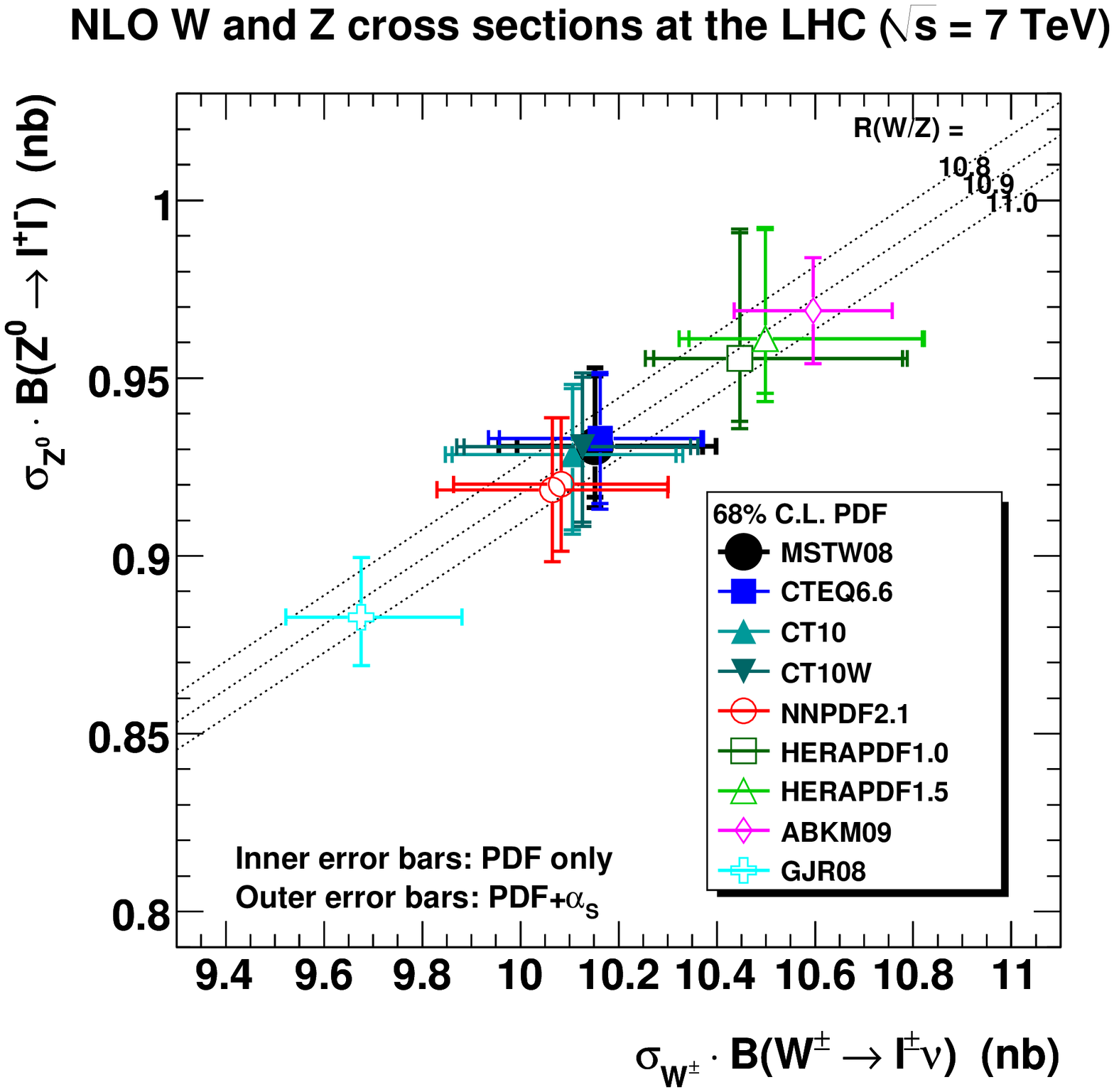}
  \end{minipage}%
  \begin{minipage}{0.5\textwidth}
    (b)\\
    \includegraphics[width=\textwidth]{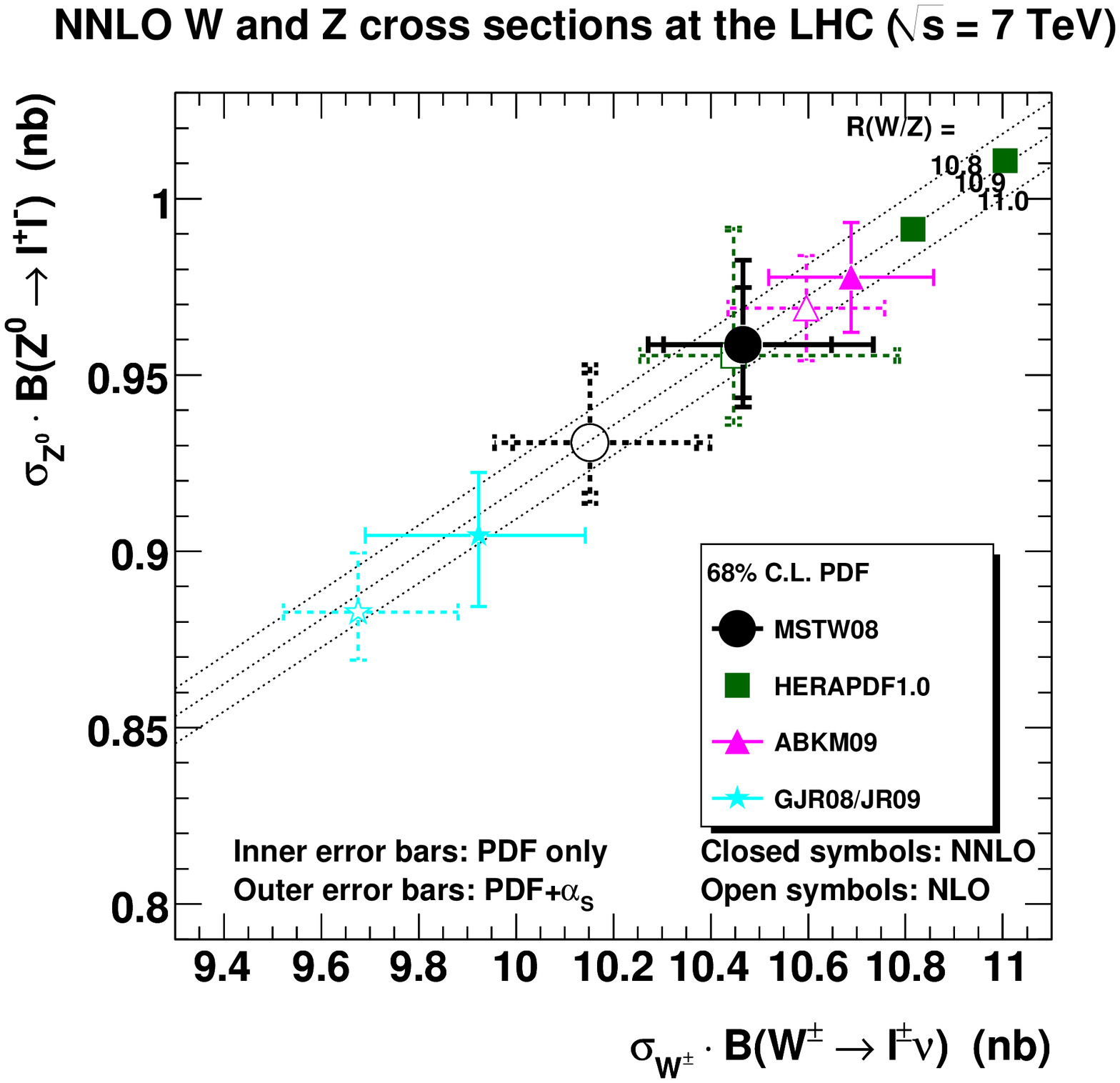}
  \end{minipage}
  \begin{minipage}{0.5\textwidth}
    (c)\\
    \includegraphics[width=\textwidth]{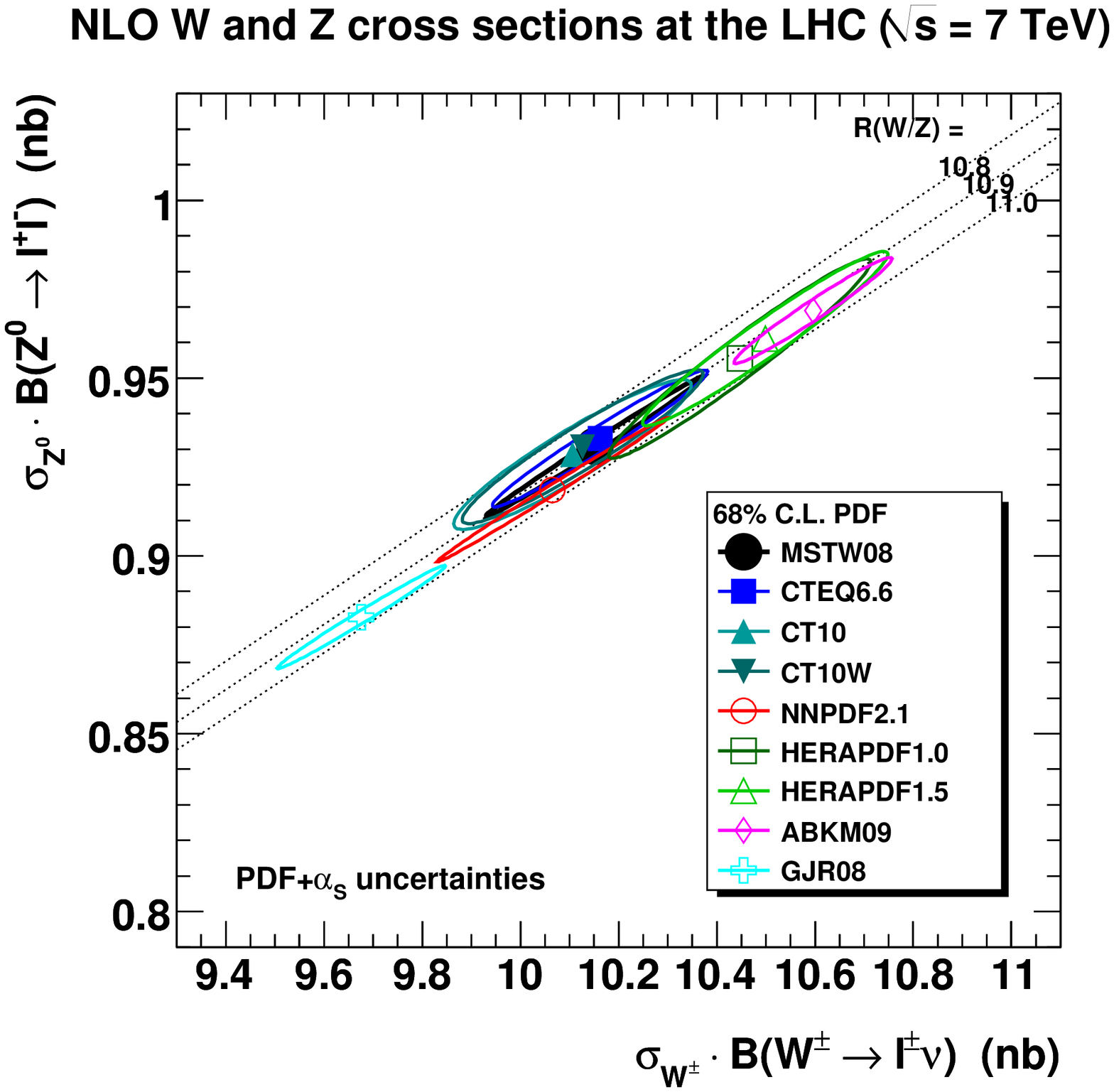}
  \end{minipage}%
  \begin{minipage}{0.5\textwidth}
    (d)\\
    \includegraphics[width=\textwidth]{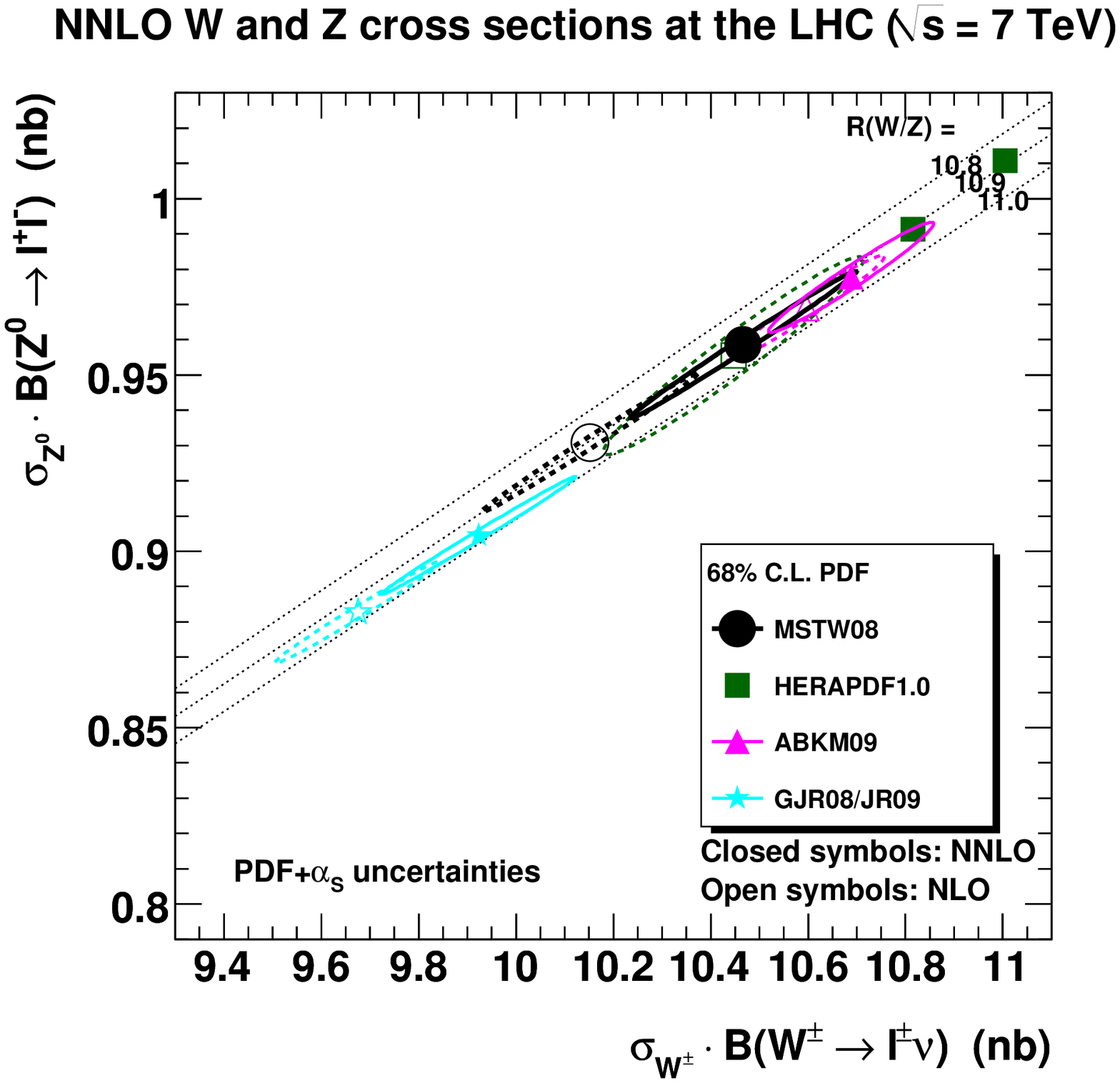}
  \end{minipage}
  \caption{$W^\pm$ ($=W^++W^-$) versus $Z^0$ total cross sections at (a)~NLO~and (b)~NNLO, then (c,d)~the same plots with ellipses accounting for PDF correlations between the two cross sections.}
  \label{fig:wandz}
\end{figure}

\begin{figure}
  \centering
  \begin{minipage}{0.5\textwidth}
    (a)\\
    \includegraphics[width=\textwidth]{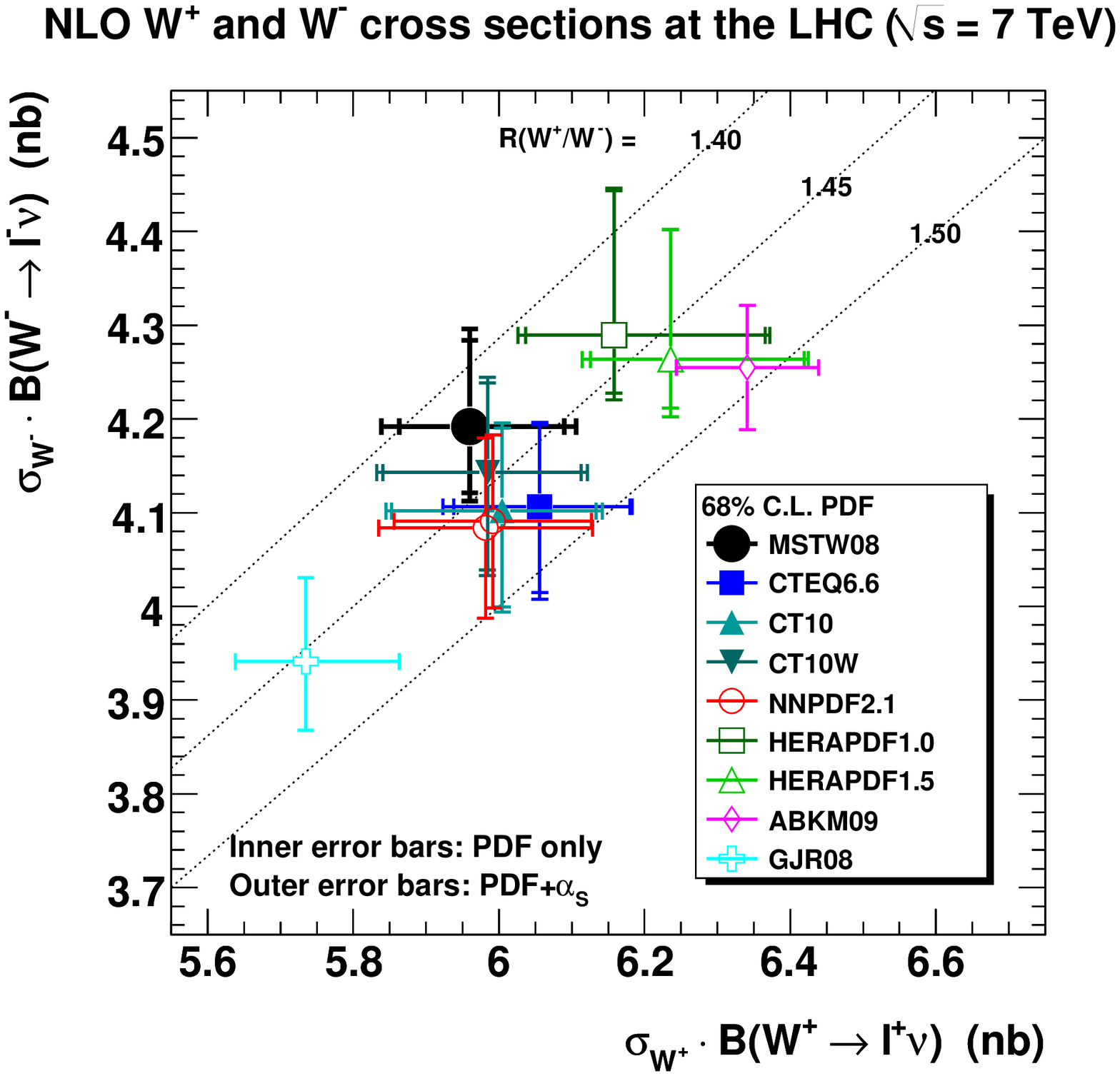}
  \end{minipage}%
  \begin{minipage}{0.5\textwidth}
    (b)\\
    \includegraphics[width=\textwidth]{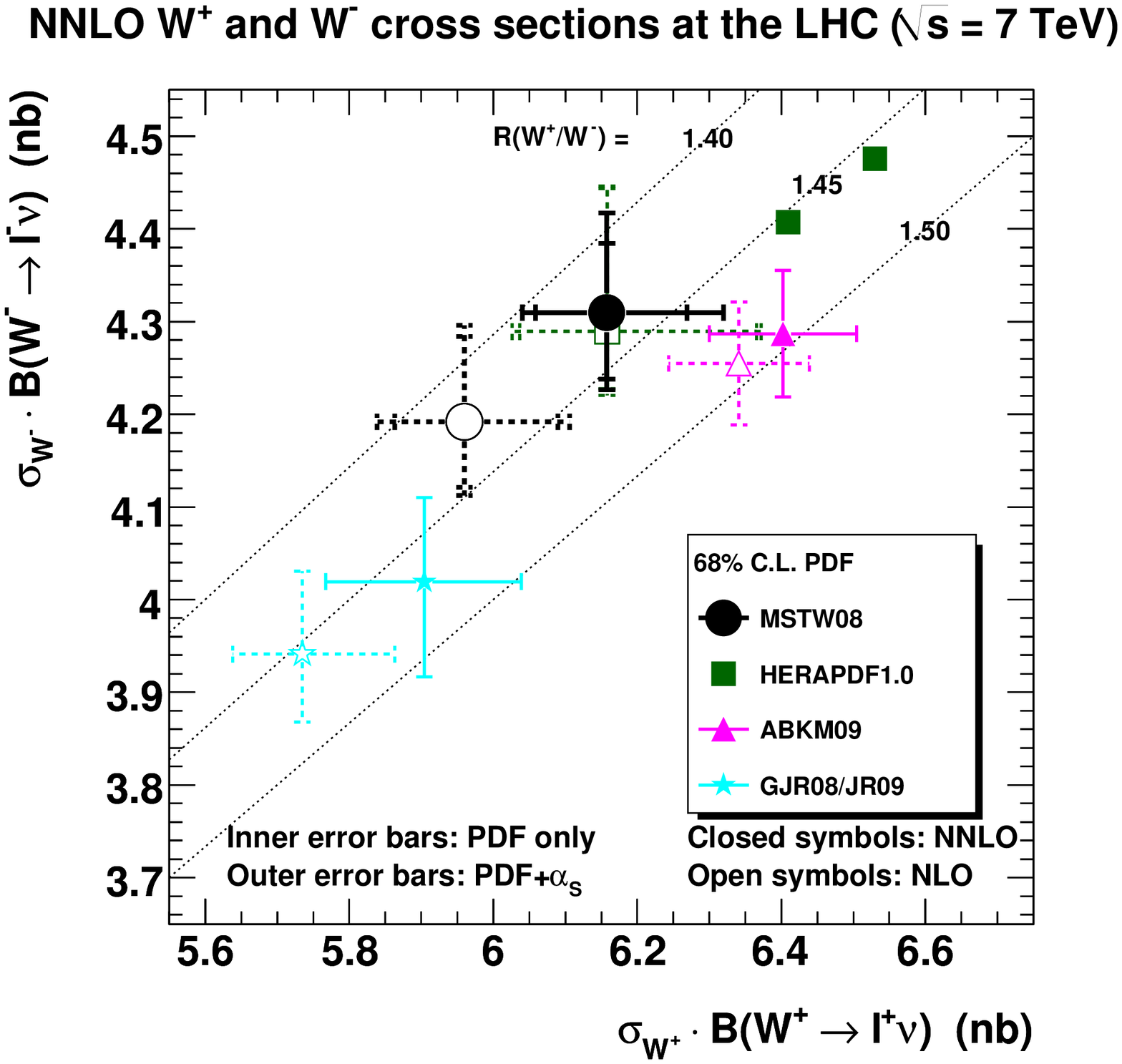}
  \end{minipage}
  \begin{minipage}{0.5\textwidth}
    (c)\\
    \includegraphics[width=\textwidth]{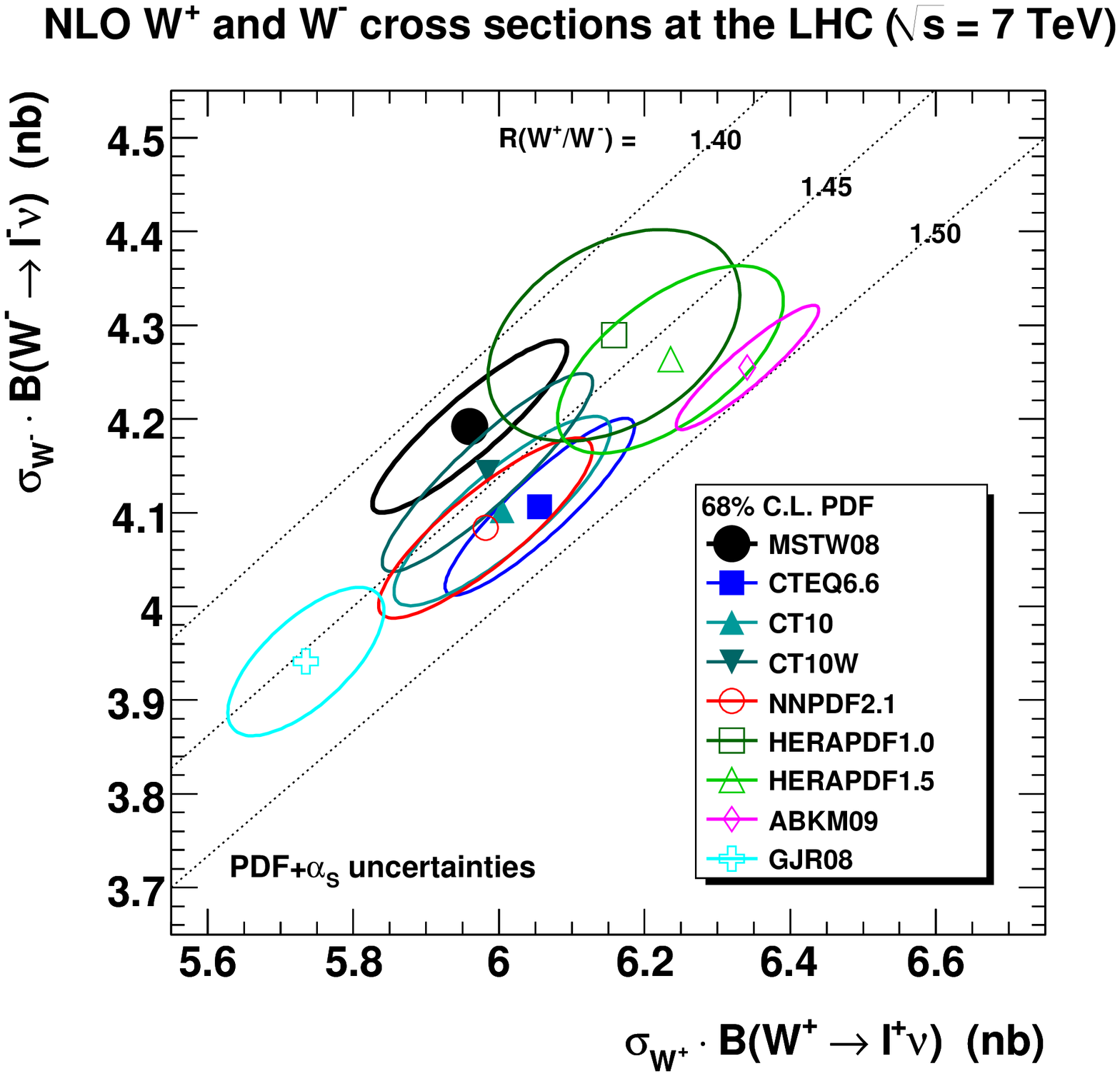}
  \end{minipage}%
  \begin{minipage}{0.5\textwidth}
    (d)\\
    \includegraphics[width=\textwidth]{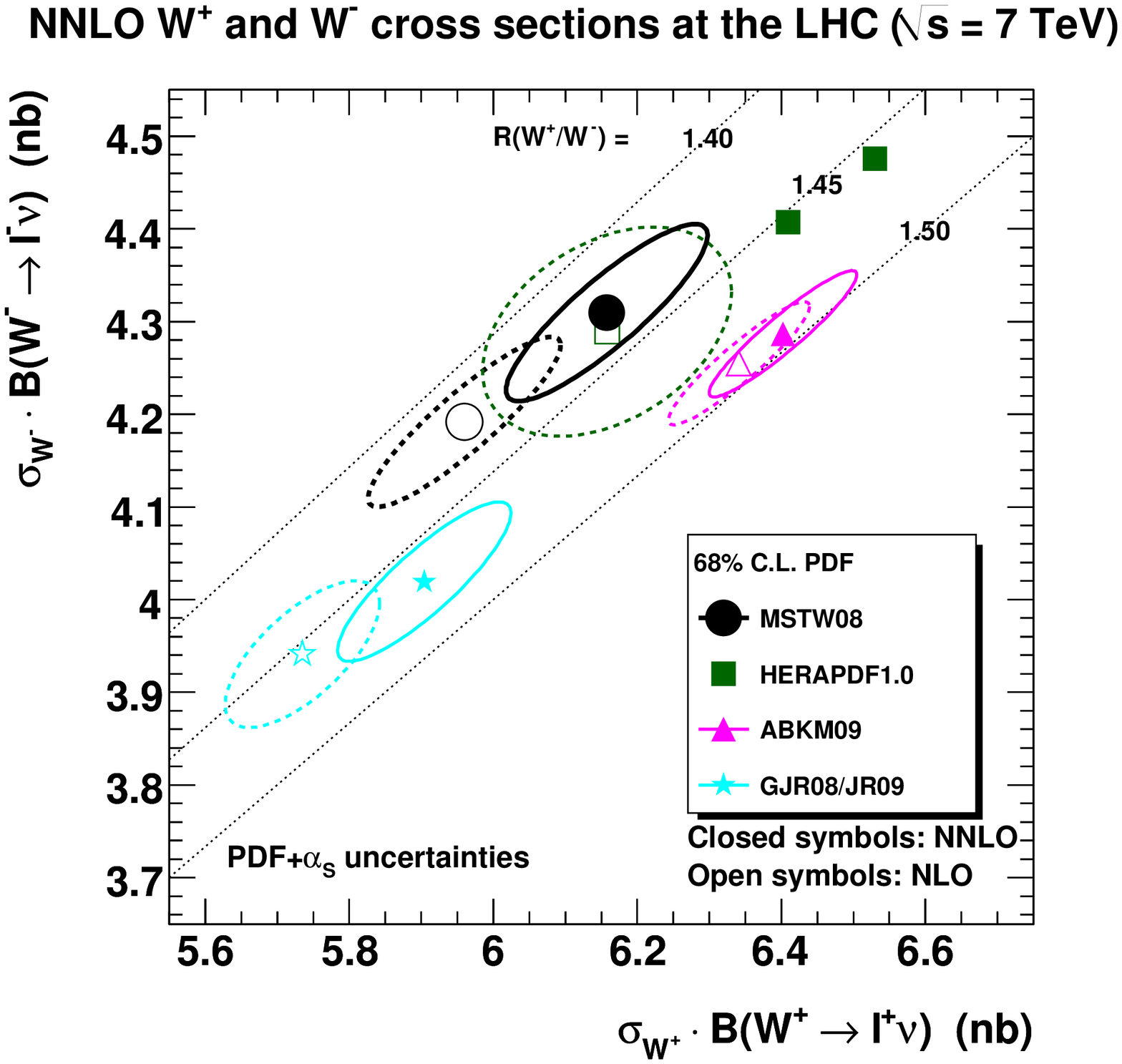}
  \end{minipage}
  \caption{$W^+$ versus $W^-$ total cross sections at (a)~NLO~and (b)~NNLO, then (c,d)~the same plots with ellipses accounting for PDF correlations between the two cross sections.}
  \label{fig:wpandwm}
\end{figure}
In figure~\ref{fig:wandz} we show $W^\pm$ ($=W^++W^-$) versus $Z^0$ total cross sections and in figure~\ref{fig:wpandwm} we show $W^+$ versus $W^-$ total cross sections.  In figures~\ref{fig:wandz}(a,b) and \ref{fig:wpandwm}(a,b) the inner error bars represent ``PDF only'' uncertainties where possible, i.e.~with the exception of ABKM09 and GJR08/JR09, and the outer error bars represent ``PDF+$\alpha_S$'' uncertainties.  The ellipses in figures~\ref{fig:wandz}(c,d) and \ref{fig:wpandwm}(c,d) are drawn using eqs.~\eqref{eq:ellipseF} and \eqref{eq:ellipseG} with the appropriate correlation cosine, where the PDF+$\alpha_S$ uncertainties have been symmetrised when necessary.  For MSTW08, CTEQ6.6, CT10, CT10W, HERAPDF1.0 and HERAPDF1.5, the correlation cosine defining the error ellipse is calculated only from the eigenvector PDF sets for the central $\alpha_S$ value, since a definite prescription is not clear for the calculation of the correlation cosine in the presence of $\alpha_S$ uncertainties (and model and parameterisation uncertainties for HERAPDF1.0 and HERAPDF1.5).  We draw dotted lines where the ratio of cross sections, $W^\pm/Z^0$ or $W^+/W^-$, is constant.  It can be seen from the plots that the uncertainty in the ratio of cross sections is closely related to the correlation of the ellipse, also seen explicitly from eq.~\eqref{eq:uncertaintyR}.  In figures~\ref{fig:wandz}(b,d) and \ref{fig:wpandwm}(b,d) we also show the NLO predictions (open symbols and dashed lines) together with the corresponding NNLO predictions (closed symbols and solid lines) to explicitly demonstrate how the size of the NNLO corrections depends on the PDF choice.

Note that the $Z$ cross sections presented in this paper are calculated in a zero-width approximation and neglecting the virtual photon ($\gamma^*$) contribution, whereas the LHC experimental measurements include the $\gamma^*$ and are defined in a finite range of the invariant mass of the lepton pair from the $Z/\gamma^*\to\ell^+\ell^-$ decay.  We can calculate correction factors using the \textsc{vrap} program~\cite{Anastasiou:2003ds} with MSTW08 PDFs.  The results are the same at NLO and NNLO to three decimal places:
\begin{align*}
\text{ATLAS cuts~\cite{Aad:2010yt,ATLAS:WandZ}:}\quad\sigma(Z/\gamma^*, 66 < M_{\ell\ell} < 116~\text{GeV})\;/\;\sigma(Z\text{-only}, M_{\ell\ell}=M_Z) &= 0.991, \\
\text{CMS cuts~\cite{Khachatryan:2010xn,CMS:WandZ}:}\quad\sigma(Z/\gamma^*, 60 < M_{\ell\ell} < 120~\text{GeV})\;/\;\sigma(Z\text{-only}, M_{\ell\ell}=M_Z) &= 1.006,
\end{align*}
i.e.~there is a 1.5\% difference in the predictions for the ATLAS and CMS kinematic cuts.  Plots comparing to the preliminary ATLAS~\cite{ATLAS:WandZ} and CMS~\cite{CMS:WandZ} data can be found in ref.~\cite{LHCEWWG:April2011}.  The (dominant) luminosity uncertainty of 3--4\%~\cite{ATLAS:WandZ,CMS:WandZ} is comparable with the PDF spread and is much improved compared to the previous luminosity uncertainty of 11\% for the published cross sections~\cite{Aad:2010yt,Khachatryan:2010xn}, but given the preliminary nature we choose not to directly compare to data in this paper, particularly due to reasons associated with acceptance corrections which we explain below.

The $W^\pm$ and $Z^0$ total cross sections in figure~\ref{fig:wandz} are highly correlated, which can be understood by considering the dominant partonic contributions arising from $u$ and $d$ quarks, i.e.
\begin{equation} \label{eq:WoverZ}
  \frac{\sigma_{W^+}+\sigma_{W^-}}{\sigma_{Z^0}} \sim \frac{u(x_1)\,\bar{d}(x_2)+d(x_1)\,\bar{u}(x_2)}{0.29\,u(\tilde{x}_1)\,\bar{u}(\tilde{x}_2)+0.37\,d(\tilde{x}_1)\,\bar{d}(\tilde{x}_2)} \sim \frac{u(x_1)+d(x_1)}{0.29\,u(\tilde{x}_1)+0.37\,d(\tilde{x}_1)},
\end{equation}
where we have neglected the contributions with $q\leftrightarrow\bar{q}$, assuming that $q(x_1)\bar{q}(x_2)$ dominates over $\bar{q}(x_1)q(x_2)$, and the numerical values in the denominator are the appropriate sums of the squares of the vector and axial-vector couplings.  In the last simplification in eq.~\eqref{eq:WoverZ} we have assumed that $\bar{u}(x_2)\approx\bar{d}(x_2)$.  Here, the momentum fractions are
\begin{align}
x_1=(M_W/\sqrt{s})\,\exp(+y_W) \quad&\ge\quad x_2=(M_W/\sqrt{s})\,\exp(-y_W), \\
\tilde{x}_1=(M_Z/\sqrt{s})\,\exp(+y_Z) \quad&\ge\quad \tilde{x}_2=(M_Z/\sqrt{s})\,\exp(-y_Z),
\end{align}
and $y_W$ or $y_Z$ should be interpreted as some ``average'' rapidity appropriate for the (rapidity-integrated) total cross section.  The combination of $u$- and $d$-quark contributions is very similar (in numerical prefactors, in $x$ values, and in $Q^2=M_{W,Z}^2$ values) in both the numerator and denominator of eq.~\eqref{eq:WoverZ}, therefore the PDF dependence almost cancels in the $W^\pm/Z^0$ ratio.  The $W^+$ and $W^-$ total cross sections in figure~\ref{fig:wpandwm} are much less correlated, since
\begin{equation}
  \frac{\sigma_{W^+}}{\sigma_{W^-}} \sim \frac{u(x_1)\bar{d}(x_2)}{d(x_1)\bar{u}(x_2)} \sim \frac{u(x_1)}{d(x_1)},
\end{equation}
and therefore the $W^+/W^-$ cross-section ratio is a sensitive probe of the $u/d$ ratio and there is more variation between different PDF sets.  The preliminary CMS measurement of the $W^+/W^-$ total cross-section ratio with 36~pb$^{-1}$ of integrated luminosity (the full 2010 data set), combining the electron and muon channels, is~\cite{CMS:WandZ}
\begin{equation} \label{eq:cmsratio}
  \frac{\sigma_{W^+}}{\sigma_{W^-}} = 1.421\pm0.006({\rm stat.})\pm0.014({\rm syst.})\pm0.030({\rm th.}).
\end{equation}
Comparing to the predictions shown in figure~\ref{fig:wpandwm}, or more easily those in figure~\ref{fig:wpwmvsasmznnlo}, we see that the preliminary CMS measurement is in good agreement with MSTW08 and is more than 2-$\sigma$ below the central value of ABKM09.

\begin{figure}
  \centering
  \begin{minipage}{0.5\textwidth}
    (a)\\
    \includegraphics[width=\textwidth]{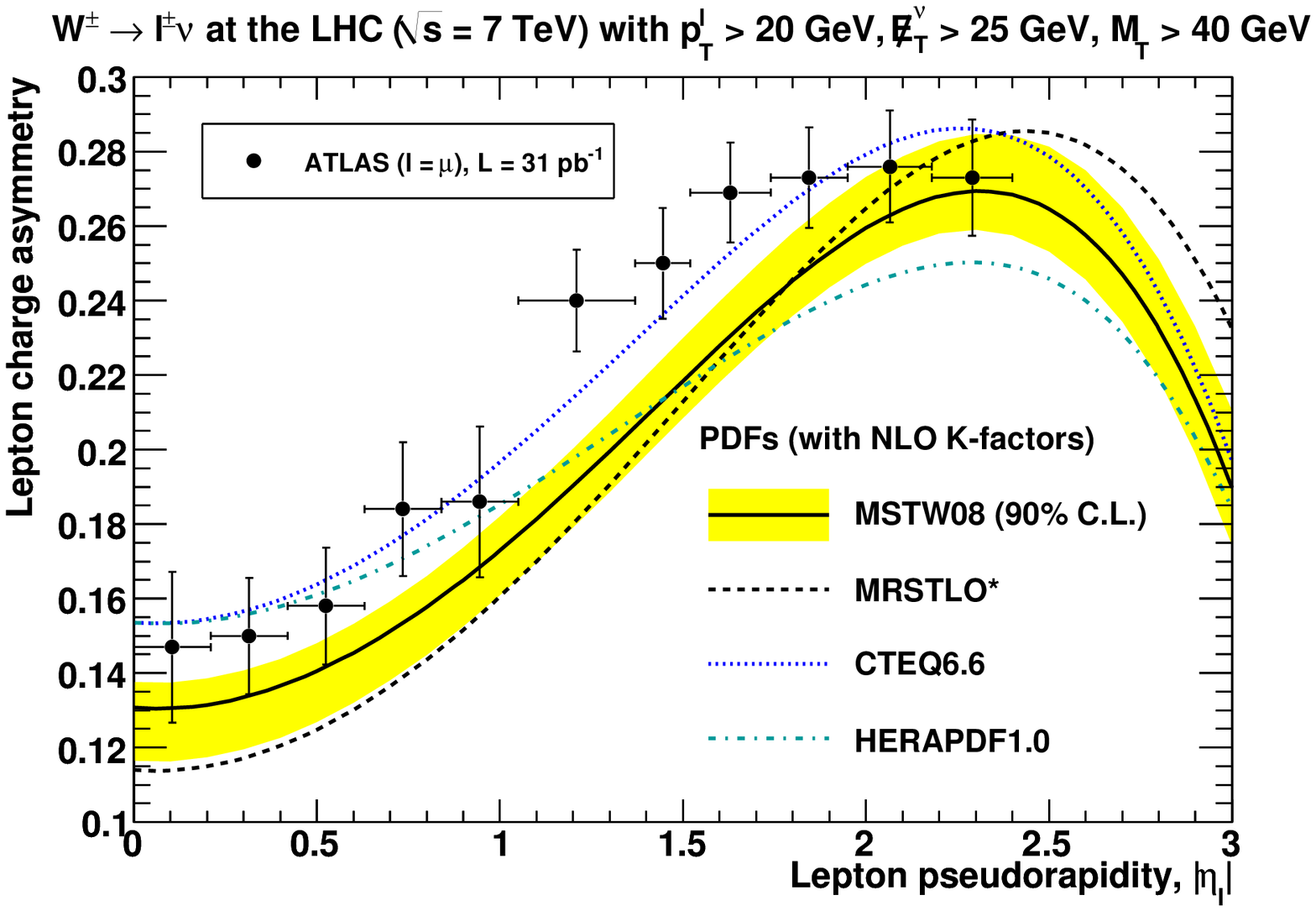}
  \end{minipage}%
  \begin{minipage}{0.5\textwidth}
    (b)\\
    \includegraphics[width=\textwidth]{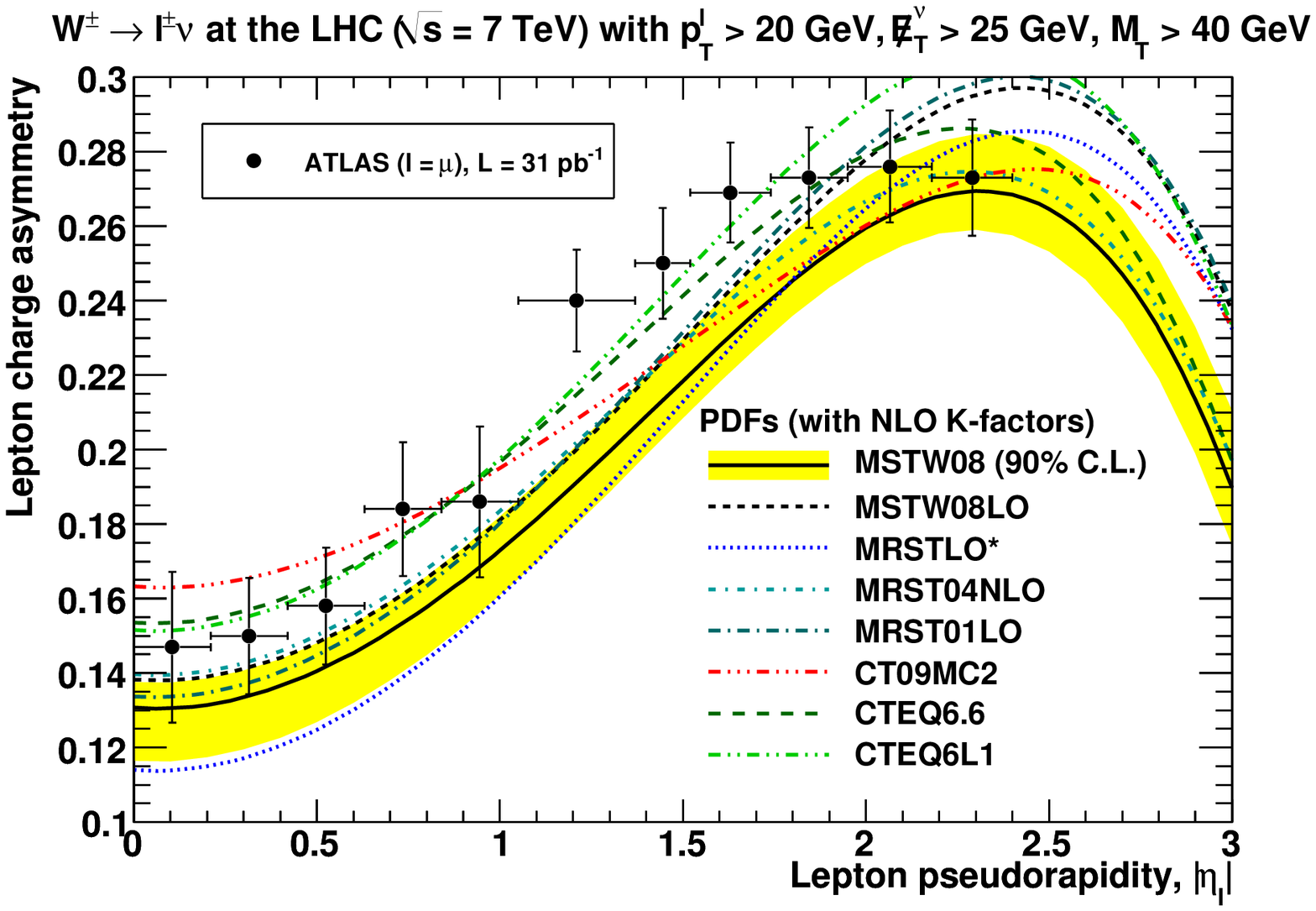}
  \end{minipage}
  \caption{ATLAS $W\to\ell\nu$ charge asymmetry~\cite{Aad:2011yn} compared to MSTW08 NLO PDFs with 90\% C.L.~uncertainty band.  (a)~Comparison also to MRST LO* (used as the central value for the ATLAS acceptance calculation~\cite{ATLAS:WandZ}), CTEQ6.6 and HERAPDF1.0 (used to evaluate the uncertainty~\cite{ATLAS:WandZ}).  (b)~Comparison to various other LO, LO* and NLO PDFs from MRST/MSTW and CTEQ/CT.}
  \label{fig:atlas}
\end{figure}

The (dominant) theory uncertainty in the preliminary CMS $W^+/W^-$ total cross-section ratio, eq.~\eqref{eq:cmsratio}, comes from extrapolating the measured (fiducial) cross section over the whole phase space, where the central value of the acceptance is obtained using the \textsc{powheg} Monte Carlo event generator with CT10 PDFs (with MSTW08 and NNPDF2.0 PDFs used to evaluate the uncertainty).  The corresponding ATLAS measurement of the $W^+/W^-$ ratio with 35~pb$^{-1}$ of integrated luminosity has not yet been explicitly presented.  The acceptance corrections for the central values of the individual preliminary cross-section measurements~\cite{ATLAS:WandZ} were obtained using \textsc{pythia} with MRST LO* PDFs~\cite{Sherstnev:2007nd}, and seem to give an anomalously low value of acceptance ratios leading to a high value of the central $W^+/W^-$ total cross-section ratio of 1.51~\cite{ATLAS:WandZ}, whereas alternative acceptance calculations would move the central ATLAS $W^+/W^-$ total cross-section ratio much closer to the CMS result~\cite{LHCEWWG:April2011}.  Moreover, the MRST LO* PDFs~\cite{Sherstnev:2007nd} give a poor description of the ATLAS data~\cite{Aad:2011yn} on the $W\to\ell\nu$ charge asymmetry inside the acceptance, worse than any other PDF set (see figure~\ref{fig:atlas}).  The $W\to\ell\nu$ charge asymmetry is defined differentially as a function of the pseudorapidity $\eta_\ell$ of the charged-lepton, i.e.
\begin{equation}
  A_\ell(\eta_\ell) = \frac{{\rm d}\sigma(\ell^+)/{\rm d}\eta_\ell-{\rm d}\sigma(\ell^-)/{\rm d}\eta_\ell}{{\rm d}\sigma(\ell^+)/{\rm d}\eta_\ell+{\rm d}\sigma(\ell^-)/{\rm d}\eta_\ell}.
\end{equation}
Here, common NLO $K$-factors have been applied for all curves in figure~\ref{fig:atlas}, but these have only a very small effect~\cite{LHCEWWG:April2011}.  The LO* notation denotes a modified LO fit with use of a NLO strong coupling and relaxation of the momentum sum rule~\cite{Sherstnev:2007nd}, the latter feature perhaps leading to less constraint on the relevant $u/d$ ratio at $x\sim M_W/\sqrt{s}\sim 0.01$.  The analogous CT09MC2~\cite{Lai:2009ne} prediction also differs substantially from the corresponding CTEQ6L1 and CTEQ6.6 predictions; see figure~\ref{fig:atlas}(b).  It is clear that the current data precision requires better acceptance calculations for the $W$ and $Z$ total cross sections and/or data to theory comparisons at the level of the fiducial cross section~\cite{LHCEWWG:April2011}.

Fiducial cross sections have been quoted by ATLAS~\cite{Aad:2010yt,ATLAS:WandZ} and CMS~\cite{Khachatryan:2010xn,CMS:WandZ}, but not for the cross-section ratios.  The LHCb experiment \emph{only} quote fiducial $W$ and $Z$ cross sections (and ratios)~\cite{LHCb:WandZ} given that extrapolation to total cross sections would not be meaningful due to the limited LHCb acceptance.  ``Fully exclusive'' calculations at NNLO, allowing direct computation of the fiducial cross section within leptonic cuts, are available in the form of the public \textsc{fewz}~\cite{Melnikov:2006kv} and \textsc{dynnlo}~\cite{Catani:2009sm} codes.  However, data-to-theory comparisons using these tools at the level of detail performed for total cross sections in this paper would be arduous (see refs.~\cite{Adam:2008ge,Adam:2008pc,Adam:2010tg} for some early NNLO \textsc{fewz} studies with leptonic cuts), due to the difficulty\footnote{This technical difficulty meant that it was not possible to compute accurate NNLO corrections using \textsc{fewz} for the Tevatron $W\to\ell\nu$ charge asymmetry data included in the MSTW 2008 NNLO fit~\cite{Martin:2009iq}.} in achieving sufficient numerical integration precision in a reasonable amount of CPU time, and the need for multiple independent runs to evaluate PDF uncertainties.  The situation has improved recently with the release of the updated \textsc{fewz} 2.0~\cite{Gavin:2010az}, with refined integration routines and automatic calculation of PDF uncertainties in a single run, but so far only for $Z/\gamma^*$ production.  After the similarly-improved \textsc{fewz} 2.0 for $W^\pm$ production is released, then detailed NNLO comparisons at the level of the fiducial cross section will be more practical, although the leptonic cuts are different for ATLAS and CMS (and LHCb), and for the electron and muon channels, so the data-to-theory comparisons would need to be performed separately for each experimental measurement.

The large spread of predictions for the $W^+/W^-$ ratio, and the related $W\to\ell\nu$ charge asymmetry, between different PDF sets is of some concern.  We recalculated the $W^+/W^-$ total cross-section ratios with only $u$ and $d$ quark (and antiquark) distributions to confirm that the same trend is seen.  To examine sources of tension we performed a variant of the MSTW08 NLO fit with a pseudo-data point added corresponding to the CTEQ6.6 value of the $W^+/W^-$ total cross-section ratio at the 7~TeV LHC (see figure~\ref{fig:wpwmvsasmznlo}) with a small uncertainty.  The data sets for which the fit quality most deteriorated were the Tevatron Run II $W\to \ell\nu$ asymmetry~\cite{Acosta:2005ud,Abazov:2007pm}, the NMC $\mu n/\mu p$ ratio~\cite{Arneodo:1996kd} and the E866/NuSea $pd/pp$ Drell--Yan ratio~\cite{Towell:2001nh}.  The CTEQ6.6/CT10 input parameterisation, and that of some other groups such as HERAPDF, \emph{assumes} the same small-$x$ behaviour of the $u$ and $d$ valence-quark distributions, whereas the MSTW08 NLO fit has independent small-$x$ powers, finding that at the input scale $Q_0^2=1$~GeV$^2$~\cite{Martin:2009iq}:
\begin{equation}
  xu_v(x,Q_0^2) \propto x^{0.29^{+0.02}_{-0.01}},\quad xd_v(x,Q_0^2)\propto x^{0.97^{+0.11}_{-0.11}}.
\end{equation}
The additional freedom in parameterisation may provide some explanation of the difference between MSTW08 and CTEQ6.6 at central pseudorapidities seen in figure~\ref{fig:atlas} and for the $W^+/W^-$ total cross-section ratio.  On the other hand, the preference of the ATLAS $W\to\ell\nu$ charge asymmetry data~\cite{Aad:2011yn}, and also of the CMS $W\to\ell\nu$ charge asymmetry data~\cite{Chatrchyan:2011jz}, seems to be for CTEQ6.6/CT10 rather than MSTW08.  This implies some tension between the LHC $W\to\ell\nu$ charge asymmetry data~\cite{Aad:2011yn,Chatrchyan:2011jz} and the data already included in the MSTW08 fit.  Other tensions have been observed with the precise Tevatron data on the $W\to\ell\nu$ charge asymmetry, and partially resolved by more flexible nuclear corrections for deuteron structure functions~\cite{Thorne:2010kj}.  Further attempts to resolve these tensions will be necessary for any future update of the MSTW08 fit and this will not be a straightforward endeavour.  Other than the LHC $W\to\ell\nu$ charge asymmetry data~\cite{Aad:2011yn,Chatrchyan:2011jz,LHCb:WandZ}, further PDF constraints from electroweak data will come from the $Z/\gamma^*$ rapidity distribution, already presented in preliminary form by CMS~\cite{CMS:Zrap} and LHCb~\cite{LHCb:WandZ}.

\section{Conclusions} \label{sec:conclusions}

We presented results on total cross sections versus $\alpha_S(M_Z^2)$ for the LHC ($\sqrt{s} = 7$ TeV) benchmark processes: $W^\pm$, $Z^0$, $t\bar{t}$, $gg\to H$~($M_H=120,180,240$~GeV).  This study was an extension and update of the results presented in ref.~\cite{Alekhin:2011sk}.  All plots (and more) are available from a webpage~\cite{mstwpdf}, which may be updated in future as more results become available.  Comparing predictions at the \emph{same} $\alpha_S(M_Z^2)$ value allows a fairer comparison, but there are still sizeable differences between results from some PDF groups, particularly those groups using more limited data sets and/or restrictive input PDF parameterisations.  Looking at partonic luminosities can often give a more global picture of PDF differences compared to total cross sections.  Similar differences are apparent by looking at rapidity distributions (see figures~11--13 of ref.~\cite{Alekhin:2011sk}).  We did not, in general, attempt to evaluate theory uncertainties other than those arising from PDFs and $\alpha_S$, which roughly decouple from other considerations such as scale dependence, therefore the results presented here will be valid also for more complete studies.

There is now reasonably good agreement between the NLO \emph{global} fits from MSTW08~\cite{Martin:2009iq}, CT10~\cite{Lai:2010vv} and NNPDF2.1~\cite{Ball:2011mu}, all using variants of a GM-VFNS to treat DIS structure functions.  There is more variation with other PDF sets, particularly at NNLO where MSTW08 is the only public \emph{global} PDF fit available.  The $t\bar{t}$ and $gg\to H$ cross sections are sensitive to the $gg$ luminosity and $\alpha_S(M_Z^2)$, but we can use Tevatron jet data to \emph{discriminate}~\cite{higgs}.  Fits not including Tevatron jet data should therefore be used with caution (if at all).  We discussed the recent PDF4LHC \emph{Interim Recommendations}~\cite{Botje:2011sn}.  A proposal for a simpler recipe is to use only MSTW08 with PDF+$\alpha_S$ uncertainties at 90\% C.L., which gives a common prescription at NLO and NNLO, with similar results to the ``envelope at 68\% C.L.'' prescription suggested by the PDF4LHC Steering Committee~\cite{Botje:2011sn}.  The 2010 Tevatron Higgs exclusion analysis~\cite{CDF:2010ar}, using $gg\to H$ cross sections calculated with only MSTW08 at 90\% C.L., is robust using exactly this prescription, although the more recent analysis~\cite{CDF:2011gs} instead uses the PDF4LHC recipe which gives similar results; see ref.~\cite{higgs} for more details.

We discussed the comparison of data and theory for $W$ and $Z$ production at the LHC.  The ATLAS and CMS data~\cite{ATLAS:WandZ,CMS:WandZ} are now reaching a level of precision where the theoretical uncertainty on the total cross sections due to the acceptance corrections is sizeable, particularly for cross-section ratios, and comparison at the level of the fiducial cross section would be desirable.  Nevertheless, the total cross section remains a convenient ``pseudo-observable'' for examining differences due to PDFs, even if it does not allow a direct comparison to be made between data and theory.  The more differential LHC $W\to\ell\nu$ charge asymmetry data~\cite{Aad:2011yn,Chatrchyan:2011jz,LHCb:WandZ}, defined only within fiducial cuts, will be addressed in more detail in a future study, together with data on the $Z/\gamma^*$ rapidity distribution~\cite{CMS:Zrap,LHCb:WandZ}.

\acknowledgments

I thank W.~J.~Stirling for providing the original cross-section codes used in this study, and A.~D.~Martin, W.~J.~Stirling and R.~S.~Thorne for useful comments at various stages.

\bibliographystyle{JHEP}
\bibliography{bench7TeV}

\end{document}